\RequirePackage{fix-cm}
\documentclass[natbib,smallextended]{svjour3}       
\smartqed  
\usepackage{graphics,graphicx,epstopdf,wrapfig,color}
\usepackage[colorlinks=true,linkcolor=blue,citecolor=blue,urlcolor=blue]{hyperref}
\usepackage{amssymb}
\usepackage{times}
\usepackage{comment}
\usepackage{caption}
\usepackage{xcolor}
\usepackage{rotating}
\usepackage{tabularx}
\usepackage{txfonts}
\usepackage{lscape}

\graphicspath{{./}{./figs/}}

\newcommand{\lsim}{\!\!\!\phantom{\le}\smash{\buildrel{}\over
 {\lower2.5dd\hbox{$\buildrel{\lower2dd\hbox{$\displaystyle<$}}\over
                                 \sim$}}}\,\,}
\newcommand{\gsim}{\!\!\!\phantom{\ge}\smash{\buildrel{}\over
{\lower2.5dd\hbox{$\buildrel{\lower2dd\hbox{$\displaystyle>$}}\over
                               \sim$}}}\,\,}

\def\msun{$M_{\odot}$}
\def\lsun{$L_{\odot}$}
\newcommand{\Lsun}{\mbox{$L_\odot$}}
\newcommand{\LTIR}{\mbox{$L({\rm TIR})$}} 
 
\usepackage{imakeidx}
\makeindex

\journalname{The Astronomy and Astrophysics Review}
\begin{document}

\title{Star formation and nuclear activity in luminous infrared galaxies: An infrared through radio review
}

\titlerunning{Star formation and nuclear activity in luminous infrared galaxies}

\author{Miguel P\'erez-Torres{}$^1$ \and Seppo Mattila$^2$ \and  Almudena Alonso-Herrero$^3$ \and Susanne Aalto$^4$ \and Andreas Efstathiou$^5$
}

\authorrunning{M. P\'erez-Torres, S. Mattila, A. Alonso-Herrero, S. Aalto, A. Efstathiou} 

\institute{$^1$ Instituto de Astrof\'isica de Andaluc\'ia, Consejo Superior de Investigaciones Cient\'ificas (CSIC), Glorieta de la Astronom\'ia s/n, E-18008, Granada, Spain\\
$^2$ Tuorla observatory, Department of Physics and Astronomy, University of Turku, FI-20014 Turku, Finland\\
$^3$  Centro de Astrobiolog\'ia (CAB, CSIC-INTA), ESAC Campus, E-28692 Villanueva de la Ca\~nada, Madrid, Spain \\
$^4$ Department of Earth and Space Sciences, Chalmers University of Technology, Onsala Space Observatory, SE-439 92, Onsala, Sweden\\
$^5$ School of Sciences, European University Cyprus, Diogenes Street, Engomi, 1516 Nicosia, Cyprus\\
\email{Miguel P\'erez-Torres: torres@iaa.csic.es, Seppo Mattila:  sepmat@utu.fi, Almudena Alonso-Herrero:
aalonso@cab.inta-csic.es, Susanne Aalto: susanne.aalto@chalmers.se, Andreas Efstathiou:
A.Efstathiou@euc.ac.cy} 
}

\date{Received: date / Accepted: date}

\maketitle

\begin{abstract}
Nearby galaxies offer unique laboratories allowing multi-wavelength spatially resolved studies of the interstellar medium, star formation and nuclear activity across a broad range of physical conditions. In particular, detailed studies of individual local luminous infrared galaxies (LIRGs) are crucial for gaining a better understanding of these processes and for developing and testing models that are used to explain statistical studies of large populations of such galaxies at high redshift for which it is currently impossible to reach a sufficient physical resolution. Here, we provide an overview of the impact of spatially resolved infrared, sub-millimetre and radio observations in the study of the interstellar medium, star formation and active galactic nuclei as well as their interplay in local LIRGs. We also present an overview of the modelling of their spectral energy distributions using state-of-the-art radiative transfer codes. These contribute necessary and powerful `workhorse' tools for the study of LIRGs (and their more luminous counterparts) at higher redshifts which are unresolved in observations. We describe how spatially-resolved time domain observations have recently opened a new window to study the nuclear activity in LIRGs. 
We describe in detail the observational characteristics of Arp 299 which is one of the best studied local LIRGs and exemplifies the power of the combination of high-resolution observations at infrared to radio wavelengths together with radiative transfer modelling used to explain the spectral energy distributions of its different components. We summarise the previous achievements obtained using high-spatial resolution observations and provide an outlook into what we can expect to achieve with future facilities.
\end{abstract}

\setcounter{tocdepth}{3} 
\tableofcontents

\section{Introduction} 
\label{sec:intro}

The first all-sky survey in the mid- and far-infrared (IR) with the Infrared
Astronomical Satellite (IRAS) led to the identification of a new class of galaxies
(e.g., \citealt{wright1984,soifer1984,houck1985}) emitting the bulk of their luminosity
in the IR, some of which were not included in previous optical catalogs because they
were too faint at optical wavelengths. Those IRAS galaxies emitted much more energy in
the infrared (IR$= 8-1000 \mu$m) than at all other wavelengths combined, and became
known as luminous and ultraluminous infrared galaxies (LIRGs: $10^{11}\, L_\odot \leq
L_{\rm IR} \le 10^{12}\, L_\odot$; and ULIRGs: $L{_{\rm IR}} \geq 10^{12}\, L_\odot$;
see e.g., \citealt{sanders03}). 

This research field has advanced enormously since the launch of IRAS in 1983, and a
number of reviews have summarized the progress over the years. For example,
\citet{soifer87} reviewed the impact of IRAS observations in the field of extragalactic
studies, while \citet{telesco88} reviewed the enhanced star formation in the centers of
galaxies from an IR perspective, \citet{sanders96} focused on the general properties of
U/LIRGs, and \citet{lonsdale06b} on local ULIRGs and their high redshift counterparts.
In this review, we focus on the star-formation and nuclear activity in nearby LIRGs, as
probed by imaging and spectroscopic observations at IR, sub-mm, and radio wavelengths,
with a special emphasis on spatially resolved observations and including also results
from time domain studies of local LIRGs.

The bulk of the energy radiated by U/LIRGs is IR emission from warm dust grains heated
by central power sources whose evolution is likely related: an active galactic nucleus
(AGN), a starburst, or both. Assuming that the dust reradiates the total bolometric
luminosity from a continuous burst for 10-100 Myr ages, with solar abundances and a
Salpeter initial mass function (IMF), the implied constant star formation rate, SFR, is
\citep{kennicutt98}

\begin{equation}
{\rm SFR} \approx 17.3 \left(\frac{L_{\rm IR}}{10^{11}\,L_\odot}\right) \, {\rm M_\odot \, yr^{-1}}
\label{eq:sfr_ir}
\end{equation}

Nearby galaxies are unique laboratories for studying the detailed physics of star
formation and AGN activity across a large range of physical conditions. They are the
workhorses for testing and applying physical models of star formation that are used in
galaxy evolution models to explain statistical studies of large populations of galaxies
at high redshifts, for which it is impossible to reach high physical resolution.  
However, even at a distance of 100 Mpc of the local U/LIRGs, a typical ground-based
seeing of $1''$ corresponds to a linear size of about 500 pc. Thus, to spatially resolve
their circumnuclear regions a sub-arcsecond resolution is a must. High-angular
resolution ($<0.1''$) observations can be currently reached either from space, e.g.,
with the Hubble Space Telescope (HST), and in the near future with the James Webb Space
Telescope (JWST\footnote{\url{https://www.jwst.nasa.gov/}}), or from the ground thanks
to adaptive optics (AO) in the IR (e.g., ESO VLT \footnote{\url{https://www.eso.org/}},
the Gemini \footnote{\url{http://www.gemini.edu}},
Keck\footnote{\url{http://www.keckobservatory.org}} and
Subaru\footnote{\url{https://subarutelescope.org/en/}} observatories). In particular,
the existing radio interferometric arrays routinely provide such sub-arcsecond angular
resolution (e.g., VLA\footnote{\url{http://www.vla.nrao.edu/}},
e-MERLIN\footnote{\url{https://www.e-merlin.ac.uk}},
ALMA\footnote{\url{https://www.almaobservatory.org/}},
NOEMA\footnote{\url{http://iram-institute.org/EN/noema-project.php}}), reaching
milliarcsecond angular resolution with the use of Very Long Baseline Interferometry
(VLBI; e.g., the EVN\footnote{\url{https://www.evlbi.org/}} and the
VLBA\footnote{\url{https://science.nrao.edu/facilities/vlba}}). The information provided
by spatially resolved observations is crucial not only to understand the physical
processes in the nearby LIRGs, but also for extrapolating our knowledge of local LIRGs
to the study of their high-redshift ($z \gtrsim 1$) counterparts. No less important is
our ability to complement high-spatial resolution studies of U/LIRGs with
state-of-the-art models of their radiation, and methods of fitting these models to the
data, as this will crucially inform similar studies at high-$z$.

\begin{figure}[htbp!]
\center
\includegraphics[width=11cm]{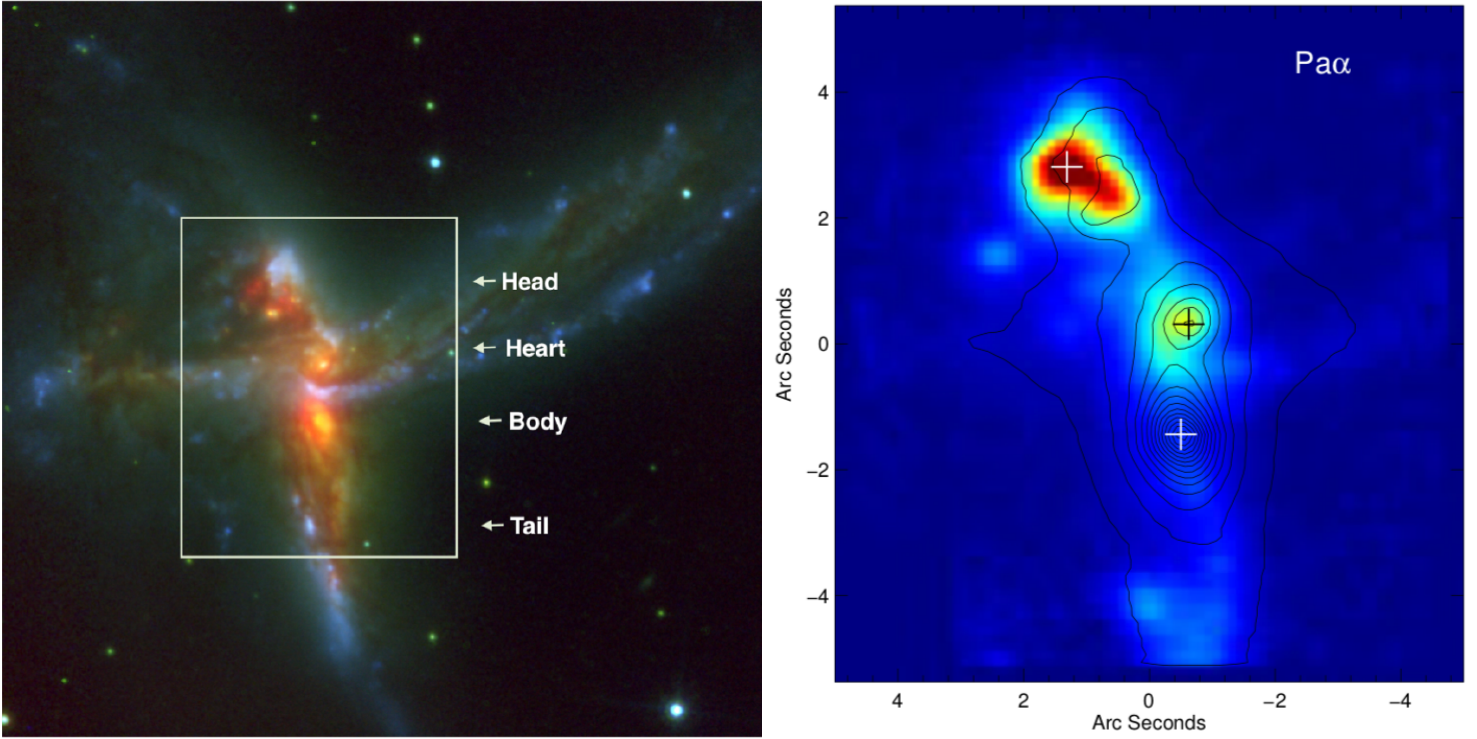}
\caption{\textbf{Left.} A colour combined near-IR VLT/NACO AO $K$-band ($\lambda \approx 2.2\,\mu{\rm m}$) + optical HST $B$ 
($\lambda \approx 430\,{\rm nm}$) and $I$ band image 
($\lambda \approx 800\,{\rm nm}$) of the LIRG IRAS 19115-2124 (dubbed the Bird) with the main components of the galaxy labeled. \textbf{Right.} VLT/SINFONI AO Pa$\alpha$ ($\lambda = 1.875\,\mu{\rm m}$) line map with the $K$-band continuum shown as contours, with the `Head' and `Heart' as the main components of the Pa$\alpha$ emission. The square in the left panel corresponds to the position of the Pa$\alpha$ line map and has a size of about 8$''$ $\times 10''$ where $1''\approx$1.0 kpc. Images from \citet{vaisanen2008,Vaisanen2017}.}
\label{fig:Birdimage}
\end{figure}

Fig.~\ref{fig:Birdimage} illustrates the power of such spatially resolved observations. The images show details of the nearby LIRG IRAS 19115-2124 with $\log (L_{\rm IR}/L_\odot)=11.91$ (for $D_L=206$\,Mpc), a triple merger whose components show signs of multiple stages of galaxy evolution, including the triggering and quenching of SF, and the appearance of AGN activity \citep{vaisanen2008,Vaisanen2017}. The authors noted that, if observed at high-$z$, the lack of the spatial information would have made this system look just like another clumpy starburst, potentially leading to a substantial misinterpretation of the ongoing processes therein.

\subsection{Surveys in the infrared and the submillimetre}
\label{back:surveys}

Following the first all-sky survey at IR wavelengths with IRAS, deeper surveys in
various fields of the sky with low galactic emission were carried out, resulting in the
discovery of LIRGs and ULIRGs at high-$z$. The two most important surveys carried out
with the Infrared Space Observatory (ISO) were  FIRBACK (Far Infrared BACKground survey;
\citealt{puget99}) and ELAIS (European Large Area ISO Survey; \citealt{oliver00}). The
ISO spectrophotometers also made a valuable contribution in understanding the physics of
luminous IRAS galaxies (e.g., \citealt{genzel98}). Almost concurrently with the ISO
surveys, important surveys from the ground were carried out at $850\,\mu{\rm m}$  with
the Submillimeter Common-User Bolometer Array (SCUBA) instrument mounted on the James
Clerk Maxwell Telescope (JCMT\footnote{\url{https://www.eaobservatory.org/jcmt/}}),
which exploited the negative k-correction in the sub-mm
\citep{smail97,hughes98,barger98,scott02,mortier05}. These surveys covered small fields
of the sky and led to the discovery of high-$z$ analogs (luminosity-wise) of local LIRGs
and ULIRGs, termed sub-mm galaxies (SMGs). 

The sensitivity of the Spitzer\footnote{\url{https://www.nasa.gov/mission_pages/spitzer/main/index.html}} Space Telescope made it possible to carry out much deeper surveys than with IRAS and ISO, such as the SWIRE (Spitzer Wide-Area Infra-red Extragalactic Survey) survey \citep{lonsdale03}. Spitzer also carried out very valuable spectroscopic surveys for a few thousand sources in the wavelength range $5$--$35\,\mu{\rm m}$ with the Infrared Spectrograph (IRS) instrument \citep{Houck2004}. The IRS spectra provided very valuable information about the physics of LIRGs and ULIRGs (e.g., \citealt{lebouteiller11}) at redshifts up to 4 \citep{riechers14}. The IR satellite AKARI\footnote{\url{https://www.cosmos.esa.int/web/akari}} carried out important surveys with a unique combination of mid-IR filters that were not available in previous missions \citep{murakami07}, and the surveys carried out with the Herschel\footnote{\url{https://sci.esa.int/web/herschel}} Space Observatory extended the wavelength coverage for millions of LIRGs and ULIRGs to the far-IR and sub-mm (e.g., H-ATLAS, \citealt{eales10}; PEP, \citealt{lutz11}; GOODS-Herschel, \citealt{elbaz11}; HerMES, \citealt{oliver12}). Finally, WISE (the Wide-field Infrared Survey Explorer, launched in 2009; \citealt{wright10}) has carried out all-sky near- to mid-IR surveys that have provided very useful data for LIRGS and ULIRGs. 

The task of combining photometry from different surveys is essential for understanding the populations of IR-bright galaxies, but it is far from trivial as we need to deal with the problem of source confusion in the far-IR and submillimetre. The Herschel Extragalactic Legacy Project (HELP\footnote{\url{https://herschel.sussex.ac.uk/}}) recently assembled the SEDs of about 170 million galaxies \citep{shirley19} by combining the data from all the main surveys carried out with Herschel plus data from a number of other surveys over 1270 square degrees. The HELP database, which is already public, promises to be a rich resource for studies of LIRGs and ULIRGs over most of the history of the Universe.

\subsection{Evolution, morphology, and environment}
\label{subsec:morphology}

\begin{figure}[htbp!]
\hspace{-2cm}
\includegraphics[width=16.cm]{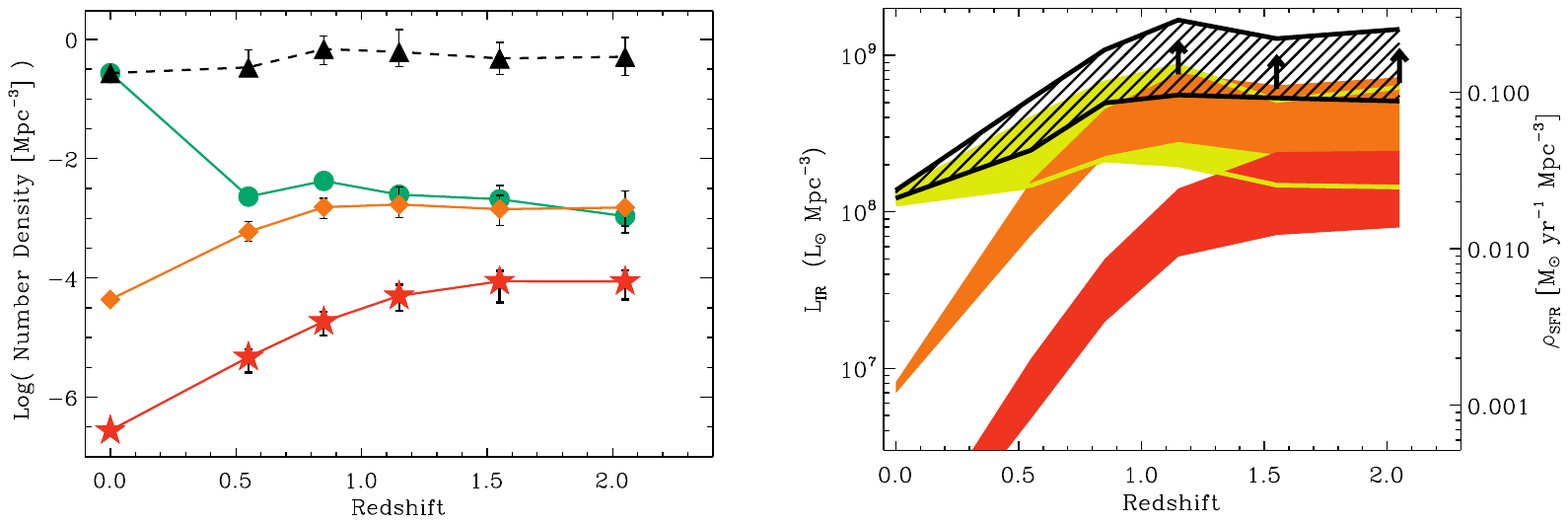}
\vspace{-15.5cm}
\caption{\textbf{Left.} Redshift evolution of the co-moving number density of ``normal'' galaxies $L_{\rm IR} <10^{11}\,L_\odot$ (black triangles), LIRGs (orange diamonds) and ULIRGs (red stars). Green circles represent the total number of galaxies which are above the 24\,$\mu$m
detection limit of the surveys presented in \citet{magnelli11}. The $z\sim0$ points are taken from \citet{sanders03}. \textbf{Right}. Co-moving IR luminosity density evolution out to $z\sim2.3$ (hatched area) and the relative contribution from normal galaxies (yellow filled area), LIRGs (orange filled area) and ULIRGs (red filled area). Image reproduced with permission from \cite{magnelli11}, copyright by ESO.}
\label{fig:magnelli11}
\end{figure}

\begin{figure}[htbp!]
\includegraphics[height=9.2cm]{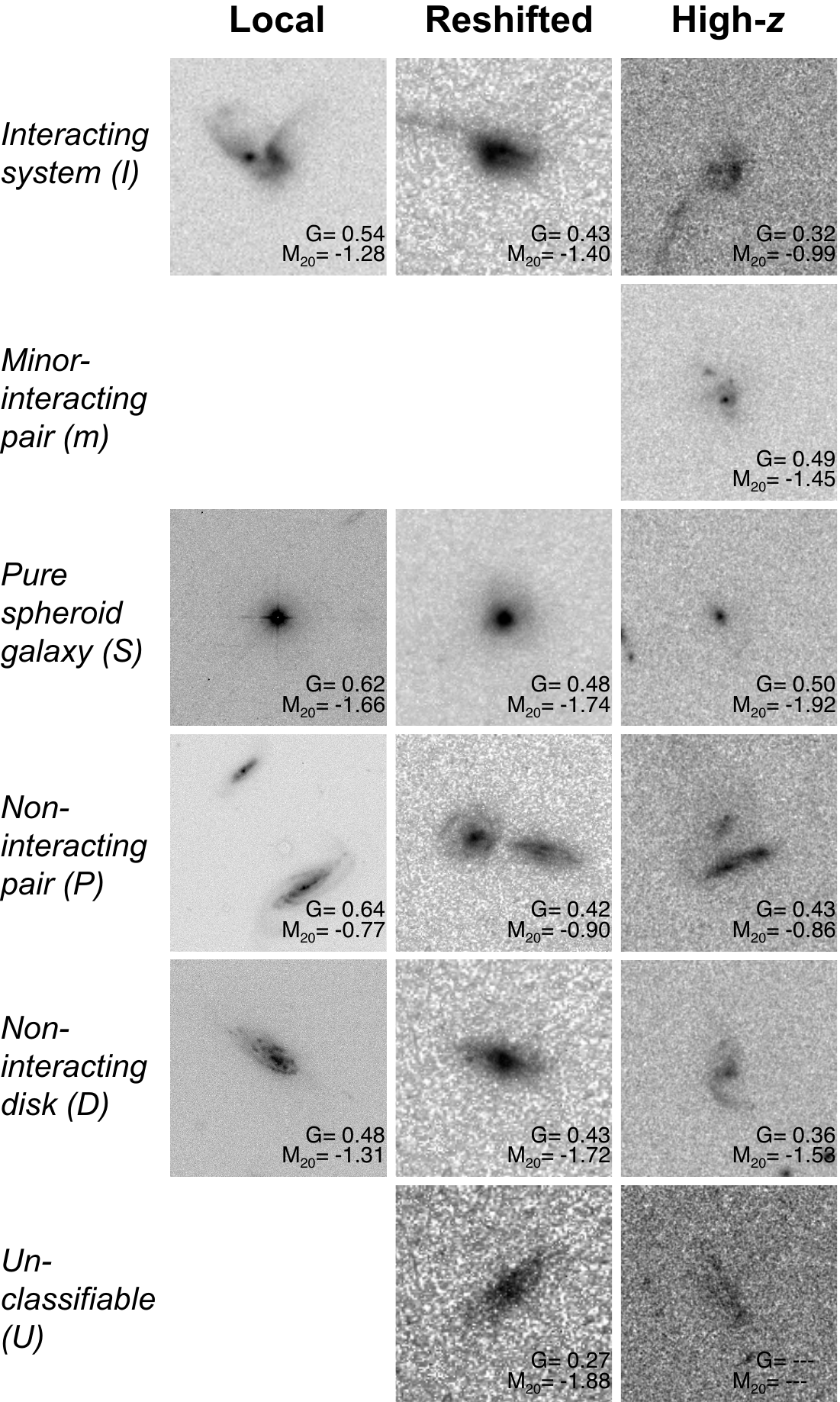}
\includegraphics[height=9.2cm]{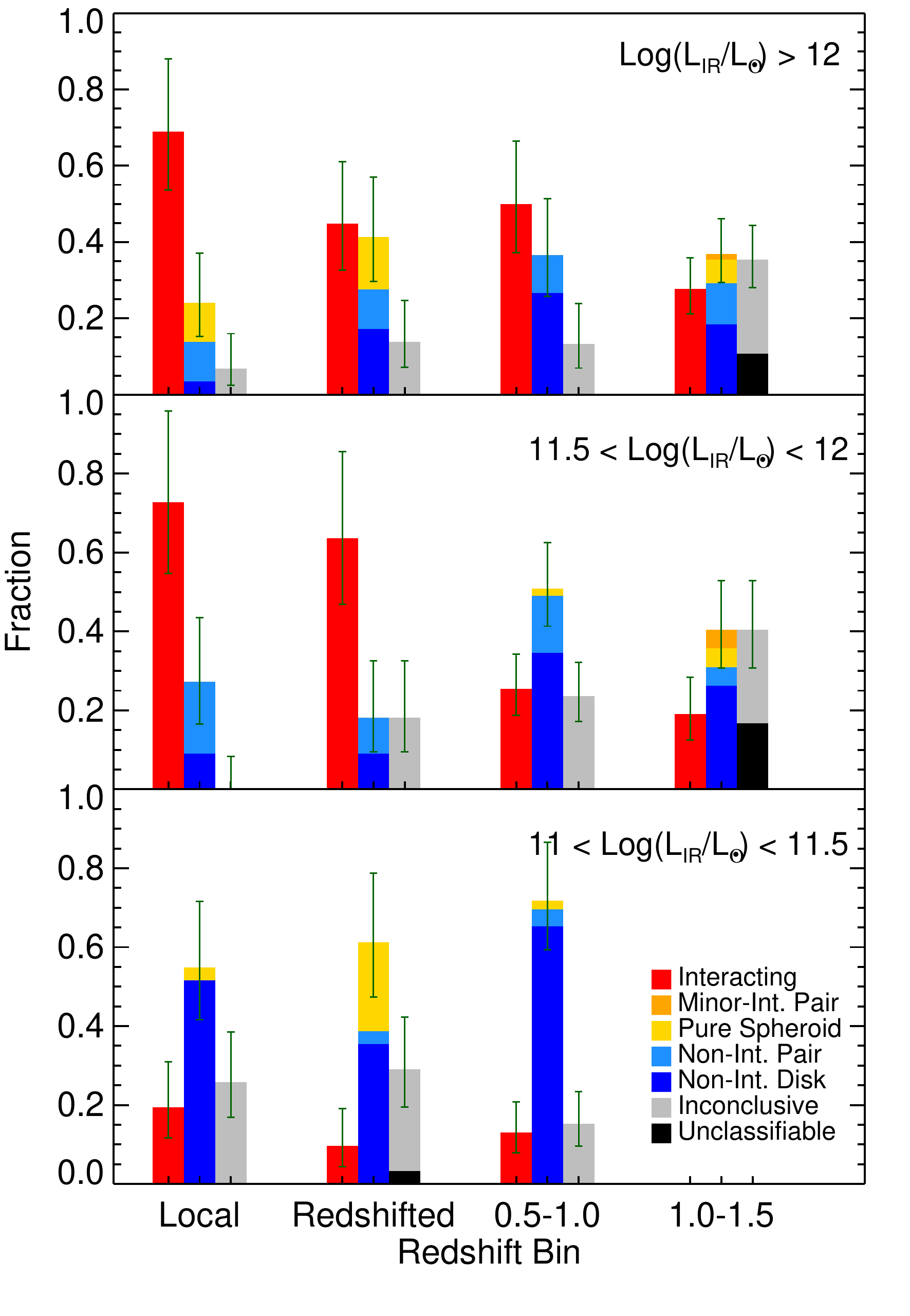}
\caption{\textbf{Left.} 
Examples of rest-frame, optical morphological types from visual classification within three (U)LIRG data sets: local ($z \lesssim 0.3$), redshifted (i.e., the local  galaxies have been artificially redshifted to $z=1$), and high-$z$ ($z\sim1$). For a similar analysis see also \citep{petty2014}. Each galaxy image encompasses a projected size of 50 kpc$\times$50 kpc. \textbf{Right.} Distribution of morphological types based on visual classification. Each sample is divided into three $L_{\rm{IR}}$ bins shown from top to bottom panels. Three vertical bars indicate the fraction of one or multiple morphological types: left bar--interacting galaxies (red); middle bar--sum of minor interacting pairs (orange), pure spheroidal galaxies (yellow), non-interacting pairs (light blue), and non-interacting disks (blue); right bar--sum of inconclusive (gray) and unclassifiable (black) cases. The error bars are determined assuming a Poisson distribution. Image reproduced with permission from \citet{Hung2014}, copyright by AAS.}
\label{fig:hung2014}
\end{figure}

The co-moving number density of LIRGs and ULIRGs in the Universe has experienced a
drastic decline, by about a factor of 40 (LIRGs) and 100 (ULIRGs) from $z\sim$1 to $z =
0$ (\citealt{Magnelli2009,magnelli11}; see also Fig.~\ref{fig:magnelli11}). LIRGs
contribute $\sim$50\% of the co-moving IR luminosity density at 0.5$< z\lesssim$1
\citep{PerezGonzalez2005,LeFloch2005,Caputi2007,Magnelli2009}, and start to dominate the
co-moving SFR density budget above $z \simeq 1.5$ (\citealt{magnelli11}; see also
Fig.~\ref{fig:magnelli11}).  Thus, there is evidence for a strong galaxy evolution, with
the IR-luminous sources dominating the star-forming activity in the Universe beyond $z
\sim$ 0.7 \citep{LeFloch2005}. There is also evidence for a strong morphological
evolution of the LIRG population: above $z \sim$0.5, roughly half of all LIRGs are
spirals, while at lower $z$ spirals account at most for one-third of the LIRG population
\citep{Bell2005,Melbourne2005}. LIRGs show also a clear transition in their
morphological properties with luminosity, independent of redshift. Below $L_{\rm IR}
\approx 10^{11.5}\, L_\odot$, most LIRGs are non-interacting galaxies, while above this
luminosity interacting/merging galaxies dominate, and this trend extends into the ULIRG
regime (\citealt{Hung2014}; see also Fig.~\ref{fig:hung2014}).

\begin{figure}[htbp]
\center
\includegraphics[width=11.5cm]{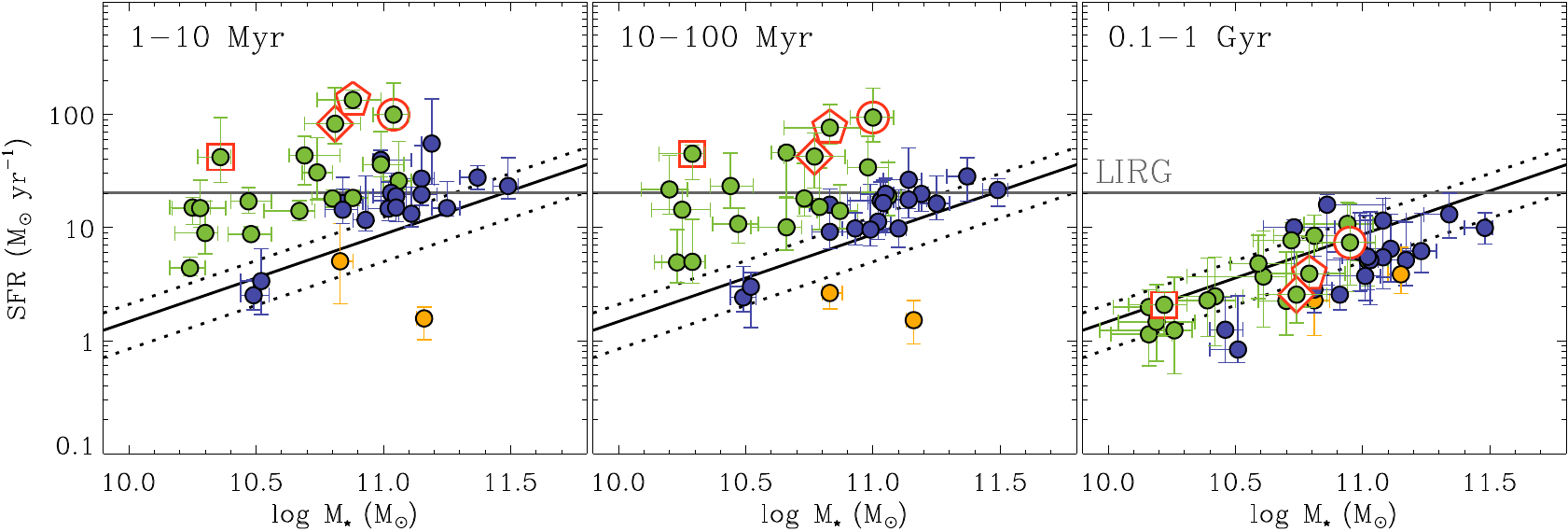}
\caption{SF properties of a sample of local galaxies with $L_{\rm IR} = 10^{10.2-11.8}\,L_\odot$. The SFR vs. stellar mass is shown in different stellar population age intervals: 1--10 Myr, 10--100 Myr and 0.1--1 Gyr. Galaxies with a constant star formation history (SFH) are shown in blue, a recent burst of SF in green and a decaying SFR in orange. The LIRGs NGC 1614, Arp 299, MCG+12-02-001 and NGC 6052 are indicated with a circle, pentagon, diamond and square, respectively. The solid and dashed black lines show the main sequence of star forming galaxies from \cite{elbaz2007} derived for local SDSS galaxies, and the horizontal line the approximate SFR needed to reach a log $L_{\rm IR} / L_\odot$ $>$ 11.0. Figure modified from \cite{PereiraSantaella2015}.}
\label{fig:pereira}
\end{figure}

Morphological studies of local LIRGs show that approximately 50\% are
interacting/mergers and the other 50\% are spiral galaxies with a high bar incidence,
indicating that bars may play an important role in triggering star formation in local
LIRGs \citep{Wang2006}. At high $z$, most large spirals have experienced an elevated SFR
similar to LIRGs, while the gas consumption results in a decline in the number of LIRGs
at low $z$ \citep{Bell2005,Melbourne2005}. Compared with high-$z$ LIRGs, local LIRGs
have already converted most of the gas into stars, and thus have lower specific SFRs,
smaller cold gas fractions, and a narrower range of stellar masses than their high-$z$
counterparts \citep{Wang2006,PereiraSantaella2015}. Fig.~\ref{fig:pereira} shows the SFR
averaged over different time intervals against the stellar mass today, 100 Myr and 1 Gyr
ago, for a sample of local galaxies in the $L_{\rm IR} = 10^{10.2-11.8}\,L_\odot$ range.
Whereas in the 0.1--1 Gyr range almost all of them follow the so called `Main Sequence'
of star forming galaxies, in the $<$ 100 Myr range those showing a burst of SF are
systematically above this relation with the most extreme cases (NGC 1614, Arp 299,
MCG+12-02-001 and NGC 6052) being mergers with high specific SFRs. While an important
fraction of local LIRGs are experiencing strong bursts of SF and are well above the Main
Sequence \citep{Wang2006,PereiraSantaella2015}, their more distant counterparts formed
stars in the Main Sequence mode \citep{elbaz2011}.

When considering local galaxies with $L_{\rm IR} > 10^{10}\,L_\odot$ \citet{hwang10}
found that the SFR, stellar mass, and specific SFR do not show significant changes with
the background density. On the other hand, \citet{tekola2012} found a strong correlation
between the surrounding galaxy densities and the IR luminosities and the resulting SFRs
of local galaxies with $L_{\rm IR} > 10^{11}\,L_\odot$ whereas below this threshold the
IR luminosity does not show any dependence on the environment. Also, for high-$z$ LIRGs,
those values increase as the background density increases, supporting the idea that
galaxy interactions and mergers play a critical role in triggering the star formation
activity in LIRGs and ULIRGs \citep{hwang10}.

\begin{figure}[htbp]
\center
\includegraphics[width=7.1cm]{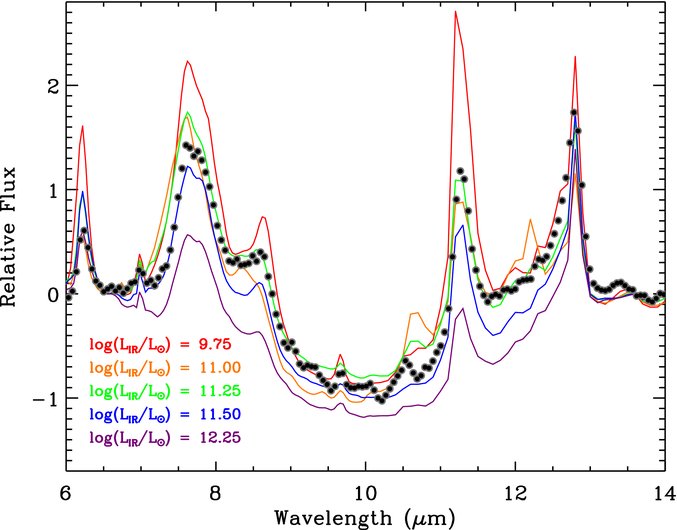}
\includegraphics[width=8.5cm]{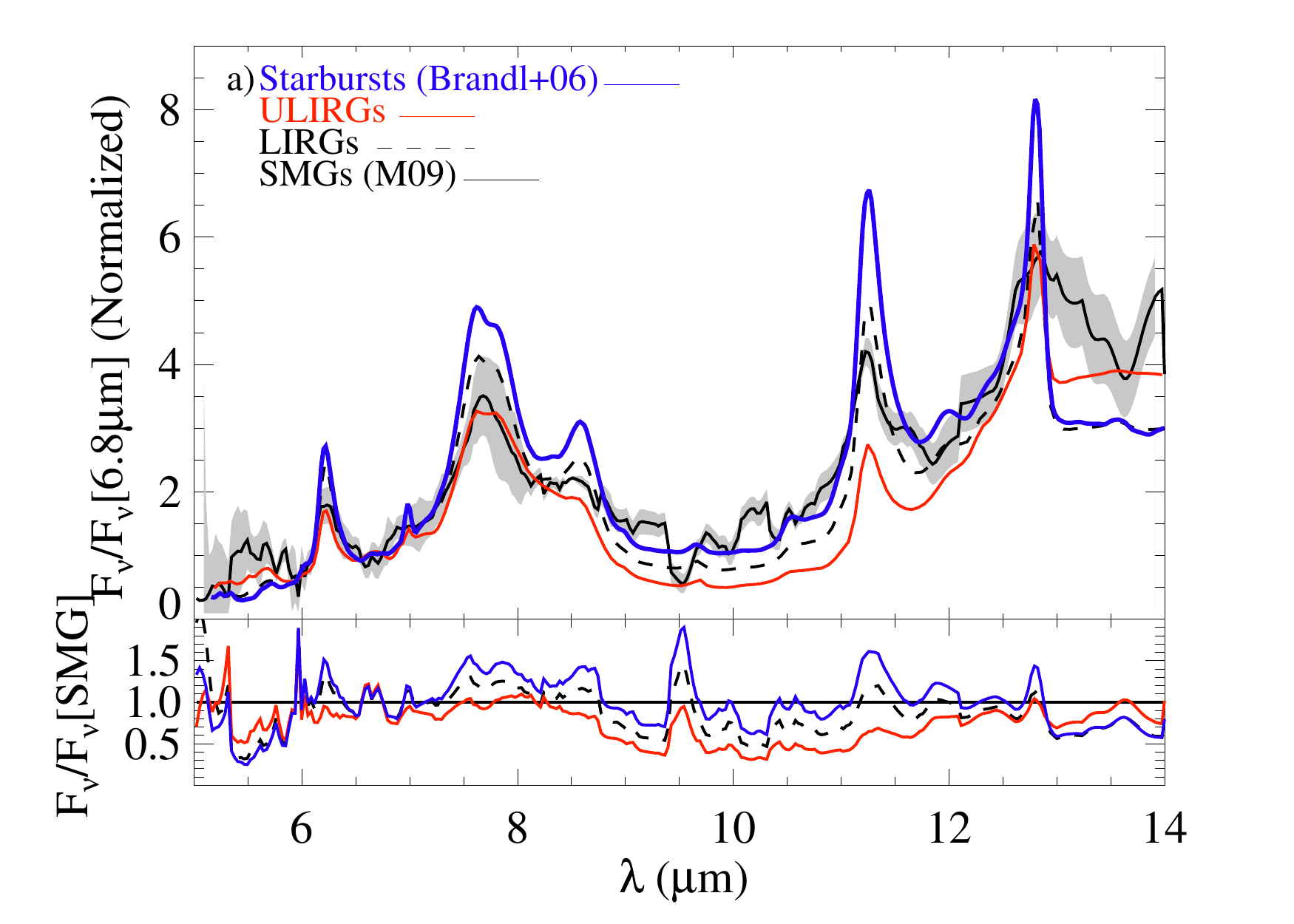}
\caption{\textbf{Top.} Stacked observed spectra of (U)LIRGs at $0.3 < z < 3.5$ with average $z = 1.1$ (circles) from
  \citet{Dasyra2009}, compared to the SED templates of \citet{Rieke2009} for different luminosities. High-$z$ ULIRGs 
  have spectral features consistent with local galaxies with
  \LTIR\  $\sim 10^{11.3}$ \Lsun where $L_{\rm TIR}$ is computed between $1$--$1000\,\mu{\rm m}$.
\textbf{Bottom.} Spitzer/IRS mid-IR spectra of local LIRGs and ULIRGs from 
GOALS compared to local starburst galaxies from \cite{Brandl2006} and $z\sim 2$ sub-mm galaxies from \cite{MenendezDelmestre2009}. Image reproduced with permission from \cite{Stierwalt2013}, copyright by AAS. In all the spectra, the most intense features are the PAH features at 6.2, 7.7, 8.7, and $11.3\,\mu{\rm m}$, and the [Ne\,{\sc ii}] emission line at $12.8\,\mu{\rm m}$. The $10\,\mu{\rm m}$ silicate feature is observed in absorption (see Sect.~\ref{sec:infrared} for more details).
}
\label{fig:MIRspectracomparison}
\end{figure}

Many works have shown that the rest-frame mid-IR spectra of high redshift ($z=0.3$ out
to $z\sim3.5$) ULIRGs and SMGs resemble those of local starburst galaxies with IR
luminosities in the range $10^{10} \le L_{\rm IR}/L_\odot \le 10^{11.5}$, (e.g.,
\citealt{Farrah2008,Dasyra2009,almudena09,Rujopakarn2011,Stierwalt2013}).  As can be
seen from Fig.~\ref{fig:MIRspectracomparison}, the observations from the Great
Observatories All-Sky LIRG Survey (GOALS\footnote{\url{http://goals.ipac.caltech.edu/}};
\citealt{Armus2009}) and the LIRG templates of \citet{Rieke2009} and high-z ULIRGs and
SMGs show  bright emission from polycyclic aromatic hydrocarbons (PAH). In local ULIRGs,
the PAH features are not as prominent and the silicate feature is deeper, indicating
higher extinctions (see Sect.~\ref{sec:infrared} for more details). This illustrates the
importance of understanding the physical processes taking place in local LIRGs, as a
first step to understanding the nature of the galaxies dominating the peak of the SFR
density.

\subsection{Modelling the spectral energy distribution of LIRGs}
\label{back:SEDmodels}

Accurate modelling of the spectral energy distributions (SED) of LIRGs is essential to
properly interpret the available multi-wavelength observations, from the ultraviolet to
the millimetre band. In particular, SED modelling is extremely useful to identify the
different sources of energy that power these galaxies in the case of unresolved
observations. However, the modelling is not trivial, as it must accurately account for
the emission of star-forming regions, AGN (often partly obscured by a dusty torus), and
of the general stellar populations of the galaxies. Furthermore, the effects of dust,
which is often found in a complex geometry, must be carefully considered.  However, the
payoff is huge: detailed SED modelling provides us with estimates of some of the most
relevant physical quantities of LIRGs: AGN fraction, stellar mass, SFR, and
core-collapse supernova (SN) rate, among others. 

\subsection{Time-domain observations}
\label{back:hiddenSNe}

It is not always realised that the luminosity and spectra of LIRGs can change on time scales as short as weeks or months. In fact, over the past couple of decades time-domain studies have revealed powerful SN factories in several nearby LIRGs where the SFRs correspond to SN rates a couple of orders of magnitude higher than observed in any ``ordinary'' spiral galaxies like the Milky Way where one SN is expected to explode in every $\sim$50 yr. LIRGs also often harbour at least one supermassive black hole (SMBH) and variability linked to the accretion onto this SMBH has been observed in the form of AGN variability and, more recently, also in the form of tidal disruptions of stars causing very luminous and long-lasting nuclear outbursts, especially prominent at IR and radio wavelengths.

\noindent 
The remainder of this paper is organized as follows: 
We concentrate on the observational IR properties of LIRGs 
in Sect.~\ref{sec:infrared},  and on the radio and on the  mm- and sub-mm properties in Sect.~\ref{sec:radio} and \ref{sec:molecular}, respectively, with an emphasis on high-angular resolution observations, which provide spatially resolved information. We then review the SED modelling of the unresolved IR emission of LIRGs in Sect.~\ref{sec:sed-modelling}. Despite using unresolved observations, SED modelling has the ability to characterize star-formation, nuclear activity, and the underlying quiescent host galaxy contribution, simultaneously, and is thus a very useful, complementary tool to resolved imaging. In Sect.~\ref{sec:timedomain}, we review time-domain studies of LIRGs with a particular emphasis on high angular resolution observations using radio interferometry and AO at IR wavelengths. In Sect.~\ref{sec:casestudies}, we focus on the nearby galaxy Arp~299 with $\log (L_{\rm IR}/L_\odot)=11.85$ (for $D_L=46\,$Mpc) as a case study of a well studied LIRG. Finally, we summarize the state of the field and provide an outlook for the studies with near-future facilities in Sect.~\ref{sec:summary}.  
In this review, when we quote specific values of the luminosity distance, $D_L$, and $L_{\rm IR}$ for a LIRG, we use the redshift, $z$, from NED, corrected to the reference frame defined by the 2.726~K CMB, and adopt $H_0 = 70$ km\,s$^{-1}$\,Mpc$^{-1}$ and a flat $\Lambda$\,CDM cosmology with
$\Omega_{\rm matter} = 0.308$ and 
$\Omega_{\rm vacuum} = 0.692$.

\section{Near-, mid- and far-IR observations of LIRGs} 
\label{sec:infrared}

\subsection{Stellar emission}\label{subsec:stellar} Near-infrared (near-IR) observations
have the advantage of reduced extinction when compared to the optical range which is
sensitive typically to $A_V\sim1$ mag. At $1.2\,\mu$m, $A_{\rm J} \sim 0.282 \times
A_{\rm V}$ and at $2.2\,\mu$m, $A_{\rm K} \sim 0.112 \times A_{\rm V}$
\citep{Rieke1985}. Therefore, near-IR observations allow the study of relatively
obscured regions in local LIRGs where the typical nuclear extinctions are $A_{\rm V}
\sim 4-5$ \citep{Goldader1997a, Scoville2000, AlonsoHerrero2006NICMOS,
PiquerasLopez2013, Tateuchi2015} and the galaxy-integrated extinctions are a few
magnitudes \citep{PereiraSantaella2015}. However, a small fraction of nuclei in local
LIRGs present higher extinctions $A_V>5\,$mag, as measured in the near-IR \citep[see
e.g.,][and also Sect.~\ref{subsubsec:silicates}]{Satyapal1999, almudena00,
AlonsoHerrero2006NICMOS}. Even higher extinctions ($A_V > 10\,$mag) are measured in some
LIRGs using mid-IR  (see Sect.~\ref{subsubsec:silicates})  and far-IR observations, and
thus these would appear completely hidden in the near-IR (see also the discussion in
Sect.~\ref{sec:con}).

Pre-HST detailed near-IR studies combined with observations at other wavelengths mostly
focused on bright, famous systems such as Arp~299, NGC~1614, NGC~3256, and NGC~7469
\citep{Doyon1994, Genzel1995, Satyapal1999, Puxley1999}. All these works found indeed
that young and intense episodes of SF are responsible for a large fraction of the IR
emission. However, an AGN, when present, can contribute a non-negligible fraction of the
IR emission (see also Sect.~\ref{subsubsec:AGN}).  In parallel, near-IR spectroscopy
studies with larger samples of LIRGs detected the bandheads of the CO absorption
features at $\simeq 2.3\,\mu$m in the majority of local LIRGs \citep{Ridgway1994,
Goldader1997b}. This was interpreted as evidence of the presence of red supergiants or
metal-rich giants produced in on-going/recent starburst episodes.  \cite{Goldader1997b}
also reproduced the IR luminosity, Br$\gamma$ emission and CO bandheads of local LIRGs
using starburst models.  The early imaging studies using NICMOS onboard the HST
\citep{Scoville2000, almudena00, AlonsoHerrero2001, AlonsoHerrero2002}  observed
\emph{again} some of the most famous local LIRGs. These observations revealed for the
first time and in great detail the nuclear morphologies of local LIRGs, including the
detection of nuclei obscured in the optical (see below) and a large number of star
clusters.

\begin{figure}
\center
\vspace{-1.5cm}
\includegraphics[width=8.cm, angle=-90]{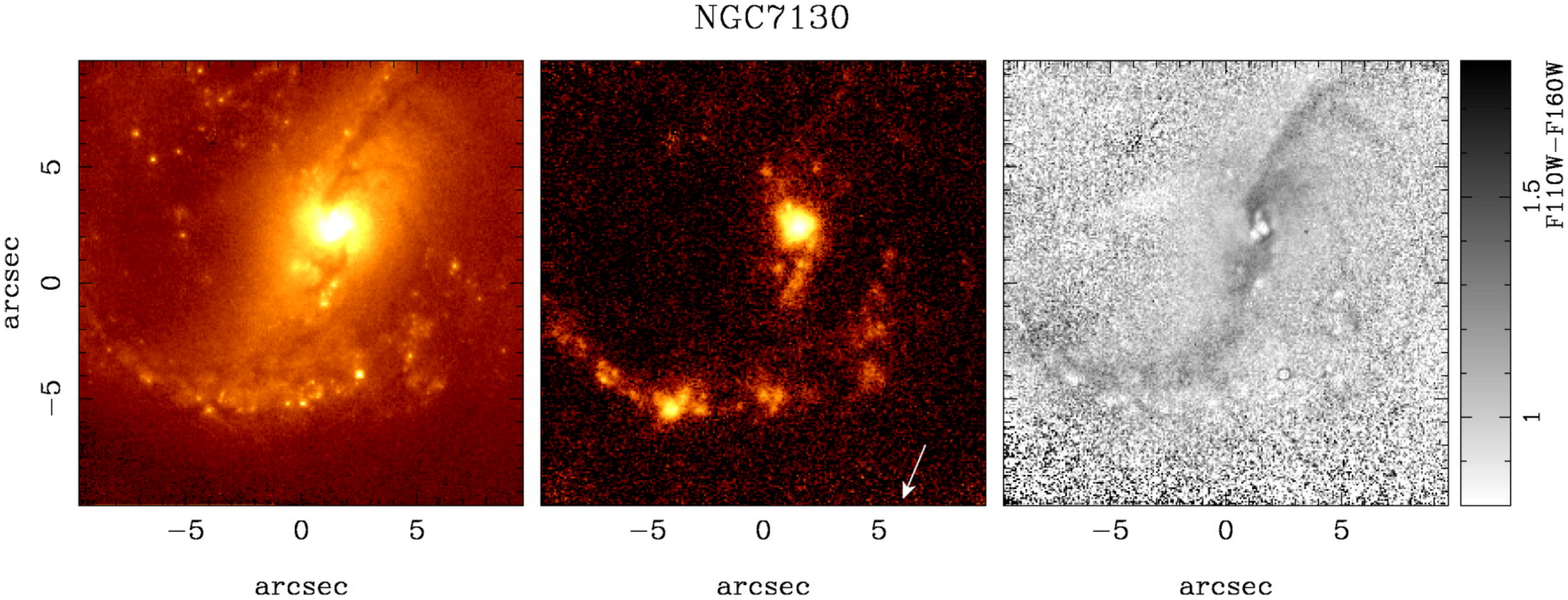}

\vspace{-3.5cm}
\includegraphics[width=8.cm, angle=-90]{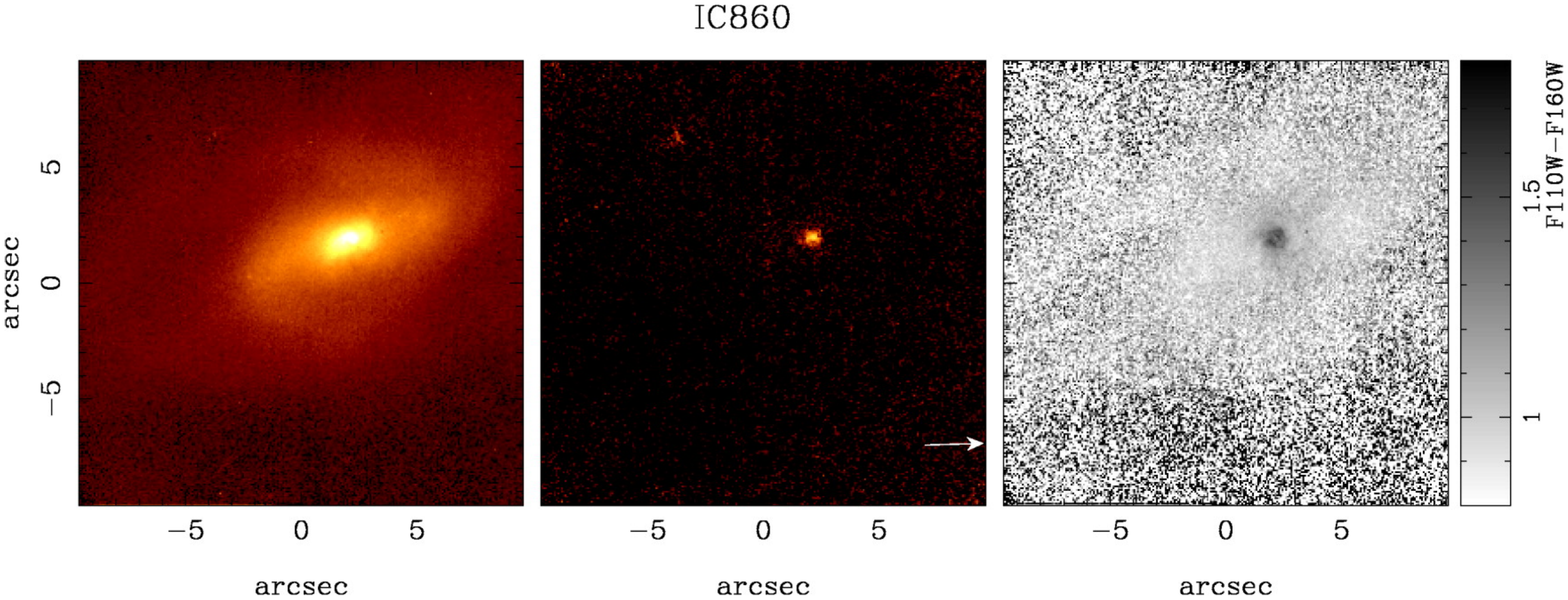}

\vspace{-3.5cm}
\includegraphics[width=8.cm, angle=-90]{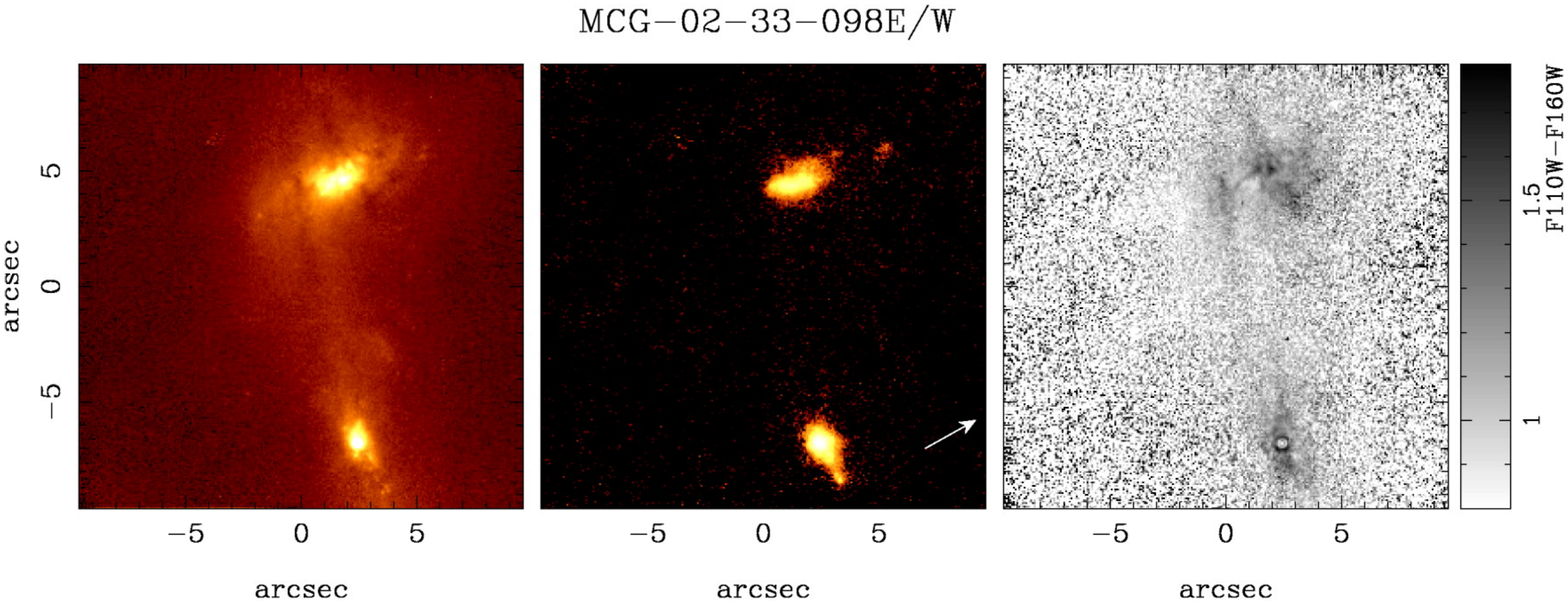}

\vspace{-2cm} \caption{Examples of different near-IR morphologies of three local LIRGs
  observed with HST/NICMOS. The images show the central $\sim 20''\times 20''$ (a few
  kpc at the distances of these galaxies) at $1.1\,\mu$m (stellar emission, left
  panels), Pa$\alpha$ emission line (middle panels), and near-IR color map (stellar
  extinction, right panels). The arrows indicate the north direction. This figure
  illustrates the diversity of Pa$\alpha$ morphologies observed in local LIRGs.
  NGC~7130 \textbf{(top)} is a spiral galaxy  classified as a Seyfert  with strong
  Pa$\alpha$ emission in the nuclear region and the spiral arms. IC~860
  \textbf{(middle)} only shows compact nuclear Pa$\alpha$ emission. MCG~-02-33-098
  \textbf{(bottom)} is an interacting system and bright Pa$\alpha$ emission arising from
  both galaxies.  As a reference, the average stellar extinctions derived from these
  color maps over the sampled fields of view are $A_V=2.9$ for NGC~7130, $A_V=3.2$ for
  IC~860 and $A_V=3.3$ for MCG~-02-33-098. Image reproduced with permission from
  \cite{AlonsoHerrero2006NICMOS}, copright by AAS.  } \label{fig:NICMOSmorphologies}
\end{figure}

\cite{AlonsoHerrero2006NICMOS} used NICMOS to obtain near-IR continuum and Pa$\alpha$
($\lambda=1.875\,\mu$m) observations (see also Section \ref{subsubsec:hydrogen}) of a
volume-limited sample (distances $\sim 35-75\,$Mpc) of local LIRGs drawn from the IRAS
Revised Bright Galaxy Catalog \citep{sanders03}. Thus, this sample is mostly composed of
LIRGs at the low luminosity end of the $L_{\rm IR}$ distribution with an average value
of $\log (L_{\rm IR}/L_\odot) = 11.25$. As discussed in the Introduction,
morphologically these galaxies are classified as  isolated, disky galaxies and galaxies
in groups, with only a small fraction being interacting/merger systems.
\cite{AlonsoHerrero2006NICMOS} observed a variety of morphologies of the
nuclear/circumnuclear near-IR continuum emission with typical physical resolutions of
tens of parsecs. They detected bright central point sources, which are more prevalent in
LIRGs classified as AGN, large numbers of star clusters in the spiral arms and rings of
SF, and prominent dust lanes as traced with near-IR color maps.
Figure~\ref{fig:NICMOSmorphologies} shows a few representative examples in this
volume-limited sample of local LIRGs. NGC~7130 with $\log (L_{\rm IR}/L_\odot)=11.33$
(for $D_L=66\,$Mpc) is a spiral galaxy that is optically classified as a Seyfert and
shows strong SF activity both on nuclear and circumnuclear scales over a few kpc. IC~860
with $\log (L_{\rm IR}/L_\odot)=11.06$ (for $D_L=52\,$Mpc) is an example of local LIRG
with compact and highly obscured nuclear activity \citep[see Section~\ref{sec:con} and
][]{aalto19}. MCG~-02-33-098 with $\log (L_{\rm IR}/L_\odot) 11$ (for $D_L=74\,$Mpc) is
an interacting pair of galaxies with optical classifications of composite (H\,{\sc
ii}/AGN) for the western nucleus and H\,{\sc ii} for the eastern nucleus.

\cite{Haan2011} also used HST/NICMOS imaging observations but focused on a sub-sample of
LIRGs and ULIRGs from GOALS\footnote{The GOALS sample contains 21 ULIRGs and 181 LIRGs
from the IRAS Bright Galaxy Sample \citep{sanders03} and is thus a large, statistically
complete sample of 202 local (distances of less than 400\,Mpc) IR-luminous galaxies.}
\citep[GOALS,][]{Armus2009}, which included the most luminous objects at $\log (L_{\rm
IR}/L_\odot)> 11.4$ at distances between 40 and 400\,Mpc. As such, this sample contains
mostly interacting and merger systems. They found a large fraction of double and even
triple nuclei merger systems. Moreover, about half of the near-IR identified double
nuclei LIRGs are classified as single nucleus LIRGs in the optical when observed at
similar resolutions with HST. Another result from this work was that the nuclear
separations in the interacting systems are larger in local LIRGs than ULIRGs (median
projected separations of 6.7\,kpc and 1.2\,kpc, respectively).

Optical and ultraviolet (UV) studies with HST have shown that the GOALS LIRGs contain a
large population of massive star clusters \citep[see][and references
therein]{Vavilkin2011} and with a fraction of them classified as super-star clusters
\citep[that is, very massive clusters, see e.g.,][]{Linden2017}. In general this
relatively unobscured population has ages ranging from a few Myr to hundreds of Myr.
The near-IR properties of (super)star clusters in local LIRGs have been studied for
instance with HST/NICMOS \citep{Scoville2000, almudena00, AlonsoHerrero2001,
AlonsoHerrero2002, AlonsoHerrero2006NICMOS} and with ground-based AO-assisted
observations \citep{Randriamanakoto2013, randriamanakoto2013b, Randriamanakoto2019}.
Constraining the ages and stellar masses of star clusters requires fitting
multi-wavelength SEDs that cover a broad spectral range. Including near-IR observations
helps constrain the age and extinction of the clusters  \citep[see for instance][for the
star clusters in the circumnuclear ring of SF of NGC~7469 a LIRG with  $\log (L_{\rm
IR}/L_\odot) =11.57$ for $D_L=65\,$Mpc]{DiazSantos2007}. Alternatively, color-color
diagrams can also provide reasonable estimates.  The typical values of the stellar
masses of the star clusters are between $10^4$ and $10^6\,M_\odot$. In general, only a
relatively small fraction of  star clusters in local LIRGs (approximately 20\%, on
average) are \emph{truly} young, with ages of less than 10\,Myr
\citep[see][]{AlonsoHerrero2002, Vavilkin2011, Linden2017}. For example,
\citet{Randriamanakoto2019} found that the majority of the star clusters in Arp~299 (see
Section~\ref{sec:casestudies} for a detailed discussion about the properties of this
galaxy) are very young, with ages $<15\,$Myr.

\citet{Randriamanakoto2013} found some evidence for a different slope of the near-IR
luminosity function of their near-IR selected star clusters in LIRGs compared to the one
observed for clusters in normal spiral galaxies. The authors interpreted this as due to
the effects of cluster disruption \citep[see also][]{AlonsoHerrero2002} in the
interaction/merger processes taking place in many LIRGs.

\subsection{Ionized gas}\label{sec:ionized}
\subsubsection{Hydrogen recombination lines}\label{subsubsec:hydrogen}

The near-IR and mid-infrared (mid-IR) spectral range contains hydrogen recombination
lines from the Paschen, Brackett and Pfund series that trace ionizing photons from AGN
and/or young, massive stars. In star-forming regions, when corrected for extinction, the
luminosities of the hydrogen recombination lines provide a direct estimate of the young
($< 10\,$Myr) SFR \citep[see e.g.,][for comprehensive reviews]{kennicutt98,
Kennicutt2012}.

\begin{figure}

\hspace{-1.5cm}
\includegraphics[width=15cm]{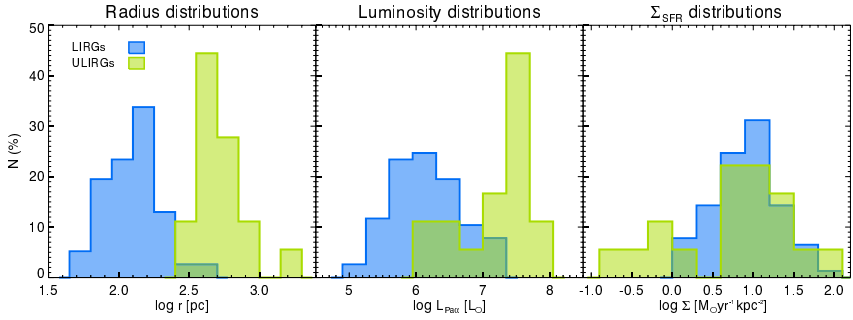}

\vspace{-15cm} \caption{Distributions of the properties of the individual star-forming
regions detected in local LIRGs (blue) and ULIRGs (green) as measured from VLT/SINFONI
Br$\gamma$ ($\lambda_{\rm rest}=2.166\,\mu$m for the LIRGs) and Pa$\alpha$ (for the
ULIRGs) observations. The {\bf left} panel shows the radii of the clumps, the {\bf
middle} panel the extinction-corrected Pa$\alpha$ luminosities, and the {\bf right}
panel the extinction-corrected SFR surface densities $\Sigma$. The observed differences
in the distributions of region radius and luminosity between LIRGs and ULIRGs are likely
due, at least in part, to resolution effects. In this work the typical physical
resolution for the LIRG sample was 200\,pc whereas it was 900\,pc for the ULIRG sample
\citep[see][for a detailed discussion]{PiquerasLopez2013}. The deficit of star-forming
regions in local LIRGs with $\Sigma < 1\,M_\odot \,{\rm yr}^{-1}\, {\rm kpc}^{-2}$ is
because the SINFONI field of view only samples the central few kpc of the galaxies where
the star forming regions tend to show high SFR surface densities. Image reproduced with
permission from \cite{PiquerasLopez2016}, copright by ESO.} \label{fig:HIIregions}
\end{figure}

The Pa$\alpha$ morphologies of local LIRGs vary with their luminosity. At the low
luminosity end, many relatively isolated LIRGs resemble spiral galaxies with a large
number of H\,{\sc ii} regions mostly distributed in the spiral arms of the galaxies. At
the high luminosity end, where interacting galaxies are more prevalent (see
Introduction), the H\,{\sc ii} regions are in some cases confined to the nuclear
regions. However, in other cases bright H\,{\sc ii} emission is also observed in the
spiral arms, in the overlap regions of interacting galaxies \citep{AlonsoHerrero2002,
AlonsoHerrero2006NICMOS, Tateuchi2015} and even in off-nuclear regions dominating the
total Br$\gamma$ ($\lambda = 2.166\,\mu$m) emission (for the case of the LIRG
IRAS~19115-2124, the `Bird', also discussed in Sect. \ref{sec:intro} and Fig.
\ref{fig:Birdimage}, see \citealt{vaisanen2008,Vaisanen2017}). The emitting regions in
LIRGs are distributed across the galaxies covering regions with sizes between
$\lessapprox$1\,kpc to approximately 10--20\,kpc. These sizes are measured directly from
Pa$\alpha$ narrow-band imaging (see a few examples in Fig.~\ref{fig:NICMOSmorphologies})
with HST \citep{AlonsoHerrero2002, AlonsoHerrero2006NICMOS, Larson2019} and from the
ground \citep{Tateuchi2015} as well as from integral field spectroscopy
\citep{PiquerasLopez2012}. The sizes of the emitting regions are thus intermediate
between those of normal star-forming galaxies and local ULIRGs. \cite{Rujopakarn2011}
measured typical sizes of the emitting regions in local LIRGs and ULIRGs of $\sim
1\,$kpc. However, their LIRG sample is not fully representative of the observed $L_{\rm
IR}$ distribution. Indeed, they did not include LIRGs at the low end of the L$_{\rm IR}$
distribution which clearly show SF extending over physical scales well beyond the
central kpc \citep[see, e.g., the H$\alpha$+NII and MIPS $24\,\mu$m images
in][]{PereiraSantaella2015}.

The individual star-forming regions in local LIRGs have sizes
 ranging from $<$100\,pc  to a few hundred parsecs (see Figure~\ref{fig:HIIregions}). They are young  \citep[ages of $<$6--7\,Myr,][]{almudena00, AlonsoHerrero2001, DiazSantos2008, Vaisanen2017, 
 Larson2019}
and luminous, with a fraction of them having luminosities similar to giant H\,{\sc ii} regions \citep[see][]{AlonsoHerrero2002,
AlonsoHerrero2006NICMOS}. They are thus
 larger and more luminous than the H\,{\sc ii} regions of local star-forming
spiral galaxies and might be related to the formation of super
star clusters \citep{AlonsoHerrero2002}.  When compared with local
ULIRGs, the individual star-forming regions of LIRGs are smaller and less
luminous. However, these star-forming regions display similar SFR surface densities in LIRGs and
ULIRGs, covering quite a
large range from $\sim 0.1$ to $100\,M_\odot \, {\rm yr}^{-1} \,{\rm kpc}^{-2}$ 
\citep[see][and Fig.~\ref{fig:HIIregions}]{PiquerasLopez2016}.

\subsubsection{Fine structure lines and H$_2$ lines}\label{subsubsec:fine}
The IR spectral range contains a large number of fine structure lines  with a broad range of
ionization potentials (IP) and critical densities \citep[see table~1 in][]{Spinoglio2017}. There are also molecular hydrogen lines tracing hot ($T\sim 2000$\,K) gas with the rot-vibrational lines in the near-IR and warm gas ($T\sim 100$--$500$\,K) with the mid-IR rotational lines.
All these lines can be used to trace (or distinguish between) SF activity, AGN emission, and shocks, as we discuss in this section.

\begin{figure}

\hspace{-2cm}
\includegraphics[width=16.cm]{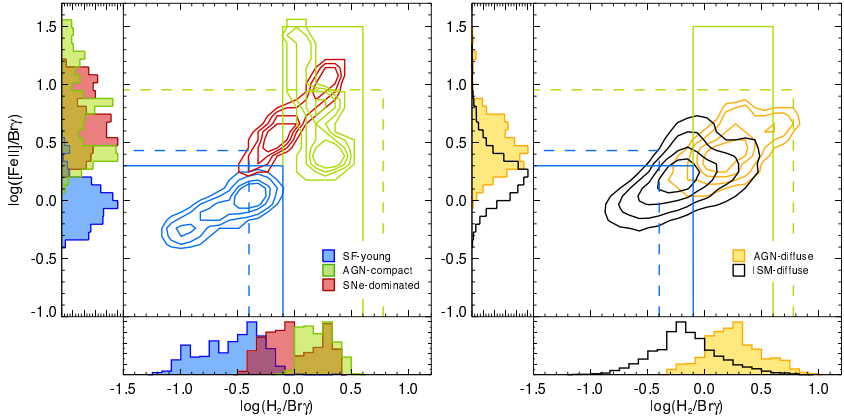}

\vspace{-15cm}
\caption{Near-IR BPT-like diagrams based on IFS observations of three representative  local LIRGs obtained with VLT/SINFONI, namely, IC~4687 a SF-dominated LIRG  with $\log (L_{\rm IR}/L_\odot)=11.44$ (for $D_L=75\,$Mpc), NGC~7130 a composite AGN-SF LIRG, and NGC~5135 another composite AGN-SF LIRG dominated by SN activity with $\log (L_{\rm IR}/L_\odot)=11.33$ (for $D_L=64\,$Mpc). \textbf{Left.} Compact, high surface brightness young star-forming clumps (blue), aged, SN-dominated clumps (red), and compact nuclear AGN (green).  \textbf{Right.} Diffuse AGN emission (yellow) and general diffuse medium (black).  Image reproduced with permission from \cite{Colina2015}, copright by ESO. The solid blue and green lines mark upper limits for young star-forming regions and Seyferts identified in this work, compared to previously defined limits by other works as dashed lines.}
\label{fig:nearIRBPT}
\end{figure}

\cite{Colina2015} used near-IR IFS observations obtained with VLT/SINFONI to map the
two-dimensional ionization structure in a sample of local LIRGs. They proposed a
BPT-like diagram (see Fig.~\ref{fig:nearIRBPT}) involving Br$\gamma$  and the [Fe\,{\sc
ii}] line at $1.64\,\mu$m \citep[although other hydrogen recombination and Fe$^+$ lines
can also be used, see e.g.,] []{Vaisanen2017} as well as the molecular hydrogen H$_2$
line at $2.12\,\mu$m.  (A BPT-diagram is a set of nebular emission line diagrams used to
distinguish the ionization mechanism of nebular gas, widely used as an AGN diagnostic,
after the classical work of \citealt{BPT81}).  In high surface brightness regions, these
line ratios can be used to determine the dominant excitation mechanism. Young (ages less
than approximately 6\,Myr) SF formation dominated regions have both low [Fe\,{\sc
ii}]/Br$\gamma$ and H$_2$/Br$\gamma$ line ratios.  AGN-dominated and SN-dominated
regions have higher [Fe\,{\sc ii}]/Br$\gamma$ line ratios with the latter mentioned
regions showing H$_2$/Br$\gamma$ line ratios that are intermediate between those of SF
regions and AGN. We refer the reader to  \cite{Colina2015} for the exact line ratio
ranges. However, local LIRGs also show regions with diffuse extended interstellar medium
(ISM) emission in the same regions of this diagram, as defined by the three excitation
mechanisms. \cite{Colina2015} concluded that in local LIRGs the radiation/heating field
is due to a combination of different ionization sources, weighted by their relative flux
contribution and spatial distribution.

The [Ne\,{\sc ii}] line at $12.8\,\mu$m and the [Ne\,{\sc iii}] line at $15.6\,\mu$m are
the brightest mid-IR lines in SF-dominated regions in LIRGs and can be used to obtain an
estimate of the SFR in galaxies \citep[see for instance, ][for a SFR calibration using
the sum of both lines]{Ho2007}. Indeed,  in the GOALS sample,  the neon line
luminosities are well correlated with other SFR indicators such as the IR luminosity and
the $24\,\mu$m monochromatic luminosity \citep{Inami2013}.  \cite{PereiraSantaella2010}
mapped  with Spitzer/IRS the [Ne\,{\sc ii}] and [Ne\,{\sc iii}] line emission over the
central $20''-30''$ regions of a sample of local LIRGs. The emission from both lines
extends over several kpc, confirming the results obtained with hydrogen recombination
lines such as H$\alpha$ and Pa$\alpha$ (Section~\ref{subsubsec:hydrogen}).  Ratios of
emission lines with relatively low or intermediate IP (e.g., [Ne\,{\sc ii}], [Ne\,{\sc
iii}],  [S\,{\sc iii}] at $18.7\,\mu$m, [S\,{\sc iv}] at $10.5\,\mu$m, [Ar\,{\sc ii}] at
$6.99\,\mu$m and [Ar\,{\sc iii}] at $8.99\,\mu$m) can be predicted with photoionization
models for young stellar populations \citep[see e.g.,][]{Rigby2004, Snijders2007} for
different ages, metallicities, ionization parameters and electron densities. Comparisons
of model predictions with mid-IR observations
\citep{almudena09,PereiraSantaella2010,Inami2013} constrained the ages (a few Myr, as
also found from hydrogen recombination lines), metallicities ($\sim 1$--$2\,Z_\odot$),
electron densities (median value of $300\,{\rm cm}^{-3}$) and ionization parameters
(range $2$--$8\times10^7\,{\rm cm \,s}^{-1}$) of star-forming regions in LIRGs
\citep[see also][for similar results using far-IR lines, and below]{DiazSantos2017}. 

 \cite{PereiraSantaella2010} showed that 
in local LIRGs, ratios involving relatively low IP emission lines (e.g., the [Ne\,{\sc iii}]/[Ne\,{\sc ii}] and 
[S\,{\sc iii}]/[Ne\,{\sc ii}]) show spatial variations on scales of a few tens of arcseconds, corresponding to a few kpc at the distances of their LIRGs. In particular, in 
SF-dominated LIRGs these ratios are lower in the nuclei than at larger
galactocentric distances. 
At a given metallicity, ionization parameter, and density,
the [Ne\,{\sc iii}]/[Ne\,{\sc ii}] line ratio is predicted to decrease as the age of the
\emph{young} stellar population increases. Thus, in local LIRGs this  would imply that the nuclear
regions have older stellar
populations than the circumnuclear regions. Another possibility owing to the
high densities in the nuclear regions in some LIRGs
\citep{almudena09} 
is that this ratio is suppressed, including
the possibility that the massive stars are hidden in ultra-compact H\,{\sc ii} regions \citep{Rigby2004}.

\begin{figure}

\vspace{-3cm}
\includegraphics[width=12.0cm]{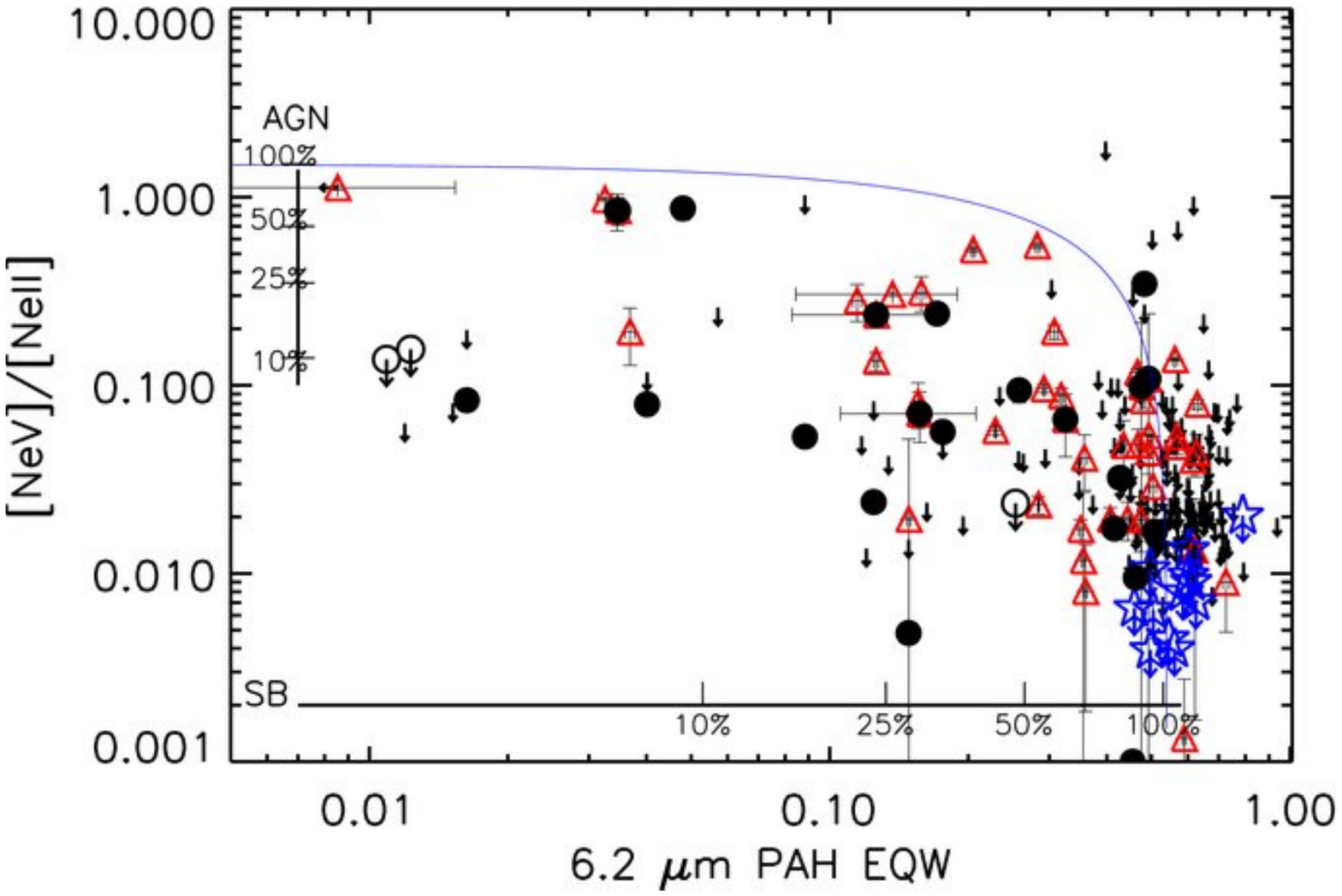}

\vspace{-9cm}
\includegraphics[width=12.0cm]{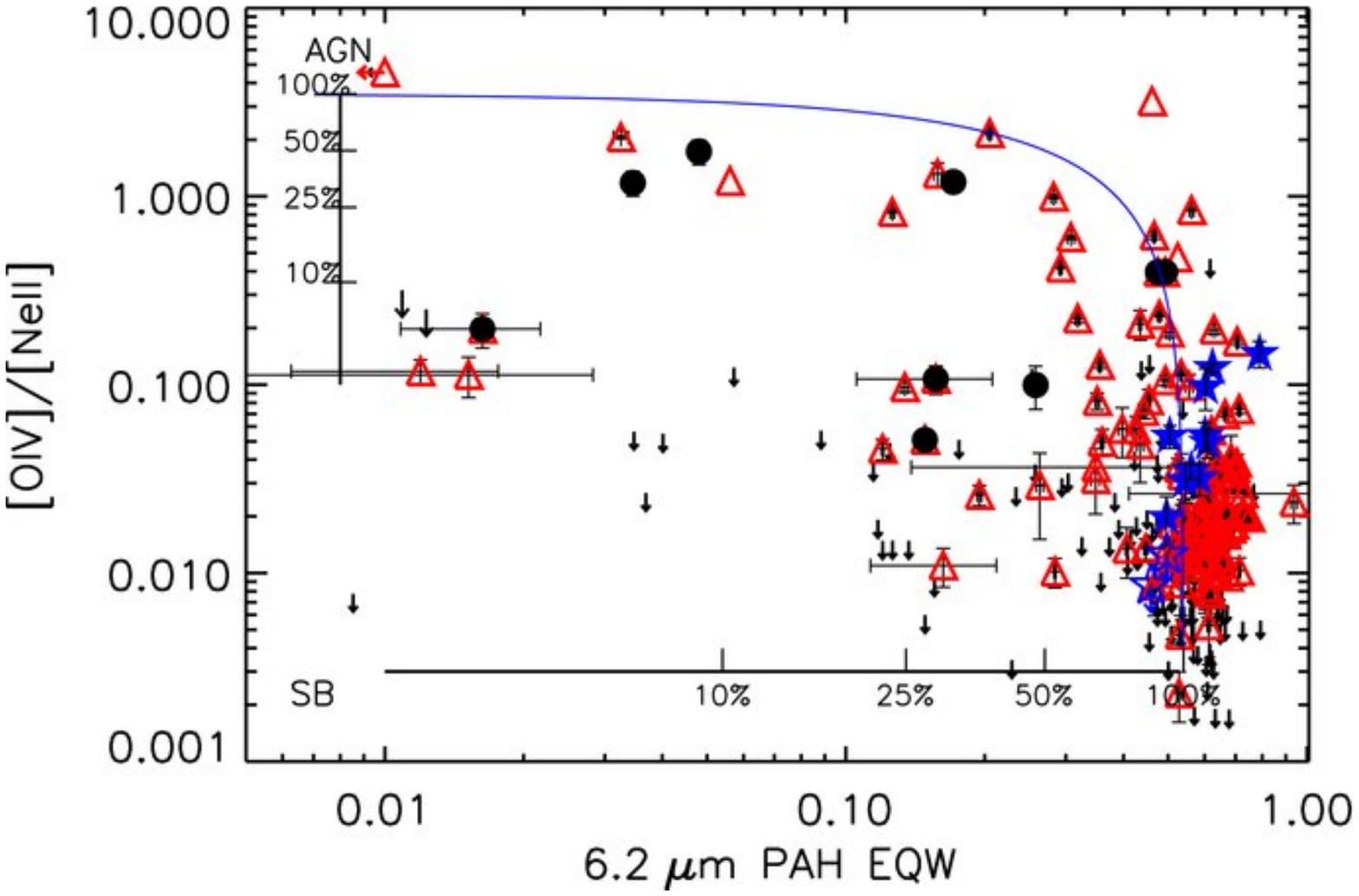}

\vspace{-5cm}

\caption{Examples of mid-IR diagnostic diagrams constructed with Spitzer/IRS observations and used to estimate the SF and AGN contributions to the mid-IR emission. Red triangles correspond to LIRGs in the GOALS sample with detections of high excitation lines, while arrows are upper limits. Black circles indicate ULIRGs, with filled symbols corresponding to detections and empty symbols to upper limits. The blue stars are starburst galaxies with the same convention for detections and non-detections. Black vertical lines indicate the fractional AGN contribution to the mid-IR emission from the line ratios and black horizontal lines the fractional starburst contribution from the EW of the $6.2\,\mu$m PAH feature, assuming a simple linear mixing model. Blue line indicates where the combined contribution of SF and AGN equals 100\%. Image reproduced with permission from \cite{Petric2011}, copyright by AAS.
}
\label{fig:MIRlinevsEWPAH}
\end{figure}

High excitation lines such as [Ne\,{\sc v}] at 14.3 and $24.3\,\mu$m
(${\rm IP}=97.1\,$eV)
can be used to identify \emph{elusive} (i.e., obscured in the optical) AGN. In local
LIRGs these lines are detected
in approximately 20\% of the nuclei \citep{Petric2011,
  AlonsoHerrero2012}. Also the [O\,{\sc iv}] line at $\sim 25.9\mu$m
(${\rm IP}=54.9\,$eV) is found 
to correlate well with the AGN luminosity in Seyfert galaxies \citep{Rigby2009}.
However, 
in a volume-limited sample of local LIRGs, the [O\,{\sc iv}] line is detected in approximately
70\% of the nuclei \citep{AlonsoHerrero2012}, whereas in the optical only
about half of LIRGs are classified as AGN or composite
\citep[see e.g.,][and references therein]{Yuan2010}. \cite{AlonsoHerrero2012}
showed this line emission is emitted by
an AGN only in those galaxies that are optically classified as Seyferts and with high luminosities
($L({\rm [OIV]}) > 10^7 L_\odot$). In LIRGs without an AGN detection, the [O\,{\sc iv}] luminosity can be accounted for
by SF activity. On the other hand, the intensity of the [O\,{\sc iv}] line (or the [Ne\,{\sc v}] lines)
relative to 
a line with a lower IP (generally the [Ne\,{\sc ii}]
line) and the equivalent width (EW)
of a PAH feature (see also Sect.~\ref{subsubsec:PAHs}) can be used to estimate the fraction of the
mid-IR emission contributed by the presence of an AGN \citep[see][and also Section~\ref{subsubsec:AGN}]{Petric2011, AlonsoHerrero2012}. Examples of these diagnostic diagrams are shown in Fig.~\ref{fig:MIRlinevsEWPAH} for the GOALS sample. From this figure, it is clear that for the majority of LIRGs the mid-IR emission is not dominated by the contribution from an AGN.

\begin{figure}
\center

\vspace{-2.5cm}
\includegraphics[width=9.cm]{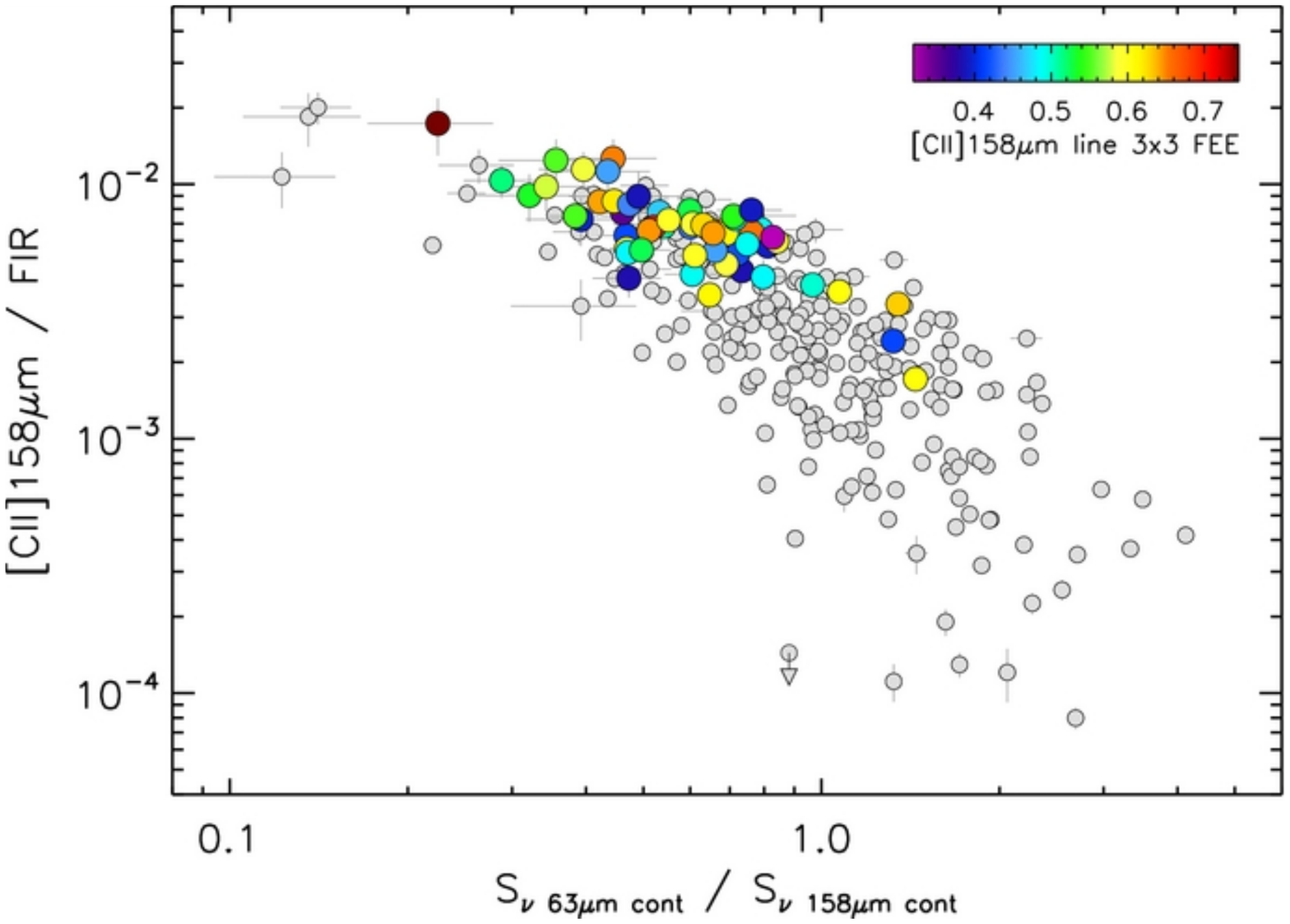}

\vspace{-3.5cm}
\caption{Diagram of the ratio of [C\,{\sc ii}]$158\,\mu$m flux to far-IR flux as a function of the $63\,\mu$m to $158\,\mu$m continuum flux density ratio for extended (colored circles) and nuclear  (gray circles) regions of spatially resolved galaxies in the GOALS sample. The color scale indicates the fraction of the [C\,{\sc ii}] emission that is extended. The extended regions of local LIRGs show smaller [C\,{\sc ii}] \emph{deficits} (that is, stronger [C\,{\sc ii}] emission relative to the far-IR emission) and lower $63\,\mu$m/$158\,\mu$m  ratios than the majority
of the LIRG nuclei. Image reproduced with permission from \cite{DiazSantos2014}, copyright by AAS. }
\label{fig:CIIdeficits}
\end{figure}

Using a variety of mid-IR diagnostics,
\cite{Petric2011} and \cite{Stierwalt2013} concluded that only in approximately 10-20\% of the GOALS sample the AGN dominates the mid-IR emission. 
\cite{Stierwalt2013} 
found that the fraction of LIRGs with a mid-IR bright AGN stays
approximately constant with the merger processes, whereas composite
sources (those in which the AGN emission does not dominate in the mid-IR) show a marked increase at later merger
stages. They suggested that the timescales for an AGN to start
dominating the mid-IR emission are longer than the merger timescales (a
few hundred million years). 

In the far-IR, some of the brightest emission lines in star-forming galaxies include
[O\,{\sc i}] at $63\,\mu$m, [O\,{\sc iii}] at $52\,\mu$m and $88\,\mu$m,  [N\,{\sc ii}]
at $122\,\mu$m, [O\,{\sc i}] at $145\,\mu$m, and [C\,{\sc ii}] at $158\,\mu$m.
In local LIRGs, the [C\,{\sc ii}] line emission as observed with Herschel/PACS is
extended on scales between 1 and 10\,kpc and is produced mostly in photo-dissociation
regions \citep{DiazSantos2014, DiazSantos2017}.  The [C\,{\sc ii}] flux to far-IR flux
density ratios in the nuclear regions of LIRGs decrease by more than a factor of ten
from the typical value  ($\sim 10^{-2}$) measured in star-forming galaxies
\citep{DiazSantos2013}. These authors also found decreasing nuclear [C\,{\sc ii}] flux
to far-IR flux density ratios for deeper $9.7\,\mu$m silicate features (that is, higher
extinction and more embedded regions, see also Section~\ref{subsubsec:silicates}),
higher mid-IR surface brightnesses and in SF-dominated LIRGs also more compact regions.
The [C\,{\sc ii}] \emph{deficits} observed in LIRG nuclei are not as extreme as those in
local ULIRGs (see for instance \citealt{Farrah2013}), as can be seen from the comparison
between Arp~299 and Arp~220 in Figure~1 of \cite{Fischer2014}. However, these deficits
might indicate different physical conditions from  those in normal star-forming disk
galaxies.  On the other hand, the extra-nuclear regions in local LIRGs show [C\,{\sc
ii}] flux to far-IR flux density ratios similar to those of the disks of normal galaxies
(see Fig.~\ref{fig:CIIdeficits}) and points to differences in the physical conditions in
the nuclear and extra-nuclear regions of local LIRGs.

The majority of local LIRGs shows molecular hydrogen emission, as traced by near- and
mid-IR H$_2$ lines, extending over several kpc scales \citep{almudena00, almudena09,
PereiraSantaella2010, PiquerasLopez2012, Colina2015, Vaisanen2017, Petric2018}. The
derived masses of molecular hydrogen are in the range $10^6 -10^9\,M_\odot$.
\cite{Petric2018} found a tendency for AGN-dominated LIRGs to have warmer temperatures
than SF-dominated LIRGs, although this appears to be a general behavior in galaxies
\citep{Lambrides2019}. A small fraction of local LIRGs present  mid-IR H$_2$ lines that
are kinematically resolved with Spitzer/IRS and are  generally associated to mergers and
AGN-dominated systems \citep{Petric2018}. This probably indicates the presence of
inflows and/or outflows in these LIRGs. Additionally, \cite{PereiraSantaella2010}
detected higher H$_2$ to PAH line ratios in the nuclear regions of some local LIRGs and
attributed them to an excess of H$_2$ emission produced by X-ray and/or shock
excitation. Indeed, \cite{Emonts2014} detected in both the near-IR H$_2$ $2.12\,\mu$m
line and the CO(3--2) transition an outflow originating from the secondary obscured
nucleus of NGC~3256 ($\log (L_{\rm IR}/L_\odot)=11.76$ for $D_L=45$\,Mpc) and extending
for hundreds of parsecs. The derived outflow rate in the molecular phase for this galaxy
is $20\,M_\odot \, {\rm yr}^{-1}$ and is likely AGN-driven. We refer the reader to
Sect.~\ref{subsec:coldmoloutflows} for a more general discussion of outflows detected in
cold molecular gas in local LIRGs.

\subsection{Dust emission}

The IR spectral range allows to study the dust emission (see also
Sect.~\ref{sec:sed-modelling}) in galaxies using several features, namely, the continuum
emission, which is sensitive to different dust temperatures depending of the chosen
wavelength, emission from PAHs, and the silicate features centered at $\sim 10\,\mu$m
and $18\,\mu$m, which are due to the Si-O stretching and the O-Si-O bending modes,
respectively, of amorphous silicate grains.

\subsubsection{Emitting regions}\label{subsubsec:dustcont}

Unlike many local ULIRGs, a
large fraction of GOALS local LIRGs show extended mid-IR
emission, including the continuum, line, and PAH emission, on
scales of up to 10 kpc \citep{DiazSantos2010Spitzer}. In LIRGs the mean nuclear size at $13.2\,\mu$m  is approximately 2.6\,kpc
at the Spitzer resolution.  This is thus in agreement with the
extended sizes ($>1\,$kpc) measured from the  Pa$\alpha$ emission and other fine structure lines in many local
LIRGs (Section~\ref{subsubsec:hydrogen} and \ref{subsubsec:fine}). However, LIRGs with IR luminosities above
$\log(L_{\rm IR}/L_\odot) \sim 11.8$ (including ULIRGs) and LIRGs with an important AGN contribution as well
as merger LIRGs tend to show more compact
mid-IR emission at $13.2\,\mu$m.

High angular resolution ($0.''3-0.''4$)
mid-IR (from $\sim 8$ to $18\,\mu$m) imaging studies using ground-based 8-10m class
telescopes \citep{soifer2001, AlonsoHerrero2006MIR,
DiazSantos2008, Imanishi2011, 
pereira15} resolve the central (few kpc) mid-IR emission of LIRGs in
individual star-forming regions with typical sizes of tens of
parsecs to a  few hundred parsecs. When an AGN is present, the mid-IR emission appears unresolved at the resolution provided by 8--10-m telescopes and is consistent with emission from dust in a torus heated by the AGN \citep[see for instance][]{AlonsoHerrero2013}. The mid-IR morphologies are similar, although
not completely 
identical, to the hydrogen recombination line morphologies which trace the
youngest stellar populations in local LIRGs \citep[see][and also Sect.~\ref{subsubsec:hydrogen}]{DiazSantos2010TReCS}. 
LIRGs hosting a known 
AGN or  a buried AGN tend to show high surface brightness in the mid-IR ($>>
10^{13}\,L_\odot \,{\rm kpc}^{-2}$) nuclear sources
\citep{Egami2006,  DiazSantos2010TReCS, Imanishi2011, Mori2014, MartinezParedes2015}. In these sources a large fraction of
the nuclear mid-IR emission is due
to dust heated by the AGN but there can also be some contribution from
nuclear SF activity which is  traced by the $11.3\,\mu$m PAH feature (see
below). 

\subsubsection{PAH emission}\label{subsubsec:PAHs}
The PAH emission in galaxies is believed to trace on-going/recent SF activity \citep[see, e.g.,][]{Peeters2004}.
SF-dominated
LIRGs show nuclear and extra-nuclear mid-IR spectra with strong emission from PAH features
\citep{Soifer2002, almudena09, DiazSantos2010TReCS,
  PereiraSantaella2010, pereira15, Imanishi2010, 
AlonsoHerrero2012, Stierwalt2013}, as can be seen from Fig.~\ref{fig:MIRspectracomparison} (bottom panel). Some
LIRGs classified as Seyferts still show PAH emission, even on nuclear scales  \citep[][]{Mori2014, MartinezParedes2015},
but with lower EW and steeper mid-IR continua. The PAH emission of LIRGs, when observed on scales of a few hundred parsecs to galaxy-integrated scales
correlates broadly with the emission from  hydrogen recombination
lines \citep[see, e.g.,][]{DiazSantos2008, Imanishi2010}. Thus, the PAH emission
in LIRGs also traces
SF activity as also found in high metallicity star forming galaxies  
\citep{Calzetti2007}. 

Using Spitzer/IRS spectral mapping \cite{PereiraSantaella2010} observed that the
6.2, 7.7 and $11.3\,\mu$m PAH feature emission extends over several kpc in a sample of local LIRGs. When compared
to the [Ne\,{\sc ii}] line emission (a proxy for the young SFR, see
Section~\ref{subsubsec:fine}), the $11.3\,\mu$m PAH emission appears to be more extended and SF-dominated LIRGs tend to show lower [Ne\,{\sc ii}]/11.3\,$\mu$m PAH ratios at larger galactocentric distances. There are also variations in the different PAH feature ratios across the galaxies (e.g., nuclear vs.\
integrated values) but these variations do not appear to be related to the hardness of the
radiation field as traced by the [Ne\,{\sc iii}]/[Ne\,{\sc ii}]
ratio \citep{PereiraSantaella2010}. On smaller physical scales (tens of parsecs to a few hundred parsecs) some differences appear between the PAH emission and the Pa$\alpha$ and [Ne\,{\sc ii}] emissions \citep{DiazSantos2008, DiazSantos2010TReCS}. These are likely related to the different ages traced by the ionized gas
and the PAH emission, as the PAH molecules are mostly excited by UV photons from B stars \citep{Peeters2004}, whereas the Pa$\alpha$ and [Ne\,{\sc ii}] emissions probe stellar ages of less than $\sim 10$\,Myr.

\begin{figure}
    \centering
    \includegraphics[width=9cm, angle=-90]{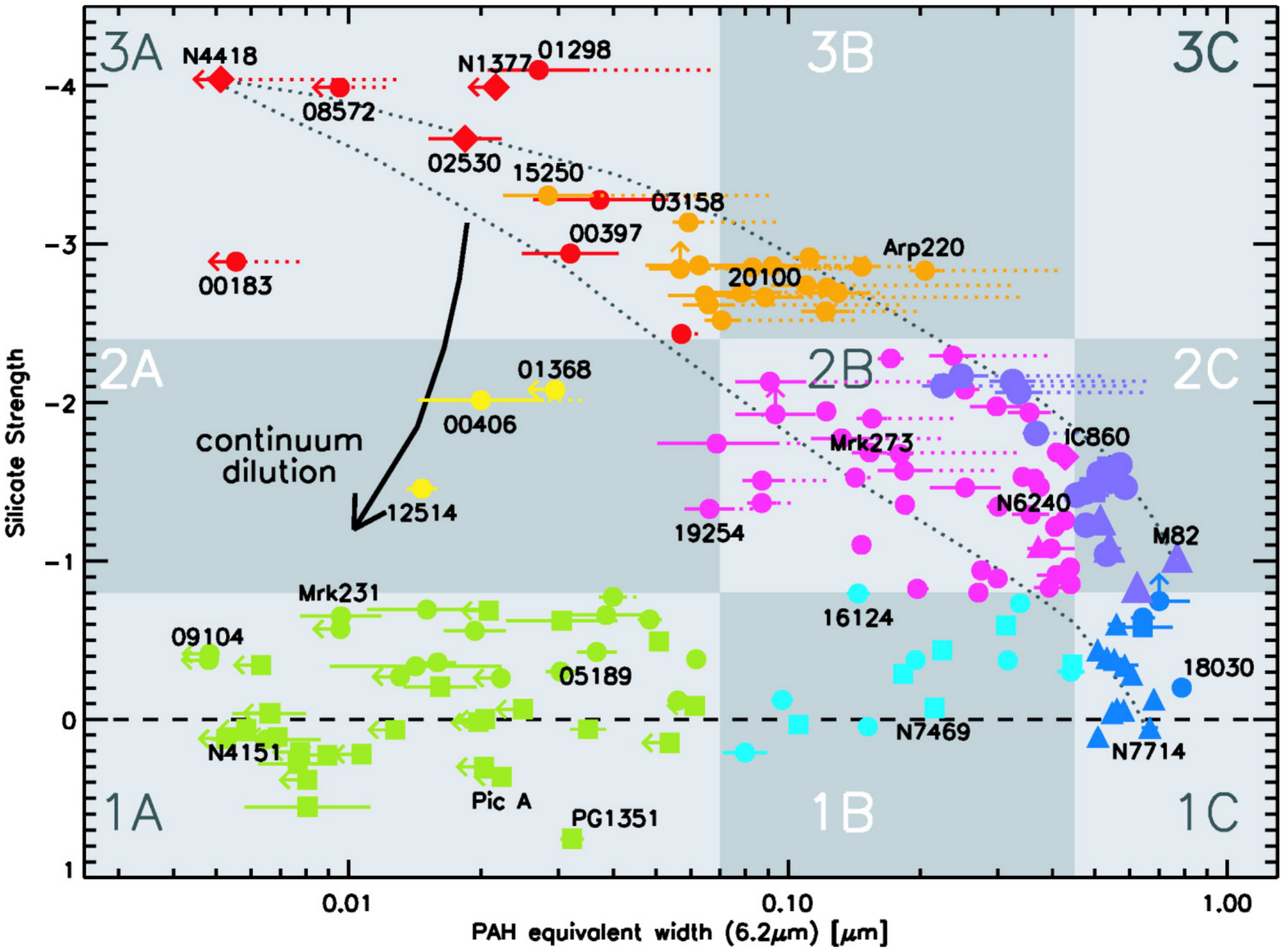}
 \vspace{-1cm}
    \caption{Diagnostic diagram showing the EW of the $6.2\,\mu$m PAH feature against the strength of the $10\,\mu$m silicate feature for different types of galaxies: ULIRGs and hyper luminous IR galaxies (circles), starburst galaxies (triangles), Seyfert galaxies and quasars (squares), and other IR galaxies (diamonds). The different regions show different classes (1A, 1B, 1C, 2A, 2B, 2C, 3A, and 3B) based on their mid-IR spectra. Class 1A (nearly featureless spectra) are AGN-dominated galaxies, and class 1C (PAH dominated spectra and moderate silicate feature absorption) are starburst galaxies, while class 1B are composite galaxies. The silicate feature becomes deeper in class 2C. In class 2B, the PAH features appear weaker than those in the class 2C spectrum.  The silicate feature depth keeps increasing for class 3B and reaches the maximum depth in absorption in class 3A where many Compact Obscured Nuclei (CONs) are located (for instance, NGC~4418 and NGC~1377, see also Sect.~\ref{sec:con}).  The EW of the PAH features decreases from class 2B to 3B and 3A. The two dotted black lines are mixing lines between the spectrum of the deeply obscured nucleus of NGC~4418 and the starburst nuclei of M~82 and NGC~7714. The curved arrow shows the  filling in of a deep $10\,\mu$m silicate feature by the presence of a mid-IR continuum produced by hot dust. The resulting effect is moving sources from  class 3A to  class 2A. Image reproduced with permission from \cite{spoon07}, copyright by AAS.}
    \label{fig:SpoonDiagram}
\end{figure}

\subsubsection{The silicate feature}\label{subsubsec:silicates}

The $10\,\mu$m silicate feature appears generally in
moderate absorption in local LIRGs, thus indicating relatively modest extinctions. Figure~\ref{fig:MIRspectracomparison} shows that the strength of the silicate feature in local LIRGs is
on average intermediate between those observed in local starbursts \citep{Brandl2006} and the deep silicate
feature (in absorption) detected in some local ULIRGs \citep[i.e.,][and see also Fig.~\ref{fig:SpoonDiagram}]{spoon07}. 
On scales of hundreds of parsecs, which are the typical physical regions probed by \textit{Spitzer}/IRS for local LIRGs, the median value of the strength\footnote{The strength of the silicate feature is measured as; $S_{\rm Si} = {\rm ln} \frac{f_{\rm obs}(9.7\mu{\rm m})}{f_{\rm cont}(9.7\mu{\rm m})}$, where $f_{\rm obs}(9.7\mu{\rm m})$ is the observed flux density at the feature (in absorption or emission) and $f_{\rm cont}(9.7\mu{\rm m})$ is the flux density at the continuum. In this definition, a positive value of the strength  means that the silicate feature is in emission and a negative value the silicate feature is in absorption.} of the nuclear silicate feature
is $S_{\rm Si}=-0.25$ \citep{PereiraSantaella2010, AlonsoHerrero2012, Stierwalt2013}. Assuming  an extinction law such as $A_V/S_{\rm Si} =16.6$ \citep{Rieke1985}, the median silicate feature strength corresponds to a typical
nuclear extinction of
$A_V=4\,$mag. This is in good agreement with the extinction values
derived from near-IR continuum and ionized gas observations
\citep[][see also Sects.~\ref{subsec:stellar} and
\ref{subsubsec:hydrogen}]{AlonsoHerrero2006NICMOS, DiazSantos2008,
  PiquerasLopez2013}. 
  
Figure~\ref{fig:SpoonDiagram} shows 
 the diagnostic diagram 
comparing the EW of the $6.2\,\mu$m PAH feature 
and the strength of the silicate feature defined by \cite{spoon07}. In this diagram different types of galaxies are plotted, including  AGN, composite galaxies, starburst galaxies as well as IR-bright galaxies, ULIRGs, and hyper-luminous IR galaxies.   \cite{AlonsoHerrero2012} and \cite{Stierwalt2013} showed that  most local LIRGs are in the 1C class, which is defined by  the spectra of local starburst galaxies with strong PAH features and moderate strengths of the silicate features (in emission or absorption). A few local LIRGs and ULIRGs are  in the 1A class defined by local Seyferts and quasars with faint PAH emission and relatively flat silicate feature, and the 1B composite class (intermediate between AGN and starburst galaxies, for instance NGC~7469 marked in the figure). The rest of local LIRGs are, in decreasing numbers, in the 2C, 2B and 3A classes. As can also be seen from this figure, a few LIRG and ULIRG nuclei
show an extremely deep silicate
feature  \citep[$S_{\rm  Si}<-1$, see Table~4 and Table~1
of][respectively, and  Sect.~\ref{sec:con}]{AlonsoHerrero2012,
Stierwalt2013}, indicating very high extinctions ($A_V>16\,$mag, in a simple dust screen model) or even completely embedded sources \citep{Levenson2007}.

There is a tendency for deeper silicate features in local LIRGs  with increasing $L_{\rm IR}$, 
especially for the most luminous LIRGs \citep{Stierwalt2013}. This is probably 
related to the presence of more compact emitting regions in the LIRGs nuclei.
Moreover, spatially resolved Spitzer/IRS observations also revealed that 
the deepest silicate features in local LIRGs are generally coincident with the location of the
nuclei \citep{PereiraSantaella2010}. This again points to the
different properties of the nuclear and extra-nuclear regions in many
local LIRGs, as found from the fine structure lines.
\\
\\
\\

\section{Radio emission from LIRGs}
\label{sec:radio}

\subsection{Synchrotron, thermal free-free and dust continuum}
\label{sec:radio-continuum}

Radio observations of LIRGs are essentially unaffected by dust, unlike optical and, to a
lesser extent, IR observations (Sect.~\ref{sec:infrared}). The bulk of the continuum
radio emission in normal galaxies, i.e., those with no significant AGN or starburst
activity, arises from thermal and non-thermal processes associated with young ($t
\lesssim$ 40 Myr), massive ($\gsim 8\,M_\odot$) stars. The thermal (free-free) radiation
of a star-forming galaxy is emitted from H{\sc ii} regions and is directly proportional
to the photoionization rate of young massive stars \citep{rubin68}.  When a massive star
ends its life, it explodes as a SN (see Sect.~\ref{sec:timedomain}). The interaction of
the fast SN ejecta with the surrounding gas results in a shocked shell where electrons
are accelerated to relativistic energies that, in the presence of a significant magnetic
field, yield copious non-thermal radio synchrotron emission.  This synchrotron emission
diffuses out into the ISM, and eventually pervades the whole host galaxy. Thus,
synchrotron emission is a tracer of the recent star formation activity in galaxies.  The
same processes are also in place in LIRGs, where powerful starbursts result in levels of
radio emission that are orders of magnitude higher than in normal galaxies.

\begin{figure}[htb!]
\includegraphics[width=0.64\textwidth]{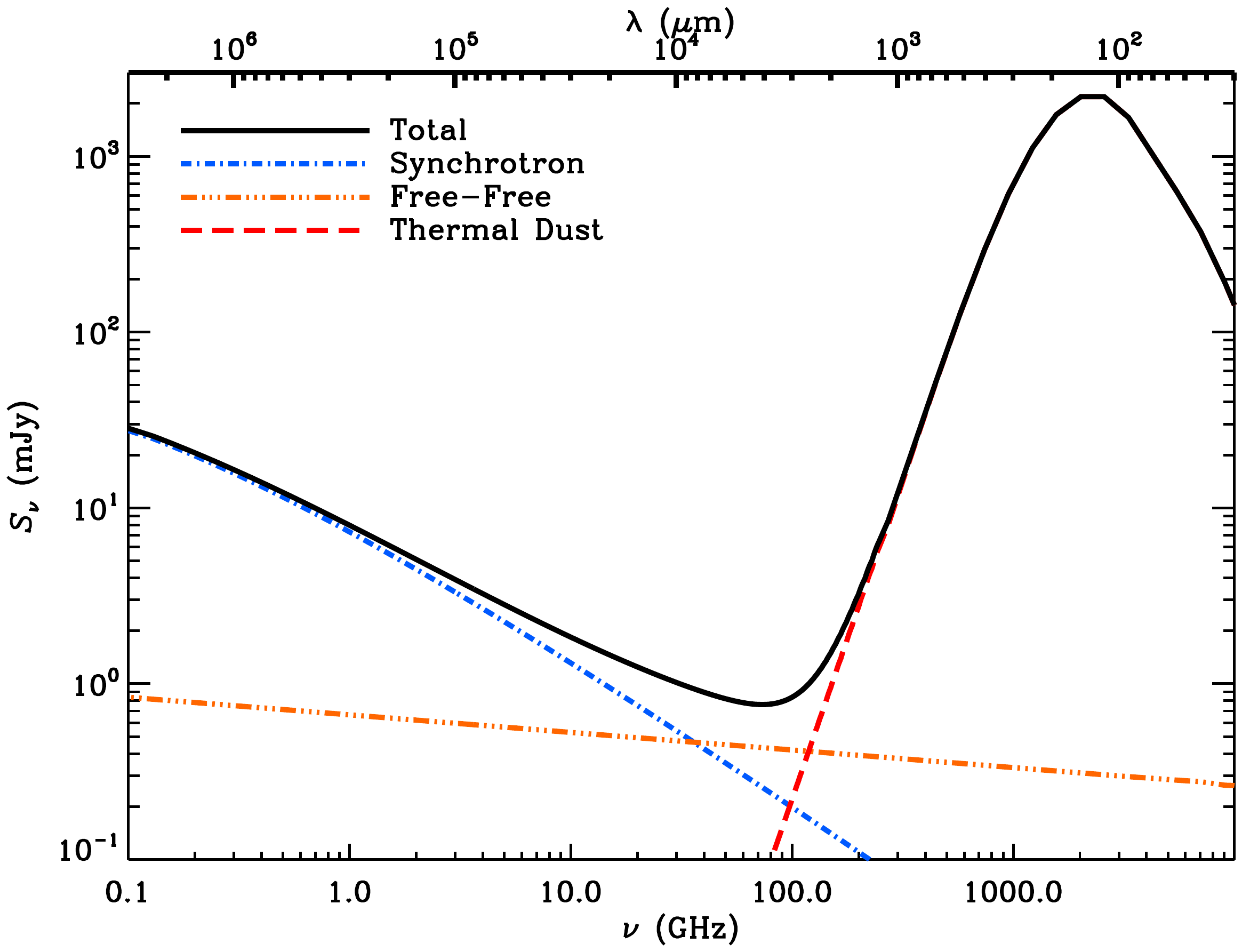}
\includegraphics[width=0.35\textwidth]{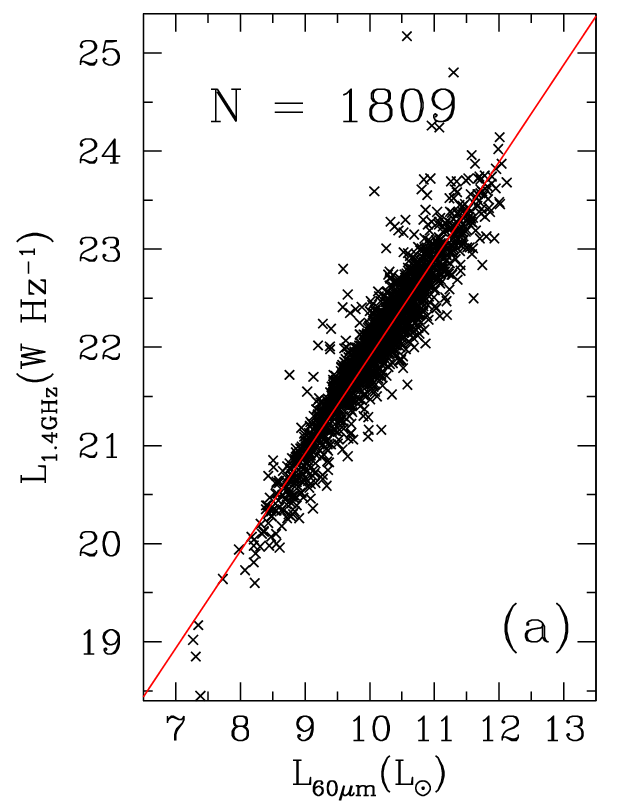} \caption{ \label{fig:radio-sed-FIRC}
  \textbf{Left.} Model radio-to-far-IR SED for a starburst-dominated LIRG.  Synchrotron
  radio emission from young SNe and SN remnants dominates at frequencies below $\sim$30
  GHz, while at higher frequencies, up to $\sim$150 GHz, thermal free-free emission from
  H~II regions dominates. Above $\sim$200 GHz, and extending into the far-IR regime,
  thermal dust radiation dominates the continuum emission.  Image reproduced with
  permission from \cite{Murphy2009}, copyright by AAS.) \textbf{Right.} 1.4 GHz radio
  luminosity versus IRAS 60 $\mu$m luminosity of galaxies, illustrating the ubiquitous
  radio-to-far-IR correlation, which holds for six orders of magnitude in luminosity for
  galaxies. The solid line corresponds to a linear relation with a constant offset (see
  Eq.~4 in \citealt{Yun2001}).  Image reproduced with permission from \citealt{Yun2001},
  copyright by AAS.  } \end{figure}

We refer the reader to \citet{Condon2016} for details on free-free and synchrotron
radiation, but give here a short description of the relevant frequency dependencies. The
centimetre ($\nu \lesssim$30 GHz, corresponding to $\lambda \gtrsim 1$ cm) emission of
normal spiral galaxies and LIRGs is dominated by synchrotron radio emission from SNe and
SN remnants; 
the emission between $\sim$30 GHz  and $\sim$100 GHz ($\lambda \sim$3 mm) is mainly
bremhsstrahlung (free-free) emission from massive stars and H~II regions, while at
frequencies above $\sim$200 GHz, corresponding to the sub-mm and far-IR regime, the
continuum  emission is dominated by thermal dust radiation (Fig.
\ref{fig:radio-sed-FIRC} left).  Since the same massive stars responsible for the
continuum radio emission in galaxies are also responsible for yielding large amounts of
thermal dust radiation at IR wavelengths, there exists a very tight correlation between
radio and far-IR emission among galaxies (e.g.,
\citealt{Harwit1975,Condon1991a,condon92}), which holds over six orders of magnitude in
luminosity (\citealt{Yun2001}; Fig. \ref{fig:radio-sed-FIRC} right).

Continuum radio observations at centimetre wavelengths trace also the absorption
processes that are responsible for the partial suppression of the observed radio
emission, most notably the free-free absorption from the ISM and the circumstellar
medium (CSM) around SN progenitor stars and, in some cases, synchrotron self-absorption.
Namely, at frequencies below $\sim$0.1 GHz, most H~II regions become opaque (optically
thick) to radio emission, since the opacity decreases strongly with frequency, $k_\nu
\propto \nu^{-2.1}$, significantly reducing both free-free and synchrotron radio
emission.  At GHz frequencies, the free-free and synchrotron opacities are hence much
smaller (optically thin), and therefore the thermal and non-thermal radio continuum
emissions from star-forming galaxies offer very good diagnostics of the SFR of massive
stars.  As shown in Fig. \ref{fig:radio-sed-FIRC}, at frequencies above  $\sim$0.1 GHz,
free-free radiation is optically thin, and  decreases very slowly with frequency $(S_\nu
\propto \nu^{-0.1}$). Synchrotron radio emission is also optically thin at frequencies
$\gtrsim$0.1 GHz, and shows a power-law dependence with frequency, $S_\nu \propto
\nu^\alpha$, where $\alpha$ is the (optically thin) radio spectral index of the
synchrotron power-law, with $\alpha \approx -0.8$ in starburst-dominated LIRGs (e.g.,
\citealt{condon92,herrero-illana17}) and extranuclear star-forming complexes
\citep{Murphy2011}.  From $\sim$0.1 GHz up to $\sim$30 GHz, the radio emission is
dominated by  non-thermal synchrotron radiation (Fig. \ref{fig:radio-sed-FIRC} left). If
this non-thermal emission is produced only by SN explosions, synchrotron emission serves
then as an indicator of the SFR over the past $\sim 10-100$ Myr \citep{Bressan2002}.
However, synchrotron emission also arises from accretion processes associated with the
supermassive black holes (SMBHs) at the centres of AGN host galaxies, so the conversion
from synchrotron radio emission to SN (and star-formation) rate requires that the galaxy
does not deviate significantly from the far-IR-to-radio correlation (\citealt{Yun2001};
see also Fig.~\ref{fig:radio-sed-FIRC}).  The thermal radio component is thus expected
to be a more direct measure of the most recent massive star formation activity.
\citet{condon90} gave an approximate expression for the ratio of the total radio
emission to the thermal free-free emission for spiral galaxies,

\begin{equation}
    \frac{S_\nu}{S_\nu^{\rm th}} \sim 1 + 10 \left(\frac{\nu}{\rm GHz}\right)^{0.1 + \alpha}
    \label{eq:thermal-fraction}
\end{equation}

\noindent The value of (optically thin) spectral index $\alpha \approx -0.8$ implies an
electron power-law index $p = 1 - 2\alpha \approx 2.6$ for the population of
relativistic electrons giving rise to non-thermal synchrotron emission, $N(E)\,dE
\propto E^{-p} \, dE$, where $N(E)$ is the  density of relativistic electrons with
energies in the interval $(E, E+dE)$.  The value $p \approx 2.6$ agrees well with the
one observed for cosmic rays \citep{Meyer69}, which are thought to be accelerated mainly
in SN remnants. 

At 1.4 GHz, Eq.~(\ref{eq:thermal-fraction}) predicts a fraction of thermal free-free emission, $f_{\rm th} = S_\nu^{\rm th}/S_\nu \sim 0.11$, so the low-frequency radio emission in LIRGs is predominantly of non-thermal, synchrotron origin. At $\nu \approx 30$ GHz, Eq.~(\ref{eq:thermal-fraction}) predicts $f_{\rm th} \sim 0.52$, while at 
$\nu \approx 200$ GHz, $f_{\rm th} \sim 0.80$, when the radio continuum peaks at sub-mm wavelengths and is powered by dust emission.
Disentangling the thermal component from the non-thermal one is difficult below $\sim$30 GHz, since the 
flat-spectrum 
($\alpha\sim-0.1$) 
thermal free-free emission is weaker than the steep-spectrum ($\alpha\sim-0.8$) non-thermal synchrotron emission found in LIRGs.
Equation~(\ref{eq:thermal-fraction}) must be used with caution, and 
multi-frequency observations above and below  $\sim$30 GHz are necessary to estimate the true thermal component.
\cite{Murphy2011} used radio to IR spectral observations of ten star-forming regions in NGC~6946 
to derive extinction-free SFRs at 33 GHz. They found that the thermal fraction, $f_{\rm th}$, at 33 GHz was significantly higher than expected from Eq.~(\ref{eq:thermal-fraction}): 85\% or higher but for the nucleus, which has $\sim$62\%.
Given the dominance of the thermal component at 33 GHz, \citet{Murphy2011} suggested that the use of the total 33 GHz flux density is a good SFR diagnostic.  Still, the most widely used radio SFR calibration is the result of the tight, empirical radio-to-far-IR correlation \citep{deJong1985,Helou1985}. The radio-to-far-IR correlation can then be combined with the empirical relation between $L_{\rm IR}$ and SFR$_{\rm IR}$ to express the SFR as a function of the 1.4\, GHz luminosity L$_{\rm 1.4}$ (e.g., \citealt{Murphy2011}):

\begin{equation}
\label{eq:sfr-radio}
{\rm SFR_{\rm 1.4\,GHz}} \approx 64 \left(\frac{L_{1.4}}{\rm 10^{30}\, erg~s^{-1}~Hz^{-1}}\right)\, M_\odot~{\rm yr^{-1}}  
\end{equation}

Equation~(\ref{eq:sfr-radio}) implicitly assumes that, if a galaxy follows the
radio-to-far-IR relation, then all of its radio emission comes from star-formation.
While radio loud AGNs clearly deviate from the L$_{1.4\rm GHz}$-to-L$_{60\mu}$m relation
seen in starburst galaxies, many radio quiet AGNs show similar L$_{1.4\,\rm
GHz}$-to-L$_{60\mu\,\rm m}$ values (e.g., \citealt{Moric2010}), and both their IR and
radio emission seem to be dominated by AGN activity, not by the host galaxy
\citep{Zakamska2016}.  Thus, even if a galaxy follows the radio-to-far-IR relation, this
does not necessarily mean that its 1.4 GHz luminosity is dominated by star-formation, so
Eq.~(\ref{eq:sfr-radio}) must be used with caution when obtaining SFR from measured
values of L$_{1.4\,\rm GHz}$.

\subsection{The central kpc of LIRGs as traced by continuum radio observations}
\label{sec:radio-high-resolution}

\begin{figure}
\includegraphics[width=0.52\textwidth]{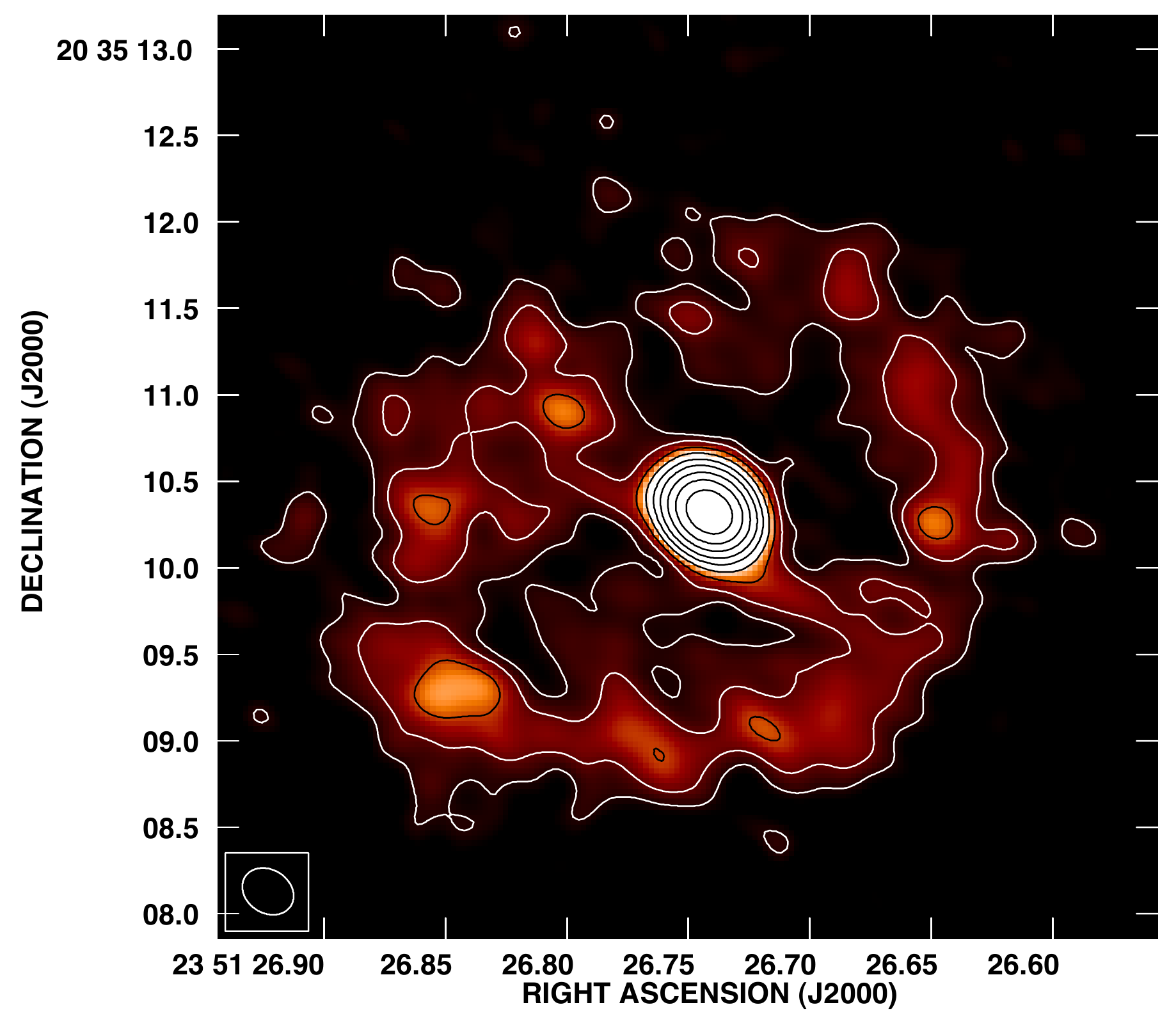}
\includegraphics[width=0.49\textwidth]{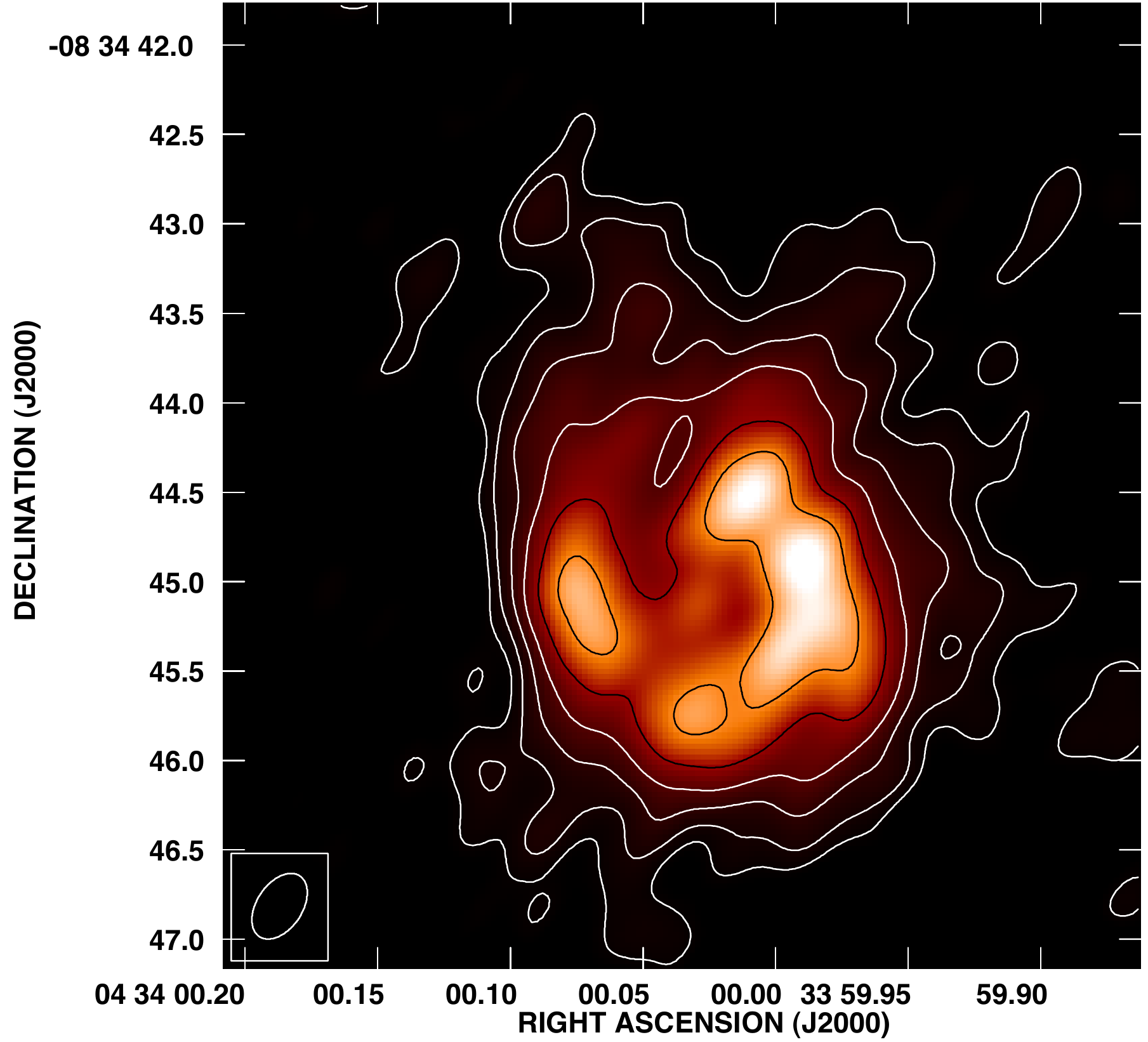}
\caption{
\label{fig:lirgs-vla}
Continuum 8.5 GHz VLA images of the central kpc region of two nearby LIRGs, Mrk~331 \textbf{(left)} and NGC~1614 \textbf{(right)} at distances of
75 Mpc and 66 Mpc, respectively.
The synthesized FWHM beams are Gaussian beams of sizes 
$0.''27\times0.''25$ (left) and $0.''42\times0.''24$ (right). 
$1''$ corresponds to 357 pc (320 pc) for Mrk 331 (NGC~1614). 
Note the existence of a prominent circumnuclear ring surrounding the nuclear region of each source, at a radial distance of $\sim 0.5 - 1.0$ kpc. Note also that the radio emission of Mrk~331 is dominated by a compact, central component, while that of NGC~1614 is dominated by the circumnuclear ring, with several regions of prominent emission. Images reproduced with permission from \citet{pereztorres13}. (For NGC~1614 see also Fig.~\ref{fig:HCN_N1614}).
}
\end{figure}

At a distance of 100 Mpc,  $1''$ corresponds to a linear size of about 500 pc.
Therefore, sub-arcsecond angular resolution is clearly needed to study in any detail the
central kpc regions of local LIRGs.  Radio interferometry  routinely provides
sub-arcsecond angular resolution images of local LIRGs. We show in
Fig.~\ref{fig:lirgs-vla} two Very Large Array (VLA) images of the nearby LIRGs Mrk~331
($\log (L_{\rm IR}/L_\odot)=11.33$ for $D_L=75\,$Mpc) and NGC~1614 ($\log (L_{\rm
IR}/L_\odot)=11.68$ for $D_L=68\,$Mpc), which illustrate how the central regions of
LIRGs look in the radio continuum.  The central region of many nearby U/LIRGs shows a
radio emission distribution with a similar pattern: a compact ($\lesssim$ 200 pc), high
surface brightness, central radio source corresponding to an AGN, immersed in an
extended low surface brightness circumnuclear ($\lesssim$ 1 kpc) ring of star-formation. 

The brightness temperature from a mixture of thermal and non-thermal emitting sources in
a galaxy can be expressed as follows \citep{condon92}: $  T_b \simeq T_e\, (1 -
e^{-\tau_{\rm ff}})/f_{\rm th}$, where $\tau_{\rm ff}$ is the average optical free-free
depth along the line of sight, $T_e$ is the electron temperature, and $f_{\rm th}$ is
the thermal fraction. In the absence of synchrotron emitting sources, $T_b \simeq T_e (1
- e^{-\tau_{\rm ff}})$. Further, at frequencies below $\sim 0.1$ GHz, $\tau_{\rm ff}$ is
usually much larger than unity, so that $T_b \simeq T_e$, i.e., the brightness
temperature directly informs us about the electron temperature of the ambient medium,
which typically is $T_e \sim (1-2)\times 10^4$ K.  On the other hand, at frequencies
where $\tau_{\rm ff} \ll 1$, $T_b \simeq T_e \,\tau_{\rm ff}$, and $T_b$ is much smaller
than $T_e$.  Therefore, if we find $T_b \gtrsim 10^3$ K at frequencies above 1 GHz, this
is a clear indication of a non-thermal process, most likely the ubiquitous synchrotron
radio emission.

Fig.~\ref{fig:lirgs-vla} provides morphological proof for the existence of circumnuclear
starburst rings in LIRGs, yet the angular resolution is  insufficient to yield
conclusive evidence about the nature of the radio emission of in the unresolved regions.
In fact, from radio observations, the brightness temperature of a source can be written
as \begin{equation} T_b = \left(\frac{S_\nu\,\nu^{-2}}{\Omega}\right) \,
\frac{c^2}{2\,k} \approx 1.6\times 10^3\,\left(\frac{S_\nu}{\rm mJy}\right)\,
\left(\frac{\nu}{\rm GHz}\cdot\frac{\theta_0}{\rm arcsec}\right)^{-2} {\rm K} \, ,
\label{eq:T_b} \end{equation} where $c$ is the speed of light, $k$ is the Boltzmann
constant, and $\Omega = \pi\, \theta_0^2/(4\, {\rm ln\,2})$ is the solid angle of a
circular Gaussian beam with FWHM diameter  $\theta_0$ \citep{Condon2016}.  This leads to
peak brightness temperatures at 8.5 GHz of $T_b \simeq 4.1\times10^4$ K and $T_b \simeq
3.7\times10^3$ K for Mrk~331 and NGC~1614, respectively, from the VLA images in
Fig.~\ref{fig:lirgs-vla}. Those values are too large for being compatible with free-free
thermal radio emission, which should be at most of a few thousand Kelvin at those high
frequencies.  The above temperatures unambiguously imply the existence of non-thermal
synchrotron radio emission from compact sources.

Since most central regions of LIRGs host a powerful starburst, an AGN, or both, one
generally expects that most, or even all, of the radio emission from those regions is
powered by the (non-thermal) synchrotron mechanism.  Brightness temperatures as large as
$T_b \lsim 10^7-10^8$ K can be explained by different compact sources of different
types, e.g., a young SN, a SN remnant, an AGN, or a combination of AGN and a transient,
compact radio emitting source, e.g., a SN. If $T_b \gtrsim 10^8$ K, the non-thermal
source is almost surely an AGN.  In this case, the radio spectral index most often shows
a power-law spectrum with $\alpha \ll -0.1$, implying the bulk of the observed radio
emission is of non-thermal, synchrotron origin.

For example, \citet{herrero-illana17} used the wide-band JVLA capabilities to determine
pixel-by-pixel radio spectral indices for a sample of local LIRGs, and found a median
value of $\alpha \approx -0.8$, which is a typical value for starburst-powered LIRGs.

\begin{figure}
\includegraphics[scale=0.307]{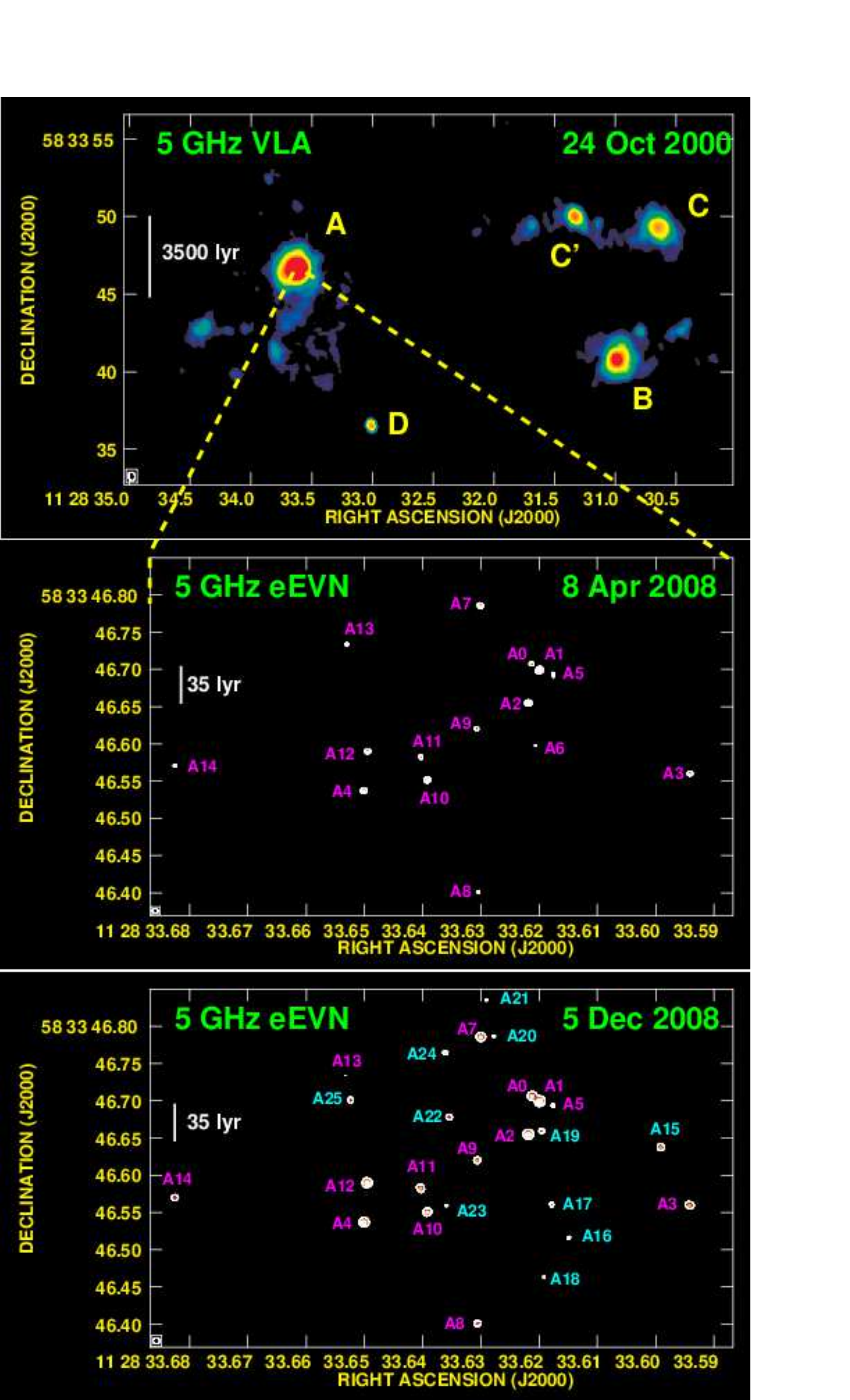}
\includegraphics[scale=0.28]{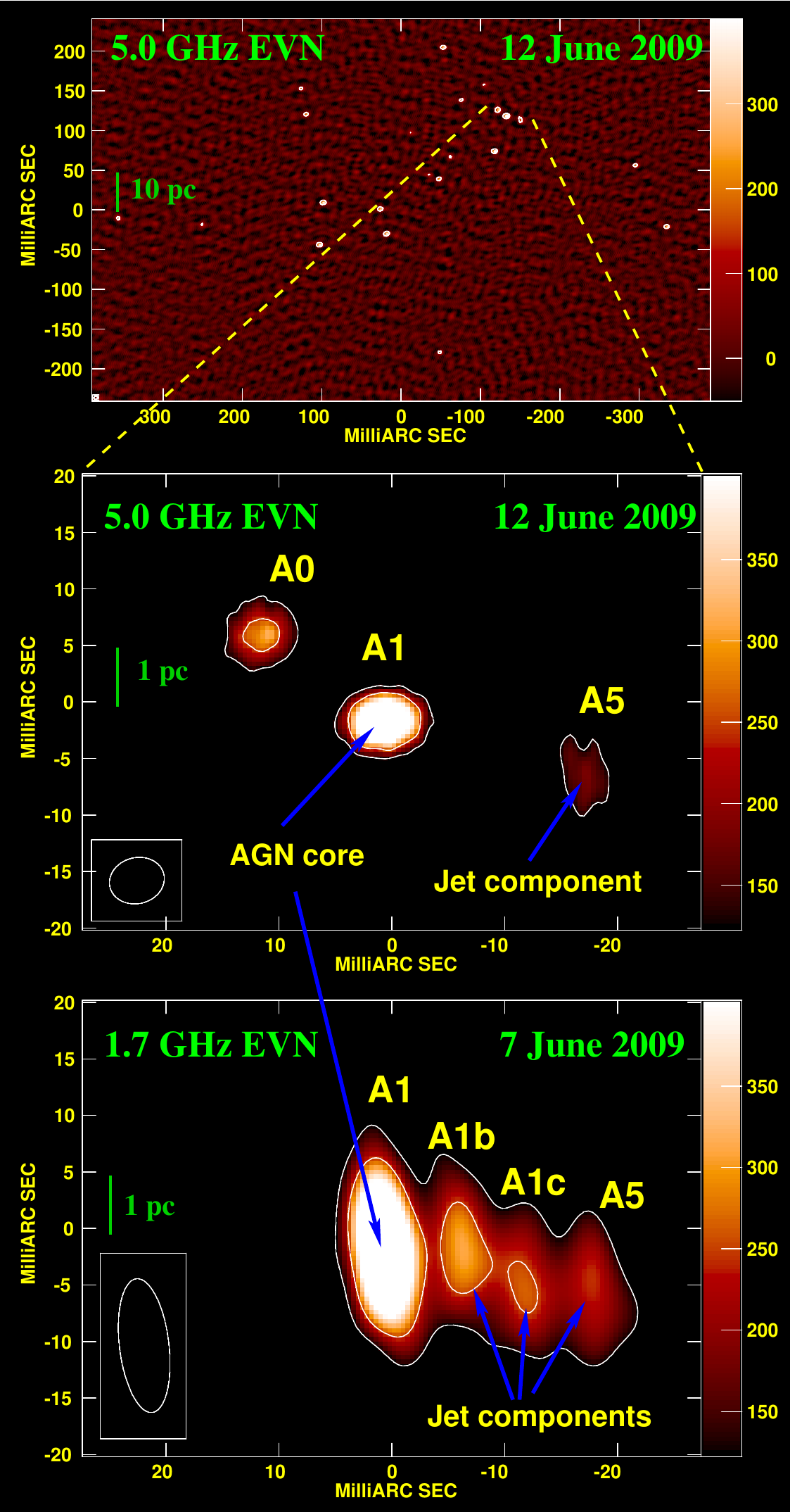} 
\caption{ \label{fig:arp299a-vlbi}
 \textbf{Left.} 5 GHz VLA archival observations of Arp 299 (top)
displaying the five brightest knots of radio emission, 
and 5.0 GHz EVN images of the central $\sim$200 pc of Arp~299-A 
on 8 April 2008 and 5 December 2008.
The use of VLBI reveals a large population of compact, non-thermal emitting sources (white contours), mostly identified with young radio SNe and SN remnmants \citep{pereztorres09a,bondi12}.
\textbf{Right.} EVN image at 5 GHz of the central $\sim200$ pc region of Arp 299-A (top), and blow-ups of the inner 8 parsecs, as imaged with the  EVN at 5.0 GHz (middle) and 1.7 GHz (bottom). The morphology, spectral index and luminosity of the A1-A5 region are very suggestive of a core-jet structure. Figures from \citealt{pereztorres09a,pereztorres10}.}
\end{figure}

The most direct evidence of non-thermal synchrotron radio emission in the central
regions of LIRGs comes from observations with Very Long Baseline Interferometry (VLBI)
arrays, like the European VLBI Network (EVN) or the Very Long Baseline Array (VLBA).
VLBI provides angular resolutions of a few milliarcseconds, or better, at
cm-wavelengths, which permits to disentangle the source, or sources, responsible for the
observed synchrotron emission.  As an illustration, we show in
Fig.~\ref{fig:arp299a-vlbi} EVN images of the innermost central region of Arp~299-A. The
angular resolution provided by those images (about 4 and 12 millarcsecond FWHM at 5.0
and 1.7 GHz, respectively) reveals a large population of compact sources that, based on
their high brightness temperatures, $\gtrsim 6.3\times10^5$ K for all sources,  were
identified with young radio SNe and SN remnants (\citealt{pereztorres09a}; see also
Sect. \ref{sec:timedomain}). Contemporaneous EVN observations at 1.7 and 5.0 GHz of the
nuclear region of Arp~299-A also revealed the existence of a bright ($T_b \gtrsim
9.0\times10^6$ K), compact, flat-spectrum source (A1 in Fig.~\ref{fig:arp299a-vlbi}),
which was identified with the long-sought AGN in Arp~299A (\citealt{pereztorres10}; see
also Sect.~\ref{sec:casestudies}).  A similar population of compact sources, thought to
be SNe and SNRs, has been studied with cm-VLBI observations of the ULIRG Arp~220
(\citealt{varenius2019} and references therein; see also Fig.~\ref{fig:RSN}).

\subsection{The magnetic field and the far-IR to radio correlation in local U/LIRGs}
\label{sec:radio-bfield}

High-angular resolution radio continuum observations are useful to estimate the total energy in magnetic fields,
electrons, and heavy particles. In addition, if equipartition between the energy in the form of magnetic fields and relativistic particles holds, the value of the intensity of the  magnetic field,  $B_{\rm min}$, can be estimated (see, e.g., \citealt{pacholczyk70,Longair2011}). 
For parameters typical of the Milky Way, 
$B_{\rm min}\approx 10\,\mu$G. On the other hand, 
values of several hundreds of $\mu$G have been inferred from VLBI observations of the central regions of, e.g., the ULIRGs IRAS~23365+3604 \citep{romero-canizales12a} and IRAS 17208-0014 \citep{momjian03}. 

\begin{figure}[htb!]
\center
\includegraphics[width=0.9\textwidth]{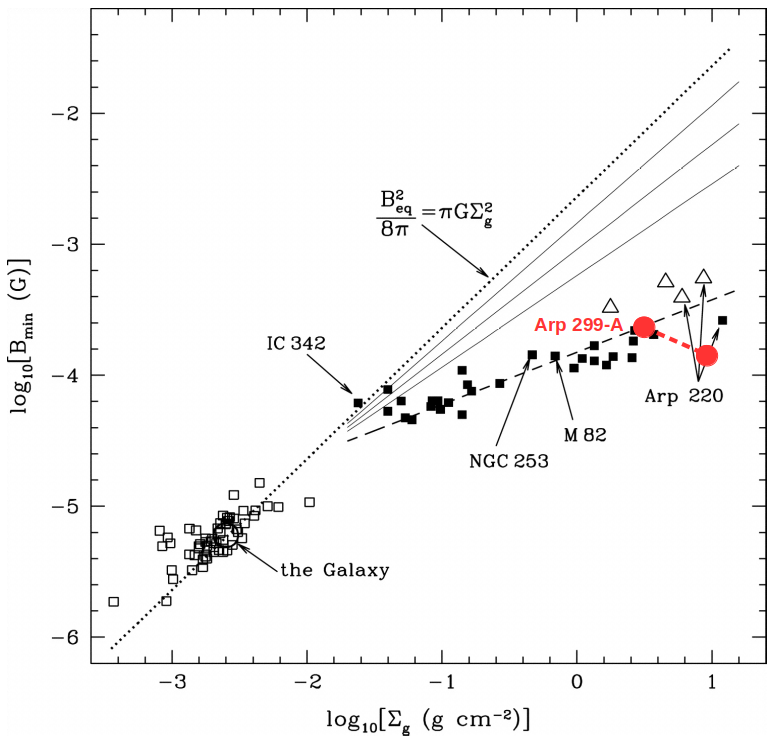}
\caption{ \label{fig:thompson06}
Minimum magnetic flux density ($B_{\rm min}$);
versus measured gas surface density ($\Sigma_g$).
Normal star-forming galaxies (open squares),
starburst galaxies (filled squares), and
the Galaxy ($B_{\rm min}=6$ $\mu$G) at the solar circle, e.g., 
\citealt{Beck2001}; $\Sigma_g\simeq2.5\times10^{-3}$ g cm$^{-2}$, \citealt{Boulares1990}) are shown.  
The dotted line corresponds to the equilibrium between the magnetic field pressure density and that of the gas. The dashed line is the scaling $B_{\rm min} \propto
\Sigma_g^{2/5}$.   The solid
lines show the scalings $B \propto \Sigma_g^a$ (with a = 0.9, 0.8, and 0.7).
The open triangles show, from left to right, $B_{\rm min}$ for IC 883 (Arp 193), Mrk 273,
and the individual nuclei of Arp 220 (West and East), as inferred using $\Sigma_g$ and
the radial size from \citet{downes98} 
and with the radio flux from \citet{Condon1991a} at 8.44 GHz. We also highlight the position of Arp 299-A, as it appears in \citet{thompson06} (left circle), and when the more accurate molecular mass and size estimates by \citet{aalto97} are considered (right circle).  Figure from \citet{thompson06}.
}
\end{figure}

\citet{thompson06} showed that these values of the magnetic field likely underestimate
their true value in U/LIRGs, if their nuclear disks are magnetically supported. In this
case, the pressure of the magnetic field should be very close to the pressure of the gas
in the nuclear disk: $B_{\rm eq} = (8\pi^2   G)^{1/2}\Sigma_g \approx 2 \, \Sigma_g
\,\,{\rm mG}.$ Here, $G$ is the gravitational constant, $\Sigma_g$ is the gas surface
density, and $B_{\rm eq}$ is the magnetic field that balances (`equilibrates') the gas
pressure of the nuclear disk, but is different from the minimum `equipartition' magnetic
field.
\citet{thompson06} estimated the minimum (equipartition) magnetic field, $B_{\rm min}$,
for a sample of galaxies with measured gas surface densities, spanning more than four
orders of magnitude in surface density, from normal spirals to luminous starbursts.
They found that $B_{\rm min}$ ranged between $\sim 60 \mu$G and $\sim 700\mu$G, and that
the ratio of the minimum energy magnetic pressure to the total pressure in the ISM,
$B_{\rm eq}^2/8\,\pi = \pi\,G\,\Sigma^2_g$, decreased substantially with increasing
surface density. 
For compact starbursts, e.g. Arp~299-A or Arp~220, this ratio is very small, $\sim (0.6
- 10) \times10^{-4}$.  \citet{thompson06} concluded that the true magnetic fields in
starbursts are significantly larger than $B_{\rm min}$ and, in the most extreme cases,
the magnetic fields can reach $\sim 20$ mG (Fig.~\ref{fig:thompson06}).
The dashed line in Fig. \ref{fig:thompson06} corresponds to the scaling $B_{\rm min}
\propto \Sigma_g^{2/5}$, which is expected if the cosmic ray electron cooling timescale
is much shorter than the escape timescale from the galactic disk, in which case the true
magnetic field is also significantly larger than $B_{\rm min}$.

The rapid cooling of relativistic electrons in compact starbursts  seems to invalidate
the minimum equipartition estimate for $B_{\rm min}$ \citep{thompson06}. In particular,
the existence of the radio-to-far-IR correlation implies that the synchrotron cooling
timescale for cosmic-ray electrons is much shorter than their escape time from the
galactic disk, which in turn implies that the true magnetic field in starbursts is
significantly larger than $B_{\rm min}$. The strongest argument against such large
fields is that one should expect starbursts to have steep radio spectra ($\alpha
\lesssim -1.0$) indicative of strong synchrotron cooling, which is not observed (e.g.,
\citealt{herrero-illana14}).  On the contrary, ionization and bremsstrahlung losses can
flatten the non-thermal radio spectra of starburst galaxies even in the presence of
rapid cooling, providing much better agreement with observed spectra
\citep{thompson06,Lacki2010a}.  
For small values of the  gas surfaces density,  $\Sigma_g \lesssim 0.1$ g\,cm$^{-2}$,
synchrotron and inverse Compton losses always dominate over bremsstrahlung and
ionization losses, and the observed radio spectrum is steep. However, for $\Sigma_g
\gtrsim 0.1$ g\,cm$^{-2}$, bremsstrahlung and ionization losses dominate over
synchrotron and inverse Compton losses, which causes a flattening of the radio spectra
(see Fig.~2 in \citealt{Lacki2010a}), in agreement with radio observations of LIRGs.
\citet{thompson06} and \citet{Lacki2010a} showed that those facts conspire to preserve
the linearity of the radio-to-far-IR correlation.

\subsection{Low-frequency radio continuum observations}
\label{sec:radio-lowfreq}

The scarcity of radio interferometric facilities operating at frequencies below 1 GHz
has been due to a number of reasons, including a poorer angular resolution of the
instrument at low-frequencies ($\nu \lesssim 1.0$ GHz) with respect to higher-frequency
radio observations, the existence of radio frequency interference,  and the severe
ionospheric effects over large distances. Fortunately,  the construction of the Giant
Metrewave Radio Telescope (GMRT\footnote{\url{http://www.gmrt.ncra.tifr.res.in/}}) in
India,  and more recently the Low Frequency Array
(LOFAR\footnote{\url{http://www.lofar.org/}}), in the Netherlands and the Murchison
Widefield Array (MWA\footnote{\url{https://www.mwatelescope.org/)}} in Australia, has
opened an avenue to study in detail the universe in the low-frequency regime (the GMRT
operates at $\sim$200, $\sim$600, and $\sim$1400 MHz, while LOFAR operates mostly at
around 150 MHz and MWA at 70-300 MHz).

The GMRT yields at most 1 arcsecond angular resolution (at 1.4 GHz), and therefore
cannot be used to obtain high-spatial resolution images of LIRGs. However, the access to
the largely unexplored regime of low-frequencies makes it useful for obtaining radio
spectra of LIRGs.  \citet{Goyal2014} observed with the GMRT a sample of 11 LIRGs,
constraining the radio spectra down to 150 MHz. They found that the median spectral
index between 150 and 325 MHz was slightly flatter ($\alpha \approx-0.34$) than the
median spectral indices between 325 and 610 MHz ($\alpha \approx-0.59$) and between 610
and 1400 MHz ($\alpha \approx-0.64$), the latter being close to optically thin
synchrotron emission. 

The extension of LOFAR to include international baselines (the International LOFAR
Telescope, ILT) yields angular resolutions of about 0.2 arcsec  at 150 MHz, a record in
angular resolution at those frequencies that will remain so until the full Square
Kilometre Array (SKA\footnote{\url{https://www.skatelescope.org/}}) is realized.  The
deep sensitivity and high angular resolution of LOFAR has had a significant impact in
the study of LIRGs. For example, the ability of LOFAR to probe free-free absorption
provides a new way to probe the geometry of the thermal ionised gas in star-forming
galaxies.

\citet{Varenius2016} obtained the first sub-arcsecond imaging of Arp~220 at $\sim$150
MHz with LOFAR, and derived thermal fractions at 1 GHz of less than 1\% for its nuclei.
The simplest explanation for this result is that a significant fraction of the ionizing
photons produced by young, massive stars are absorbed by dust before they are able to
produce any ionization. This will effectively lower the free-free fraction.
Ram\'irez-Olivencia et al. (submitted to A\&A) finds similarly small values of the
thermal fraction at 1 GHz for all four nuclei in Arp~299, also suggesting that UV-photon
absorption by dust plays a relevant role.

Recently, \citet{ramirez18} observed the LIRG Arp~299 with LOFAR at 150 MHz and found an
intriguing two-sided, filamentary structure emanating from the A-nucleus,
up to about 5 kpc from the A-nucleus in the N-S direction (see Fig.~\ref{arp299a-nicmos}
and Sect. \ref{sec:casestudies}).  From energy arguments, \citet{ramirez18} found that
the low-luminosity AGN in Arp~299-A could not drive the outflow, but that the powerful,
compact starburst in the central regions of Arp 299-A could provide enough mechanical
energy to sustain an outflow, and  concluded that the intense SN activity in the nuclear
region of Arp 299-A was driving the outflow.

\subsection{OH megamaser emission}
\label{sec:OH-masers}

It is not uncommon for LIRGs to show OH megamaser emission in their inner regions.  Most
OH megamasers are found in the 18 cm ground state main lines (1665 and 1667 MHz). Those
masers are suggested to be pumped by IR emission from the dust, with some aid by the
radio continuum.  A simple picture suggests that bright IR radiation may transfer OH
population from the ground state to excited states in the $^2$$\pi_{1/2}$ and
$^2$$\pi_{3/2}$ ladders. The ground state levels may then become inverted after
background continuum is absorbed and there is a spontaneous decay into the ground states
\citep[e.g.,][]{henkel90}. The actual pumping process is likely more complex.

The OH maser luminosity appears to correlate with the infrared luminosity
\citep{martin88,baan89a}: $L_{\rm OH} \simeq c_1 \times L_{\rm IR}^2 $, where
$c_1=10^{-21.6}$ \lsun$^{-1}$.  There are suggestions that the correlation is not
exactly quadratic, but instead closer to linear \citep{kandalian96} and/or that the $
L_{\rm IR}$ dependency may vary with luminosity. As for example discussed in
\citet{baan89a}, a quadratic relationship is expected for unsaturated amplification of
centrally concentrated radio continuum by foreground molecular gas.

The OH masing regions are largely believed to be star forming, although further studies
are still required.  For instance, some of the OH maser emission of the ULIRG Arp~220 is
believed to be emerging in a collimated molecular outflow \citep{baan89b,rovilos03}. The
occurence of OH megamaser activity is also linked to IR colours
\citep[e.g.,][]{henkel86,henkel90}.

\section{Spectral line studies and mm and submm wavelengths}
\label{sec:molecular}

\subsection{Surveys}

\subsubsection{CO surveys and probing molecular mass}
\label{subsec:molecular_line}

The $\nu=115$ GHz (3~mm) 1--0 rotational transition of carbon monoxide (CO) is a
fundamental tracer of molecular gas\footnote{For a discussion of the rotational spectra
of diatomic molecules, see chapter 15 in \citealt{wilson13}.}. Since cold H$_2$ has no
permanent dipole moment we need to resort to a tracer molecule to chart the distribution
and kinematics of cold H$_2$ in nearby and distant galaxies. The CO line is also used to
estimate molecular mass under the assumption of self-gravitating clouds with normal
metallicity \citep[e.g.,][]{bolatto13b}. The work to calibrate the CO to M(H$_2$)
conversion factor has been going on for a long time involving studies of the impact of
e.g. metallicity, gas temperature, and level of turbulence. A recent study of U/LIRGs
suggest a CO-to-H$_2$ conversion factor \citep{herrero19}:

\begin{equation}
\alpha({\rm CO})=1.8^{+1.3}_{-0.8}\, M_{\odot} ({\rm K}\,{\rm km}^{-1}\,{\rm pc}^{2})^{-1} 
\end{equation}

\noindent This is a factor of  $\sim$2.5 lower than $\alpha$(CO) for molecular clouds in
the Milky Way (see Fig.~2 in \citealt{bolatto13b}). The result by \citet{herrero19}  is
based on comparing the median gas-to-dust ratio for U/LIRGs to that of a control sample
of local, normal galaxies. The error bars include an earlier, lower value of
$\alpha({\rm CO}) \approx 0.8\pm0.3$ for U/LIRGs \citep{downes98,papadopoulos12} based
on studies considering dynamics and turbulent gas. These studies suggest that a
significant fraction of the CO emission is emerging from a continuous,
non-self-gravitating medium, and that the entire molecular medium is likely a
multi-phase medium (see also \citealt{aalto95}).  A standard, Milky Way $\alpha({\rm
CO})$, which is calibrated for a medium of self-gravitating clouds, will therefore
overestimate the molecular gas mass for such a medium.

This serves to illustrate that there is still a significant uncertainty in $\alpha({\rm
CO})$ even following the recent surveys. To improve values of the conversion factor it
is necessary to study the molecular clouds in LIRGs at a spatial resolution that allows
us to determine their properties, and how they vary with position in the galaxy. It is
also important to use additional probes of the molecular gas mass.

Examples of alternative tracers of the molecular gas include the 850 $\mu$m dust
continuum \citep{scoville17} and the fine structure lines of atomic carbon, [CI] (1--0)
$^3P_1 - ^3P_0$ and $^3P_2 - ^3P_1$ transitions (at 609 $\mu$m (492 GHz) and 370 $\mu$m
(809 GHz)) \citep[e.g.,][]{papadopoulos04}.  Also the [CII] 157.737 $\mu$m
fine-structure line may probe molecular gas not readily traced by CO ( `CO-dark' gas),
for example in low metallicity dwarf galaxies \citep[e.g.,][]{cormier14,langer17}.

There are large numbers of single dish surveys of the CO line emission in LIRGs. Many
studies date back to the early 1980s when the first surveys of CO emission in IR
luminous galaxies were carried out, finding a correlation between $L$(CO) and $L_{\rm
IR}$ \citep[e.g.,][]{young86}. These correlations were suggested to be indications of
both that  CO is a good tracer of the molecular gas mass and that the origin of the IR
luminosity was closely linked to the molecular gas -- in the form of embedded star
formation and/or AGN activity. Since then, many more surveys and studies of CO in LIRGs
have been carried out.  Further studies started to show deviations with increasing
$L_{\rm IR}$,  e.g. \citet{yao03}, who found this to be consistent with the Schmidt law
relating star formation rate to molecular gas content. The properties of the molecular
gas content in high redshift galaxies have recently been reviewed by \citet{combes18}.

\subsubsection{Surveys of dense gas}
\label{subsec:dense_surveys}

As instruments and telescopes have improved and become more sensitive, it has also
become possible to search for other lines such as HCN, HCO$^+$
\citep[e.g.,][]{rickard77,rieu92} and CS \citep[e.g.,][]{henkel85}. These molecules have
large dipole moments and their ground state transitions have high critical densities,
$n_{\rm cr}$. For example, the $\nu$=88.6 GHz 1--0 line of HCN has $n_{\rm cr}$ in
excess of $10^5$ cm$^{-3}$. This has to be compared with the CO(1--0) transition, where
$n_{\rm cr}$ is two orders of magnitudes lower. Therefore, emission lines of  HCN,
HCO$^+$ and CS are often used to probe the dense gas (gas densities $n > 10^4$
cm$^{-3}$) content. This rests on the assumption that collisions with H$_2$ is the main
excitation mechanism. High-dipole moment molecules may also become excited by collisions
with electrons \citep[e.g.,][]{goldsmith17}, while remaining unimportant for CO.
Infrared pumping may also impact the rotational excitation of molecules
\citep[e.g.,][]{carroll81,aalto07}.  It is also the case that the critical density is
affected by the abundance of the molecule \citep[e.g.,][]{mangum15}.

In a large survey of LIRGs, \citet{gao04}, found that $L$(HCN) correlates better with IR
luminosity than CO (Fig.~\ref{fig:HCN_N1614}). They proposed that this is due to the
IR-emission being more closely tied to the dense, presumably star forming, gas.  Some
studies of the efficiency of star formation in the dense molecular gas (SFE$_{\rm
dense}$) suggest that the SFE$_{\rm dense}$ of extreme starbursts in LIRGs is a factor
3--4 higher compared to normal galaxies \citep{garcia12}.

\medskip More studies of the dense gas, for example of CS and HCO$^+$, showed in some
cases a more complex picture. Most notably, an elevated HCN/HCO$^+$ 1--0 line intensity
ratio has been suggested to indicate the presence of an AGN
\citep[e.g.,][]{kohno01,krips08} while \citet{privon15} found no correlation between
elevated HCN/HCO$^+$ ratios and the nature of the embedded activity. The line ratio is
also often found to be elevated in ULIRGs, but perhaps more importantly to vary
significantly among U/LIRGs \citep[e.g.,][]{gracia08}. The behaviour of the HCN/HCO$^+$
line ratio and its diagnostic value has been extensively studied at high spatial
resolution (see Sect.~\ref{sec:high_res_mm}). To probe the activity and evolution of the
dense gas in galaxies it is also important to study less abundant species and
astrochemistry such as for example the isomer of HCN - HNC, HC$_3$N and CN (e.g.,
\citealt{aalto02,loenen08,baan10,costagliola11}; see also Sect.~\ref{sec:high_density}).

\subsubsection{Surveys of isotopic lines}
\label{subsec:isotope_surveys}

Molecular gas properties can also be probed using isotopomers, e.g. $^{12}$C$^{16}$O,
$^{13}$C$^{16}$O, and $^{12}$C$^{18}$O \citep[e.g.,][]{young86b}. Their line ratios are
useful to study the physical conditions in the molecular gas, e.g., temperature and
density distributions in the ISM. Elevated values of the $^{\rm 12}$CO-to-$^{\rm 13}$CO
line ratio ($\mathcal{R}$) have been found in luminous merging galaxies
\citep[e.g.,][]{aalto95}. This can be caused by reduced line opacities due to warm,
turbulent, high-pressure gas (and gas flows) in the merger centre
\citep[e.g.,][]{aalto95,aalto97,aalto10,davis14,koenig16}. There is strong evidence that
$\mathcal{R}$ is higher in galaxies with higher dust temperatures
\citep[e.g.,][]{young86b,aalto95,costagliola11,herrero19} and it can also be linked to
SFR surface density \citep{davis14}. The relative abundances of $^{\rm 13}$CO and $^{\rm
12}$CO can also be impacted by photo dissociation \citep[e.g.,][]{aalto95} where $^{\rm
13}$CO is being selectively destroyed since it is less able to self-shield than the more
abundant $^{\rm 12}$CO). 

Isotopic line ratios are also potential probes of enrichment and initial mass functions
(IMF).  Elevated $^{12}$C/$^{13}$C abundance ratios are suggested to be caused by the
presence of low-metallicity gas and/or a top-heavy IMF
\citep[e.g.,][]{casoli92,henkel93,sliwa17,zhang18}. The notion is that $^{13}$C is
produced in intermediate-mass  stars ($<8\,M_{\odot}$) while $^{12}$C is generated in
both intermediate and high-mass stars ($>8\,M_{\odot}$). Thus, in low-metallicity gas
the stars have not yet had time to go enrich the gas in $^{13}$C. In top-heavy IMFs, an
excess of $^{12}$C is produced in a relatively higher number of massive stars
\citep[e.g.,][]{wilson92}.  It is also useful to study carbon isotopes in other species
than CO, for example CN \citep[e.g.,][]{tang19}.  An elevated oxygen isotopic ratio
$^{\rm 18}$O-to-$^{\rm 16}$O is proposed to trace the selective chemical enrichment by
massive stars and effects of IMF \citep[e.g.,][]{gonzalez14, henkel14, falstad15,
falstad17, koenig16, sliwa17}.  However, using isotopic lines to study the rise of
metals and IMFs requires that the impact of line opacity and the potential presence of a
multiphase molecular medium can be accounted for. High-resolution and multiple-line
studies are therefore important in separating the effects of abundance from those of
cloud structure and excitation (see Sect.~\ref{subsec:isotopes}).

\subsection{High-resolution studies}
\label{sec:high_res_mm}

\subsubsection{CO imaging} A full understanding of what activity the molecular gas is
involved in (and feeding) requires higher resolution studies so that molecular line
intensities and kinematics can be resolved and associated with specific physical
processes.

\smallskip

High-resolution CO studies of LIRGs show a variety of morphologies and dynamics where
the interacting and merging LIRGs have the most complex and intriguing structures
\citep[e.g.,][]{wilson08}. Molecular gas is found in spiral structures, rings, bars
(lining them, and in orbit crowding regions, at bar ends), and in circumnuclear disks
(CNDs). Molecular gas is also found to be inflowing on small and large scales -- and in
outflows of various types (see Sect.~\ref{subsec:coldmoloutflows}).  High resolution CO
imaging of major merger LIRGs reveal massive gas concentrations in their nuclei as well
as in the regions where disks overlap -- for example in Arp~299
\citep[e.g.,][]{aalto97,sliwa12}, in NGC~6240
\citep[e.g.,][]{feruglio13,saito18,treister20}, VV114 \citep[e.g.,][]{yun94,sliwa13},
and in ``The Antennae'' (NGC~4038/39) \citep[e.g.,][]{wilson00,ueda12,herrera12}.

\begin{figure}[h] \begin{center} \includegraphics[width=5.9cm, trim = 1cm 2.5cm 11cm
  1cm, clip]{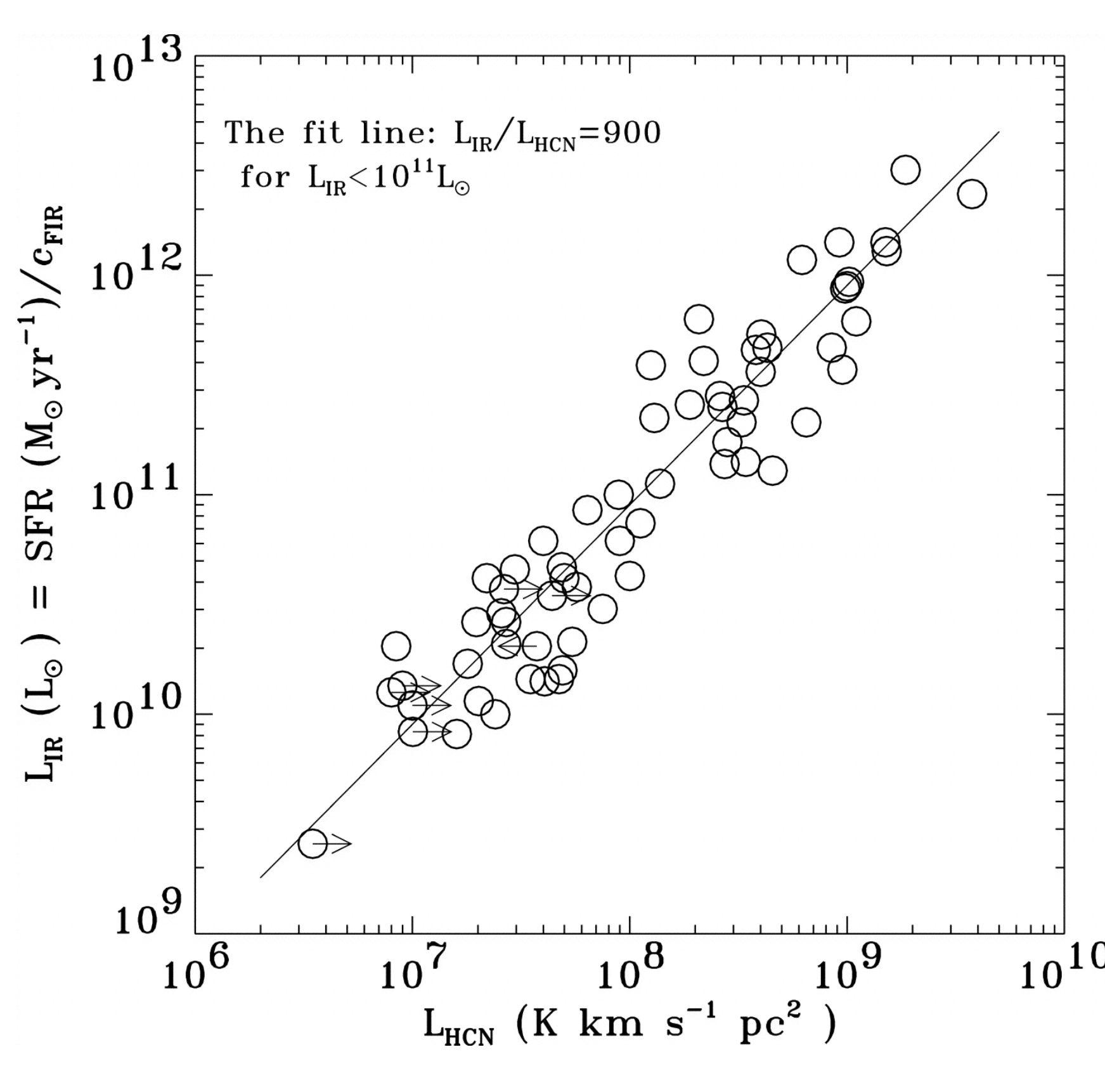}
\includegraphics[width=5.3cm, trim = 4cm 0.5cm 5cm 1cm, clip]{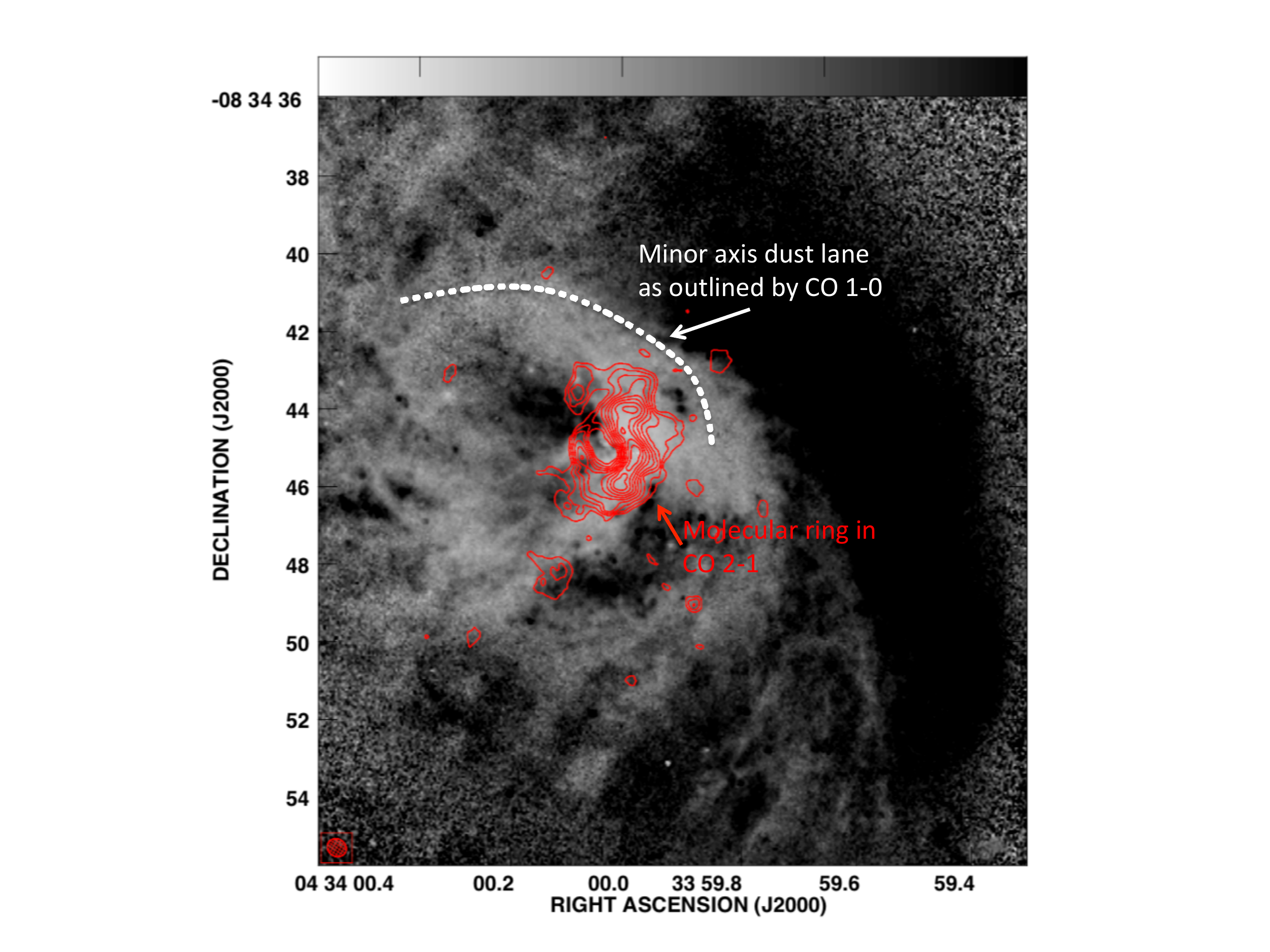}
\end{center} \caption{ \textbf{Left.} Correlation between the ground state HCN 1-0 line
  emission and IR luminosities in 65 galaxies (limits in HCN luminosities are indicated
  with arrows). The sample is divided into U/LIRGs and less luminous, ``normal'' spiral
  galaxies.  A single slope fits HCN data for both low and high IR luminosities. (From
  \citealt{gao04}).  \textbf{Right.} Overlay of the SMA CO 2--1 integrated intensity
  emission (red contours) on the HST F435W/F814W filter colour-map image of NGC~1614.
  The CO 2--1 beam size  ($0.''5$ by $0.''44$) is shown in the lower left corner. The CO
  2--1 is distributed in a ring-like structure of radius $r=230$ pc. The molecular ring
  is uneven with most of the mass on the western side, which also contains giant
  molecular associations extending into a pronounced minor axis dust lane. The molecular
  ring traced by CO 2--1 is also very prominent in radio emission (e.g.,
  \citealt{olsson10,herrero-illana14}; see also Fig.~\ref{fig:lirgs-vla}). In contrast,
  the lower excitation CO 1--0 line emission correlates well with the more extended,
  dust lane (marked with a dashed white line; \citealt{olsson10}), which is likely
  fuelling the central activity \citep{koenig13}.  } \label{fig:HCN_N1614} \end{figure}

In minor mergers there is also evidence of a variety in molecular morphology. One
example is NGC~1614, 
which is an apparent collision between a large and small spiral galaxy. The interaction
has led to the formation of a spectacular circumnuclear starburst ring
\citep{alonso-herrero01, pereira15}, which is also prominent in the radio continuum
(e.g., \citealt{olsson10,herrero-illana14}; see also Fig.~\ref{fig:lirgs-vla}) as well
as in molecular lines \citep[e.g.,][]{koenig13,xu15,garcia15}. There is a molecular ring
with radius $r=230$ pc \citep{koenig13} which is fed by a large-scale, minor axis inflow
of cold molecular gas (Fig.~\ref{fig:HCN_N1614}). The inflow occurs along a dust lane
which is likely a polar ring induced by the interaction
\citep[e.g.,][]{koenig13,koenig16}. The inner region of the molecular ring appears
devoid of gas which may be partially an effect of the starburst evolution
\citep{alonso-herrero01} and partially caused by gas dynamics \citep{koenig13}.  The
massive minor axis in NGC~1614 appears to contain mostly diffuse, unbound molecular gas
that is unable to form stars. This way large amounts of gas can be transported to the
centres of mergers, without being consumed in star formation along the way
\citep{koenig16}. Since the gas is likely unbound, a standard CO-to-H$_2$ conversion
factor (see Sect.~\ref{subsec:molecular_line}) does not apply and would overestimate the
molecular mass.

The central regions of LIRGs often harbour an intense and compact starburst region --
either in the form of a ring as in NGC~1614 (see above), or in even more compact
structures (see also Sect. \ref{sec:radio-high-resolution}).  The molecular gas surface
densities are often very high, sometimes exceeding $10^4$ \msun\ kpc$^{-2}$ and early on
this was linked to warm IRAS colours \citep[e.g.,][]{bryant99}.  Compact nuclear
molecular gas structures are also strongly linked to AGN activity. The most well studied
case is the iconic Seyfert LIRG NGC~1068
\citep[e.g.,][]{sternberg94,tacconi97,garcia14,gallimore16,imanishi18,
impellizzeri19,garcia19}.

\begin{figure}[h]
\begin{center}
\includegraphics[width=\textwidth, trim = 0cm 8cm 0cm 0cm, clip]{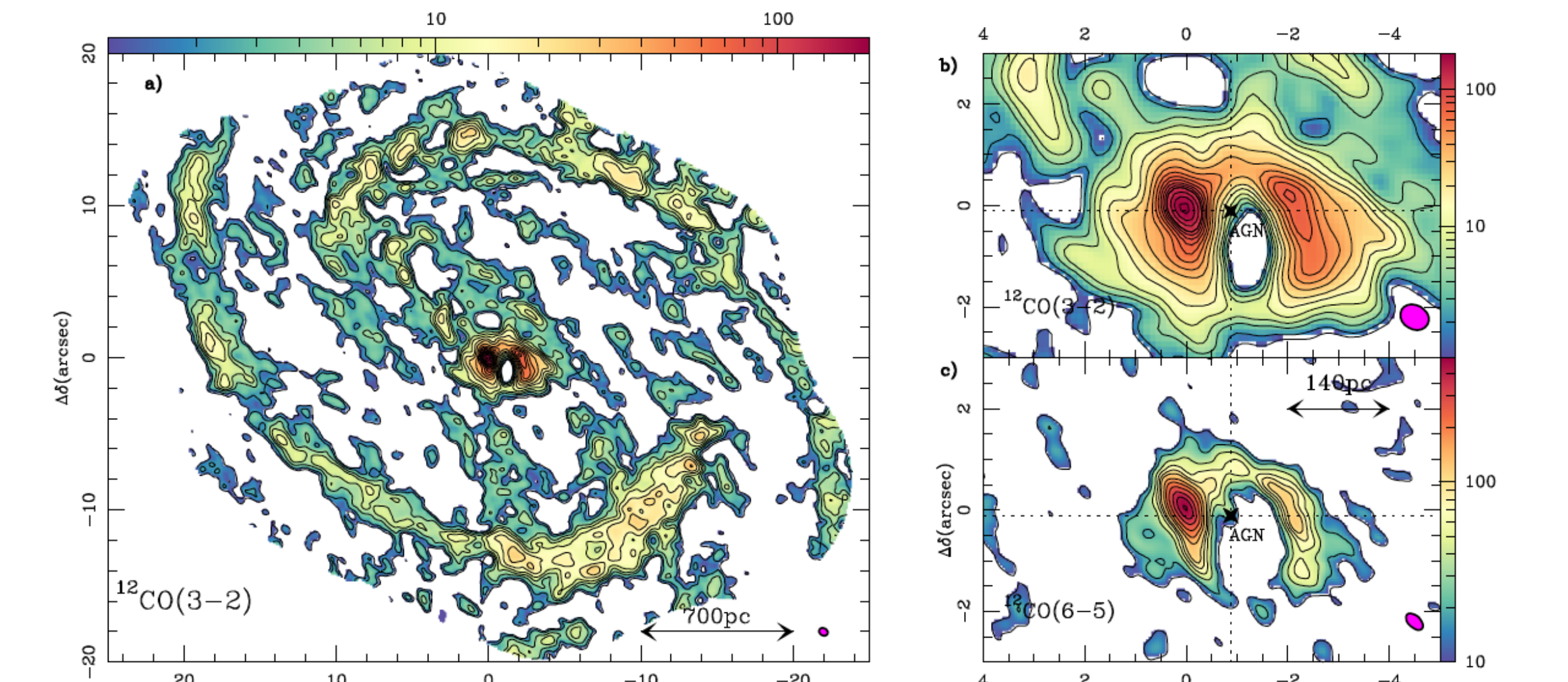}
\includegraphics[width=\textwidth, trim = 0cm 10cm 0cm 0cm, clip]{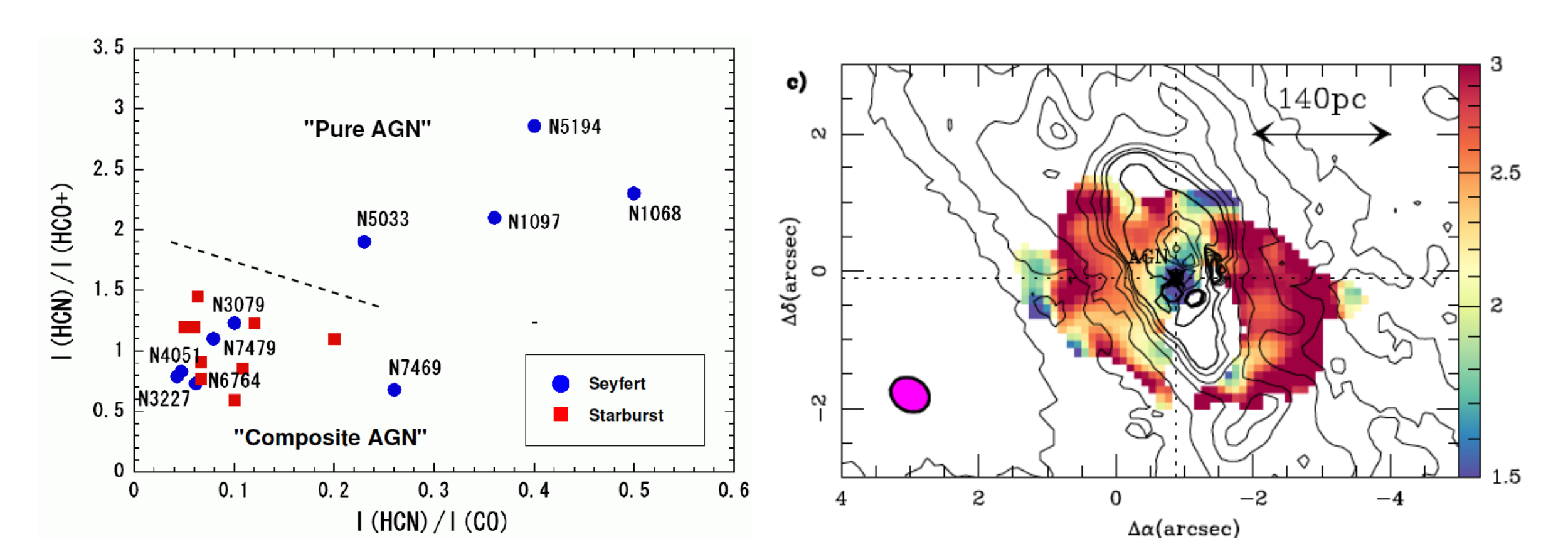}
\end{center}
\caption{\textbf{Top left.} ALMA CO 3--2 integrated intensity map (mosaiced) of the disk and nucleus of NGC~1068. The filled ellipse at the bottom right corner indicates the $0.''6$ by $0.''5$ beam size; \textbf{Top right.} CO 3--2 zoom-in on the circumnuclear disk (CND); Single field CO 6--5 integrated intensity map of the CND, the beam size is $0.''4$ by $0.''2$  \citep{garcia14}
\textbf{Bottom left.} Plot of the ground state 1--0 $I$(HCN)/$I$(HCO$^+$) vs $I$(HCN)/$I$(CO) brightness temperature line ratios for a sample of luminous galaxies -- AGNs, Starbursts and Composites \citep{kohno05}. \textbf{Bottom right.} HCN/HCO$^+$ 4--3 brightness temperature ratio map at $0.''6\times0.''4$ resolution 
\citep{garcia14}.}
\label{fig:N1068_HCN}
\end{figure}

NGC~1068 is a nearby ($D_L = 14$ Mpc) LIRG with $\log (L_{\rm IR}/L_\odot)=11.29$ that
is often viewed as the prototypical Seyfert~2 galaxy.  It has a prominent starburst ring
of $\sim$1--1.5 kpc radius and a 200 pc molecular circumnuclear disk surrounding the AGN
\citep[e.g][]{schinnerer00,garcia14}  (Fig.~\ref{fig:N1068_HCN}). CO emission is tracing
non-circular motions that can be attributed to in- and outflow motions. A radio jet
emerges from the nucleus, entraining and pushing on the molecular gas in the disk (see
Sec~\ref{subsec:coldmoloutflows}). The nuclear gas is dense ($n \gtrsim 10^{5-6}$
cm$^{-3}$) and luminous in HCN emission. The chemistry and physical conditions of the
gas is strongly impacted by the AGN, either by X-rays or shocks
\citep[e.g.,][]{tacconi94,sternberg94,usero04,aalto11, viti14,garcia17}. Recent
high-resolution ALMA studies allow us to image dust and molecular gas in the central
region of NGC 1068 at unprecedented spatial resolution.  Its dusty torus has now been
imaged at sub-mm wavelengths with ALMA for the first time \citep{garcia16,gallimore16}.
On small scales, the nuclear dynamics is very complex for example showing evidence of
apparent counter-rotation, which is either due to actual counter-rotating gas, infall or
an outflowing torus \citep{imanishi18, impellizzeri19,garcia19,imanishi20}.

\subsubsection{High-density tracers and astrochemistry}
\label{subsubsec:high_density}

Imaging emission from high-dipole moment molecules (see
Sect.~\ref{subsec:dense_surveys}) reveals how the dense gas is distributed, which can be
linked to star formation, SMBH feeding and dynamics. In general, the dense molecular gas
is concentrated in the inner region of LIRGs, although interacting systems may have
significant dense gas masses also in the regions where the galaxies overlap (e.g.,
Arp~299).  The dense gas is also where a large fraction of astrochemical processes are
occuring, and astrochemistry is an important, new tool for diagnosing embedded star
formation and AGN activity.

There are a number of standard scenarios often referred to when discussing extragalactic
astrochemistry \citep[e.g.,][]{aalto13,aalto15c}:

\textit{Photon (or Photo) dominated regions (PDR).}  These regions are affected by
far-ultraviolet photons (h$\nu$=6--13.6~eV) and are dominated by photo-chemistry
\citep[e.g.,][]{hollenbach97}. They have layered structures where the surface
temperatures can be quite high (300--1000~K) but bulk temperatures remain relatively
modest (20--50~K).

\textit{X-ray dominated regions (XDR)} are impacted by X-rays with h$\nu$=1--100~keV
which have large penetration depths resulting in larger bulk temperatures ($>$ 100~K)
compared to PDRs.  The chemistry is characterized by ion-neutral chemistry stemming from
irradiation by X-rays \citep[e.g.,][]{maloney96,meijerink05}.

\textit{Cosmic ray dominated regions (CDR or CRDR)} are regions of elevated ($>10^3 \, \times$ Galactic value) cosmic ray energy density \citep[e.g.,][]{suchkov93,meijerink11} likely originating from SNe. This results in ion-neutral chemistry that sometimes can be difficult to distinguish from that of XDRs.

\textit{Mechanically dominated regions, MDRs.}   Here, the chemistry is impacted by the
dissipation of shocks and outflows and it reflects the speed of the shock and thus the
level of grain processing \citep[e.g.,][]{viti11,kazandjian12}.  Fast shocks lead to the
destruction of the grain core (releasing silicates) while slower shocks result in the
evaporation of icy mantles (e.g. releasing H$_2$O, H$_2$S and CH$_3$OH). For a
discussion of fast and slow shocks see, e.g., \citet{lehmann16}.

\textit{Dense, shielded regions.} Warm, shielded regions with large H$_2$ column density. If they are warm (50 to 500~K) the chemistry can be ``hot-core'' like \citep[e.g.,][]{nomura04,viti05}. Evaporation of icy grain mantles impacts chemistry and also allows for the survival of relatively complex molecules (such as some carbon chains). Large columns of dust and high temperatures lead to  intense infrared radiation fields which may excite vibrational transitions of molecules such as HCN and HC$_3$N \citep[e.g.,][]{salter08,costagliola10,sakamoto10,aalto15a,falstad19,aalto19}.

\medskip
\noindent

To identify which scenario is dominating the astrochemistry of a region usually requires
multiple molecular lines
\citep[e.g.,][]{kazandjian12,aalto15b,viti14,viti16,harada18b}. Spectral scans that
include many species in one tuning allow us to study groups of species together. One
such example is the obscured LIRG NGC~4418 \citep{costagliola15} where 317 emission
lines from a total of 45 molecular species were detected. However, such scans are still
rare and it is more common to use line ratios of fewer species to study the chemical
impact of buried activity and dynamics on the molecular gas. Interferometric studies of
molecular line ratios provide both spatial resolution and sufficient astrometric
accuracy to allow us to confidently separate and identify regions of different dominant
chemical processes \citep[e.g.,][]{meier05,meier12,watanabe14,viti14,harada18a}.  Below
are some examples of diagnostic lines, and line ratios, and the challenges we still face
in their interpretation.

\medskip \noindent \textit{HCN and HCO$^{+}$.} As discussed in
Sect.~\ref{subsec:molecular_line}, elevated HCN/HCO$^{+}$ 1--0 intensity ratios have
been found around some AGNs (e.g., \citealt{kohno01,kohno03,imanishi09}; see also
Fig.~\ref{fig:N1068_HCN}). The HCO$^{+}$ abundance has been suggested to become both
suppressed and enhanced (relative to HCN) in XDRs \citep[see, e.g.,][]{maloney96,
meijerink05}, and HCN is also expected to be enhanced in warm and shocked environments
\citep[e.g.,][]{aalto12a,kazandjian12} in MDRs.  High-resolution ($0.''6$ by $0.''5$)
observations of the HCN/HCO$^+$ 4--3 line ratio in the LIRG Seyfert NGC~1068 show that
the highest ratios ($\sim$3) are found in the  circumnuclear disk (CND), according to
\citet{garcia14}, but at some distance away from the AGN. Within the CND, the ratio
drops with decreasing radius to a value of 1.3 at the locus of the AGN \citep{garcia14}.
The ratio then increases again on even smaller scales \citep{imanishi16}. This shows
that the interpretation of the line ratio is complex and that it may be governed by
multiple processes.  Indeed, elevated ($>$1) HCN/HCO$^+$ line ratios are also found in
starburst galaxies \citep{privon15} and the detection of self-absorbed HCN and HCO$^+$
in compact obscured nuclei with $N$(H$_2$)$>10^{24}$ cm$^{-2}$ (Sect.~\ref{sec:con})
\citep{aalto15b,martin16,aalto19} shows that more work is required in interpreting this
line ratio.

\smallskip \noindent \textit{HCN and HNC:}\, The isomer of HCN, HNC, is expected to be
more abundant than HCN in cold ($T< 24$~K) gas, while in dense warm and/or shocked gas
$X$(HCN)$>>X$(HNC) \citep{schilke92}. However, when the chemistry is dominated by
ion-neutral processes $X$(HCN)$\simeq X$(HNC) \citep{meijerink05} independent of
temperature, hence complicating the use of HNC as a ubiquitous temperature tracer.
Global HCN/HNC 1--0 intensity ratios in LIRGs generally lie between 1 and 6
\citep[e.g.,][]{aalto02,baan10}. Cases where the HNC/HCN intensity ratio $>1$  exist
\citep{aalto07,tunnard15} but have been suggested to be caused by mid-IR pumping of HNC.
It is easier to pump HNC than HCN since the bending mode for HNC occurs at a wavelength
$\lambda=21\,\mu$m while for HCN the corresponding mode is at $\lambda=14\,\mu$m (see
discussion in \citealt{aalto07}). It is also possible that in very obscured galaxies the
HCN emission suffers from self-absorption, resulting in a reduced line intensity
compared to that of HNC.

\smallskip \noindent \textit{HC$_3$N and CN: }\, Some LIRGs have unusually bright
emission from cyanoacetylene (HC$_3$N) $J$=10--9 compared to HCN 1--0
\citep{lindberg11,costagliola11}. HC$_3$N is expected to be abundant in dense and
shielded gas since it is destroyed by UV and particle radiation. HC$_3$N has a set of
bending transitions that can be excited by mid-IR emission and vibrational transitions
(e.g., \citealt{costagliola10,aalto19}). Interestingly, bright HC$_3$N emission has also
been found near AGN nuclei such as Mrk~231 \citep{aalto12a}, NGC~1097 \citep{martin15}
and NGC~1068 \citep{takano14}. The strong radiation from the accreting SMBH is expected
to destroy the HC$_3$N molecule. The fact that it is detected suggests the presence of
gas and dust columns large enough to shield the molecule, possibly in the form of a
torus-like structure.  CN is a radical that is (for example) chemically linked to HCN
through photo destruction.  CN enhancements are expected in regions of moderate
extinction with strong radiation fields. The expected abundance enhancement of CN over
HCN is greater in an XDR (factors 40--1000) than in a PDR (CN/HCN abundance ratio range
from 0.5 to 2) \citep{meijerink05} and future high resolution studies will show how good
an XDR tracer CN is. Global surveys (so far) show relatively modest CN enhancements in
LIRGs. Some molecular outflows show elevated CN emission which suggests that they may be
subject to strong radiative impact \citep{sakamoto14,cicone20}.

\smallskip \noindent \textit{SiO, H$_2$O, HNCO, CH$_3$OH, and HCN:} \, These molecules
are often used as shock tracers. Slow shocks result in mantle evaporation releasing
H$_2$O, HNCO, CH$_3$OH and also H$_2$S. The LIRG merger VV114 has spectacularly bright
CH$_3$OH emission in the merger overlap region, likely as a result of large scale shocks
\citep{saito18}. Luminous lines of H$_2$O are found in many LIRGs
\citep{yang13,gonzalez14,falstad15} and shocks may play a role here, but also the
evaporation of ices through irradiation or gas-phase formation of H$_2$O. Stronger
shocks can get Si off the grains allowing it to react with O$_2$ or OH to form SiO
\citep{guillet09}. \citet{meijerink13} also suggest that an elevated CO
line-to-continuum ratio is a useful diagnostic of shock excitation. This rests on the
assumption that shocks do not heat the dust (that produces the continuum) as effectively
as UV radiation does.

\subsubsection{Mapping of isotopic ratios}
\label{subsec:isotopes}

High-resolution maps of CO isotopic line ratios are important to our understanding of
how the ratios probe physical conditions and stellar enrichment in galaxies (see
Sect.~\ref{subsec:isotope_surveys}). Common tracers include the $^{\rm 12}$CO-to-$^{\rm
13}$CO line ratio ($\mathcal{R}$) and $^{\rm 13}$CO/$^{12}$C$^{18}$O (for various
rotational transitions $J$) but it is also becoming more common to study isotopic ratios
for HCN, CN and HCO$^+$ and to also use nitrogen and sulphur isotopes.

To separate effects of excitation and opacity from those of abundances, it is essential
to resolve the emitting regions and also to use multiple lines and transitions.
Correcting for effects of opacity, radiative-transfer and photo dissociation, isotopic
line ratios are potential probes of enrichment and stellar IMF (e.g.,
\citealt{gonzalez14, henkel14, falstad15, falstad17, koenig16, sliwa17}).

There is tentative evidence that inflows in merger LIRGs impact the isotopic composition
in their centres, despite ongoing powerful starbursts
\citep{aalto10,falstad17,koenig16}. High resolution imaging of the minor merger LIRG
NGC~1614 shows that $^{\rm 12}$CO and $^{\rm 13}$CO probe different gas phases where
$^{\rm 13}$CO is primarily tracing self-gravitating gas and giant molecular complexes
(GMCs), while $^{\rm 12}$CO is also probing diffuse, unbound molecular gas
\citep{koenig16}.  Faint or undetected lines of $^{18}$O are inconsistent with the
notion that the $^{\rm 13}$CO/$^{12}$C$^{18}$O line intensity ratio is linked to
luminosity \citep{zhang18} and shows that the behaviour of the lines in relation of IMF
and gas inflows requires further study. Temperature and density effects on isotopic
molecular fractionation will also impact how the line ratios are interpreted (see, e.g.,
Sect.~4.4.1 in \citealt{alatalo15} for a discussion).


\subsection{The most enshrouded objects: Compact Obscured Nuclei (CONs)} 
\label{sec:con}

LIRGs are often highly dust-enshrouded objects. The dust hides a very active
evolutionary phase of nuclear growth in the form of accreting SMBHs and/or powerful
bursts of star formation (see also Sect.~\ref{sec:sed-modelling}). Obscured objects can
be diagnosed and classified using mid-infrared silicate absorption features. Amorphous
silicate grains have peaks in opacity caused by Si--O stretching and the O--Si--O
bending modes which are located at 9.7 and 18\,$\mu$m (see, e.g., \citealt{spoon07}).
The silicate strength can be used to determine level of obscuration (see
Sect.~\ref{subsubsec:silicates}) and lines of H$_2$, CO, CO$_2$, C$_2$H$_2$ , CO$_2$ and
ice absorption can be used as additional tools to study these objects.

There is a population of LIRGs that have extremely obscured nuclei with visible
extinction $A_{\rm V} \gg 1000$ (corresponding to $N$(H$_2$)$>10^{24}$ cm$^{-2}$).  At
these high levels of extinction, mid-IR diagnostic methods suffer from opacity effects
and one has to resort to far-infrared or even sub-mm/mm lines to probe the enshrouded
activity.  The obscuring columns appear to be compact ($r<15$--$100$ pc) and both gas
and dust are warm with temperatures in the range of 100--300 K
\citep{sakamoto08,costagliola10,
sakamoto13,gonzalez12,aalto15b,falstad15,scoville17,sakamoto17,barcos18,aalto19,falstad19}.
These Compact Obscured Nuclei (CONs) may be previously undetected accreting SMBHs and/or
extraordinarily compact and luminous young stellar clusters, possibly also with
top-heavy stellar IMFs \citep{aalto19}. Radiatively excited molecular emission of, e.g.,
H$_2$O and OH \citep{werf10,gonzalez12,veilleux13,gonzalez14,falstad15} are good probes
of the nuclei as long as column densities stay below $N$(H$_2$)$=10^{25}$ cm$^{-2}$. 

However, several CONs have $N$(H$_2$) in excess of $10^{25}$ cm$^{-2}$ and in some cases
column densities may reach extreme values of $N$(H$_2$)$>10^{26}$ cm$^{-2}$
\citep{scoville17,sakamoto17,barcos18, aalto19}.  This is likely also the case for the
most nearby CON found so far, NGC~4418 (Sakamoto et al in prep.).  To probe the most
deeply enshrouded CONs we need to go to mm and even cm wavelengths, for example using
vibrationally excited lines of HCN \citep{salter08,sakamoto10,
aalto15a,aalto15b,imanishi16,martin16,aalto19,gonzalez19}. CONs are also very rich
sources of molecular emission that can be used to diagnose the nuclear activity
\citep{martin11,gonzalez12,costagliola15}.

CONs have Compton Thick (CT) obscuration on scales of tens of pc, in contrast to X-ray
identified Compton Thick AGNs (XCTs) where the obscuration may occur on scales one order
of magnitude lower \citep[e.g.,][]{risaliti99}. CONs may therefore be galaxy nuclei in a
very different stage of evolution (e.g., SMBHs in a high accretion rate) than XCTs (see,
e.g., discussion in \citealt{Ricci17}). Note that the obscuration in all XCTs may not
necessarily be explained by small scale obscuration. Interacting systems may for example
have obscuration on larger scales \citep{ricci17b,blecha18} which is also a possible
link to the CONs.
 
Even though highly embedded XCTs may be detected in nearby systems
\citep[e.g.,][]{marinucci16}, the extreme column densities found for CONs make X-ray
detections difficult. In particular, the CONs often show dense, opaque and collimated
outflows (see Sect.~\ref{subsec:coldmoloutflows}) that obscure the nucleus in the polar
direction.  Recent studies suggest that the centres of extremely obscured systems are
potential sources of PeV and higher energy cosmic rays that interact in dense nuclear
environments that may yield high energy cosmic neutrinos (see, e.g.,
\citealt{yoasthull17}).   On-going ALMA surveys of, for example, vibrationally excited
emission of HCN and mm/submm continuum, are investigating how common CONs are among
galaxies and what their true nature is.


\subsection{Cold molecular outflows and jets}
\label{subsec:coldmoloutflows}

Feedback from starbursts and AGNs can take the form of a mechanical outflow or wind. The
feedback may help prevent overgrowth of massive galaxies and can (at least partially)
explain the so-called `red-and-dead' properties of present-day ellipticals. In addition,
there are some suggestions that outflows/winds/jets may also regulate SMBH growth and
link it to the host galaxy (the Magorrian relation).

\begin{figure}[h]
\begin{center}
\includegraphics[width=\textwidth, trim = 0cm 10cm 0cm 0cm, clip]{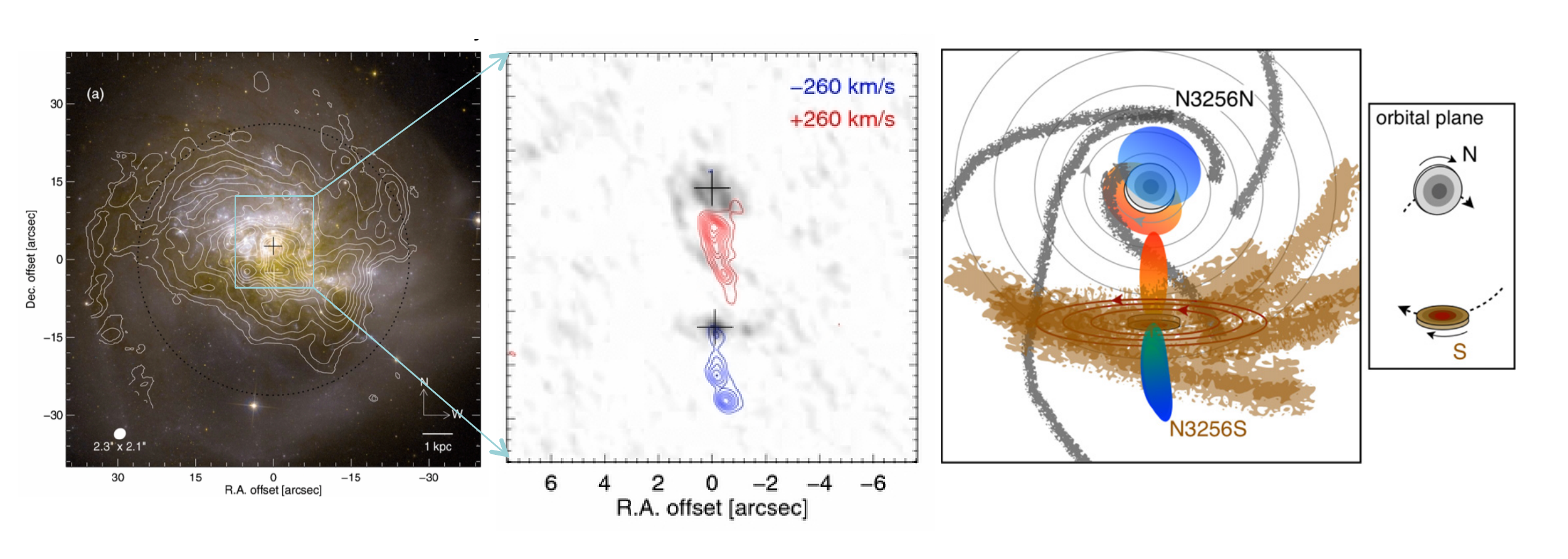}
\end{center}
\caption{Collimated molecular outflow emerging from the southern nucleus of the LIRG merger NGC~3256 \citep{sakamoto14}. The radio nuclei are indicated with plus-signs. \textbf{Left.} CO 3--2 contours on the HST B and I composite image. The dark dotted circle shows the 50\% contour of the ALMA primary beam response, while the beam size of $2.''3$ by $2.''1$ is shown in the lower left corner. \textbf{Middle.} CO 3--2 channel map of high-velocity emission. \textbf{Right.} Cartoon of the NGC~3256 system sketching its outflows and nuclear disks.
}
\label{fig:N3256}
\end{figure}

Many LIRGs have outflows of various morphologies and power. There is evidence that many
are multi-phase, where the cold molecular gas often is responsible for ejecting the bulk
of the ISM mass
\citep[e.g.,][]{fischer10,feruglio10,sturm11,alatalo11,aalto12a,aalto12b,veilleux13,cicone14,morganti15,fluetsch19,lutz20}.
In a recent review on cool winds \citep{veilleux20} the properties and driving forces of
neutral and molecular winds are discussed in detail, in particular from LIRGs and
ULIRGs. Therefore we give a short account here and refer the reader to the review for
more information.

Fast (500--1500 km\,s$^{-1}$) molecular outflows in LIRGs and ULIRGs have been detected
by the \textit{Herschel Space Telescope} \citep[e.g.,][]{fischer10, sturm11,veilleux13}.
The outflows are primarily detected in the 119\,$\mu$m OH lines (as P-Cygni profiles)
and are suggested to be wide-angle ($\sim145^{\circ}$) and fastest in luminous AGNs with
log ($L_{\rm AGN}$/\lsun) 11.8$\pm$0.3 (\citealt{veilleux13}; note that the study mostly
targeted ULIRGs and not LIRGs). Molecular outflows are primarily thought of as either
energy driven or momentum driven, where important pressure terms include ram pressure
from hot gas, UV radiation absorbed by dusty gas, cosmic rays, and impact by radio jets
(see, e.g., the review of driving mechanisms in \citealt{veilleux20}).
\citet{veilleux13} propose UV--IR radiation pressure acceleration as an important
driving force and the fastest molecular outflows are in the dustiest systems
\citep{veilleux13, gonzalez17}.  Radiation pressure is expected to be important in
driving dusty molecular outflows but energy driven processes may play an equal role
(see, e.g., \citealt{fluetsch19}).  Careful, high-resolution studies of the molecular
outflows are essential in determining the driving mechanism \citep[e.g.,][]{sakamoto14}
and that more than one process may be powering the outflow.

Molecular outflows can be mapped at mm/submm wavelengths through imaging its CO line
emission \citep[e.g.,][]{nakai87,walter02,sakamoto06,tsai09,feruglio10,
alatalo11,aalto12b, cicone14, sakamoto14, garcia14,combes13, garcia15,
gallimore16,pereira16, saito18, falstad18, fluetsch19, lutz20,fernandez20}. The presence
of an AGN can boost the outflow rate, and therefore the mass-loading factor, i.e., the
total outflow rate to SFR ratio, by up to  two orders of magnitude \citep{cicone14}.

Recent developments have opened for the study of physical conditions and chemistry of
outflows through imaging other molecules such as HCN, HCO$^{+}$, CN, CS in both AGN and
starburst driven outflows
\citep[e.g.,][]{aalto12a,sakamoto14,aalto15a,matsushita15,lindberg16,harada18a,michiyama18,barcos18,aalto19}.There
is some evidence that gas densities are enhanced in the outflows with respect to the gas
properties in the disks, suggesting that gas compression in outflows may be important.
Extreme conditions with optically thin CO emission are also found in some jet-driven
outflows \citep{dasyra16}.

Examples of well studied LIRG molecular outflows include the twin outflows of the
luminous merger NGC~3256 \citep{sakamoto06,sakamoto14,michiyama18, harada18a}.  The
high-velocity molecular gas in the system consists of two molecular outflows from the
nuclei (Fig.~\ref{fig:N3256}). The one from the northern nucleus is part of a
starburst-driven superwind ($v_{\rm out}>750$ km\,s$^{-1}$) seen nearly pole-on, while
the outflow from the southern nucleus is a highly collimated bipolar jet seen nearly
edge-on and likely powered by a (now dormant) AGN. The southern outflow shows increasing
velocity out to 300 pc from the nucleus where the velocity peaks at a staggering $v_{\rm
out}=2000$ km\,s$^{-1}$. The mass outflow rate, from the combined outflows, exceeds 100
\msun\ yr$^{-1}$.

Collimated (but radio quiet) molecular outflows have been detected in a number of LIRGs
to date \citep[e.g.,][]{sakamoto14,pereira16,falstad17,barcos18}. They appear different
from outflows powered by radio jets and their true nature requires further study. A
recent study by \citet{falstad19} suggests that the outflows of the most obscured
objects (the CONs, see  Sect.~\ref{sec:con}) also have outflows that are collimated
instead of the wide-angle outflows seen for many dusty systems \citep{veilleux13}. These
dense and collimated molecular outflows appear undetectable in the far-IR due to the
obscuration within the outflow itself, and/or because of orientation effects. The
driving mechanisms of the radio-quiet, collimated molecular outflows are not clear. 

Finally, we note that it is now possible to study the chemistry) (see
Sect.~\ref{subsubsec:high_density}) the dense gas in outflows. Chemical differentiation
between HCN, HNC and HCO$^{+}$ in the Mrk~231 outflow (for example) can be partially
caused by shock chemistry \citep{lindberg16}. Enhanced abundances of HCN, SiO and
CH$_3$OH in the outflow regions of the merger NGC~3256 are also attributed to shocks
\citep{harada18a}. 

\section{Models for the spectral energy distribution of LIRGs} 
\label{sec:sed-modelling}

    \subsection{The need for a panchromatic view of LIRGs}
    \label{sec:sed-panchromatic}

Understanding the SED of galaxies in general, and of LIRGs in particular,  requires
careful consideration of the source of their radiated energy. There are three main
sources of energy in LIRGs: a pre-existing stellar population, a generation of newly
formed young stars, and an accretion disk around the supermassive black hole of the
putative AGN. The  stellar population  that pre-existed the starburst phenomenon is
mixed together with the ISM of the galaxy. The  young stars that formed during the LIRG
phenomenon in the last few tens of millions of years, and which usually still reside in
the GMCs from which they formed, or in their immediate vicinity. As discussed in
Sect.~2.3 these star-forming regions are observed in high-resolution mid-IR images.
Finally,  the accretion disk around a supermassive black hole that is associated with an
AGN may be also a powerful energy source.
    
All three types of energy source are associated with large amounts of gas. In the case
of the AGN, the gas is usually assumed to be concentrated in the toroidal structure
predicted by the unified model for AGN \citep{antonucci93} and recently observed with
ALMA \citep{garcia-burillo16,alonso-herrero18,combes19}. A very small fraction (by mass)
of these clouds consists of solid material in the form of dust grains. Dust grains are
much more efficient than gas in absorbing the optical and UV radiation emitted by the
stars or the accretion disc by 4 or 5 orders of magnitude \citep{malygin14}, so the
radiative transfer is dominated by the dust, which absorbs the optical/UV radiation and
reradiates it in the IR and sub-mm ($\lambda \approx 1$--$1000\,\mu$m).
    
To make sense of the phenomena that take place in LIRGs, we need to keep track of all of
the energy emitted and reprocessed. In turn, this requires panchromatic observations of
galaxies ranging from the far-UV to the millimetre. We also need theoretical models for
the emission of the three different environments discussed above, i.e. the general
interstellar medium of a disc or spheroidal galaxy, a starburst consisting of GMCs
centrally illuminated by young stars and AGN tori. Finally, we need methods to compare
models with data.
    
In this section, we discuss panchromatic observations of LIRGs, models of emission, and
methods of fitting them to the data in order to extract meaningful physical quantities
such as star formation rates, stellar masses, and SN rates that help us understand the
LIRG phenomenon.
    
    \subsection{Images and SEDs: two complementary ways of studying LIRGs} 
 
 In nearby galaxies we can study the morphology of galaxies through high-resolution
 imaging (see Sect. \ref{subsec:stellar}, \ref{sec:ionized},
 \ref{sec:radio-high-resolution} and \ref{sec:high_res_mm}) and draw conclusions about
 their nature. However, in general it is not possible --even with the aid of
 high-resolution images-- to infer the contribution of different energy sources in the
 galaxy to the total luminosity, or estimate other physical quantities of interest. The
 main reason is that in systems like LIRGs, and especially for the more distant and
 luminous galaxies, the spatial resolution at mid- and far-IR wavelengths --where most
 of the energy is emitted-- is significantly worse than in the near-IR, and thus of very
 limited use. 
 
 NGC~1068 was the first LIRG where it was possible to study the SEDs of regions
 associated with the starburst and the AGN, separately. The SED of the central few
 hundred parsec nuclear region, which is associated with the AGN, was observed by
 \citet{rieke75}, and the kpc-scale star formation ring associated with the starburst by
 \cite{telesco85}. The SEDs of these regions in NGC~1068 were the first to be fit with
 both starburst \citep{rre93,ks94} and AGN torus models \citep{pk93,gd94,err95}. The AGN
 environment in NGC~1068 has been studied in very good detail in the mid-IR
 \citep{raban09} and sub-mm with ALMA (\citealt{garcia-burillo16,garcia19}; see also
 Sect. \ref{sec:high_res_mm}) and these observations have been used to test radiative
 transfer models for the dusty AGN environment and constrain the torus properties
 \citep{lopez-rodriguez18}. Arp~299 is one of the rare LIRGs where it was also possible
 to study the SEDs of the distinct regions in the galaxy \citep{charmandaris02} and
 carry out decomposition of the SED including  Spitzer spectroscopy \citep{mattila2012}.
 
 In the high-$z$ universe, the spatial resolution achieved even with facilities such as
 ALMA and HST is of  limited use. Indeed, essentially at $z > 1$, $1''$ corresponds to a
 linear size of approximately 8 kpc. Therefore,  SED analysis proves to be a  powerful
 tool  for the study of the nature of galaxies at any redshift. This does not mean
 resolved imaging (and spectroscopy) is not needed. Rather the opposite: the knowledge
 and models developed for, and tested in, nearby galaxies imaged with high resolution
 are very helpful for this analysis.
    
    \subsection{SED de-composition with radiative transfer models and other methods}
    
    The task of carrying out SED decomposition to determine the relative contributions
    from starbursts, AGN tori and the host galaxies is similar for galaxies of any
    luminosity. The methods described in this section are therefore general for galaxies
    of all luminosities. Some of the applications discussed here are for ULIRGs and
    hyperluminous IR galaxies (HLIRGs) with $L_{\rm TIR} > 10^{13} L_\odot$.
  
    The methods used to date for de-composing the SEDs of galaxies can be divided into
    three main categories:

\begin{itemize}   
    \item Methods using radiative transfer models
    \item Methods using energy balance
    \item Hybrid methods
\end{itemize}
    
    In the next two subsections we review the development of radiative transfer models
    for different environments, as well as energy balance methods. We then discuss
    various SED fitting de-composition methods employed to date. 
    
  \subsection{Radiative transfer models}

    The first study that employed radiative transfer models to interpret the spectral
    energy distributions of galaxies was that of \cite{rrc89} who fitted the SEDs of the
    227 IR bright galaxies detected by IRAS in all four bands. They used radiative
    transfer models for a `starburst' and `Seyfert' component computed with the
    spherically symmetric radiative transfer code of \cite{rr80}. They also used an
    approximate model for the interstellar medium emission as a `disc' component. The
    best fit resulting from the combination of a single template for each component was
    sought. \cite{lawrence91} de-composed the SEDs of the AGN NGC~4151 and MKN~590, two
    of the first radio quiet AGN detected in the sub-mm using similar models.
    \cite{rr91} used similar methods to study the SED of the apparently hyperluminous,
    but gravitationally lensed and therefore intrinsically ultraluminous $z=2.286$ IRAS
    galaxy IRAS~F10214+4724. \cite{rr93a} also made the first application of the AGN
    dusty disc models of \cite{err95} to interpret the SED of IRAS~F10214+4724. These
    and other attempts to determine the energy source of IRAS~F10214+4724 with radiative
    transfer models remained inconclusive until a model that also included the Spitzer
    spectroscopy was developed by  \cite{efstathiou06}. As discussed below, this model
    required `polar' dust in addition to the torus and starburst.
    
    \cite{err94} developed a multi-grain model for massive star-forming regions and
    \cite{rre93} used it to fit the starbursts in NGC~1068 and M~82. The same authors
    also presented the first model for a ULIRG, demonstrating that the SED of Arp~220
    can be fitted with a similar model to that used to fit M~82 and NGC~1068, but
    assuming an optical depth a factor of 4 higher. \cite{rigopoulou96} modeled the 10
    most luminous ULIRGs in the IRAS Bright Galaxy Survey sample using the models of
    \cite{rre93}. \cite{pearson96} and \cite{rr01} used this model and its later
    developments to make predictions of source counts at IR and sub-mm wavelengths which
    were used for designing surveys with SCUBA, ISO, Spitzer and Herschel. \cite{ks94}
    presented the first starburst model that included PAHs and small grains and fitted
    the SED of M~82. This model actually described the combined emission of star-forming
    regions called `hot spots' and diffuse dust in a spheroidal geometry. 
    
    \cite{silva98} developed the GRASIL model which also modelled the combined emission
    of star-forming regions and diffuse dust. GRASIL can treat either a spheroidal or a
    disc geometry, and incorporated for the first time a stellar population synthesis
    model. \cite{errs00} incorporated PAHs/small grains and a stellar population
    synthesis model in the model of \cite{err94}, and treated in an approximate way the
    evolution of GMCs due to HII region expansion and SNe. \cite{errs00} also introduced
    the idea that the age of the starburst can be constrained by fitting the SED of a
    galaxy. \cite{sk07} developed further the model of \cite{ks94} and computed a large
    library of models.

    Almost in parallel to the development of the first starburst models, the first
    models for the torus were developed and explored by \cite{pk92}, \cite{pk93},
    \cite{rr93a}, \cite{gd94} and \cite{err95}. \cite{pk93}, \cite{gd94}, and
    \cite{ehy95} presented the first models for the nucleus of NGC~1068 \citep{rieke75}
    with torus models. \cite{alexander99} carried out a decomposition of the SED of Cen
    A with `starburst', AGN torus and `disc' models and \cite{ruiz01} carried out a
    similar analysis for Circinus. \cite{verma02}, \cite{farrah02} and \cite{farrah03}
    decomposed the SEDs of HLIRGs and ULIRGs with the starburst models of \cite{errs00}
    and the torus models of \cite{err95}. \cite{fritz06} developed another widely used
    smooth torus model whereas \cite{nenkova02}, \cite{dullemond05}, \cite{honig06},
    \cite{nenkova08}, \cite{schartmann08} and \cite{hk17} developed clumpy torus models
    of increasing complexity and \cite{stalevski12}, \cite{siebenmorgen15} and
    \cite{stalevski16} two-phase models. 
  
\begin{figure}
\center
\includegraphics[width=\textwidth]{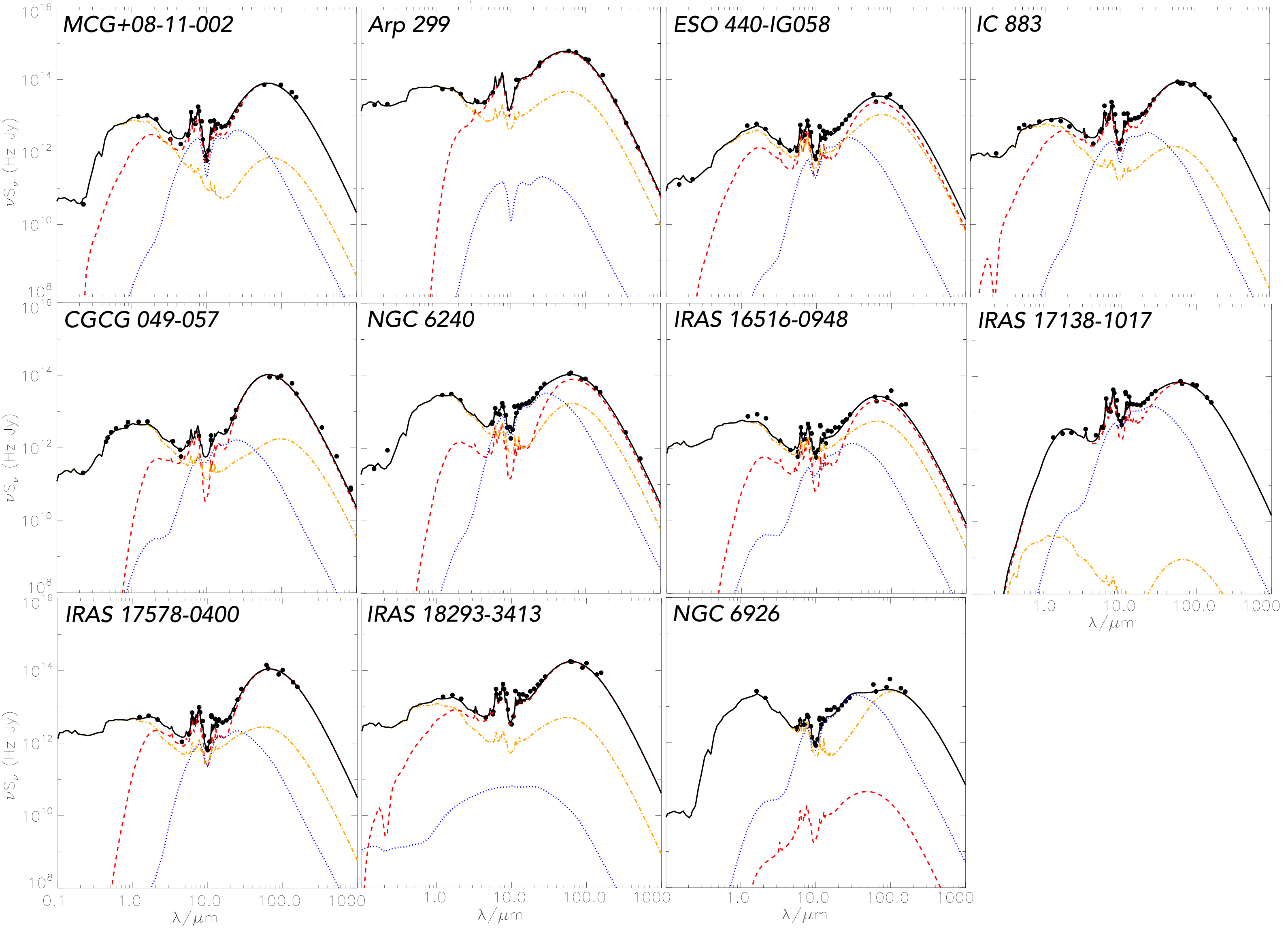}
\caption{Fits to the SEDs of local LIRGs \citep{herrero-illana17} with the CYGNUS radiative transfer models: starburst (red), AGN torus (blue), spheroidal host (orange), total (black). The MCMC SED fitting code SATMC \citep{johnson13} was used for the fits. The models show that the IR luminosity is dominated by the starburst with the AGN torus making a small contribution in the mid-IR in most of the LIRGs. The models predict a median AGN contribution of 10\%, a median SFR of $86.1\,M_\odot$/yr and a median SN rate of 1.07 yr$^{-1}$.}
\end{figure}

     Evidence that the geometry of the dust obscuration in AGN is more complex than the
     `classical' torus of \cite{antonucci&miller85}, which assumed that the polar
     regions above and below the torus are dust-free, was gathering from the early
     1990s. \cite{braatz93} and \cite{cameron93} discovered that a significant fraction
     of the $10\,\mu{\rm m}$ emission from NGC~1068 was actually coming from the
     ionization cones. \cite{ehy95} incorporated a component of `conical' or `polar'
     dust in their model of the nuclear spectrum of NGC~1068 \citep{rieke75}.
     \cite{alonso-herrero01,alonso-herrero03} carried out SED fitting of a number of
     Seyfert~1s and Seyfert~2s, including NGC~1068 with the models of \cite{err95} and
     discussed the impact of adding emission from polar dust adopting a similar approach
     as in \cite{ehy95}.   \cite{efstathiou06} and \cite{efstathiou13} also interpreted
     the SED of IRAS~F10214+4724 with a combination of `polar' dust, AGN torus and
     starburst emission. The ubiquity of `polar' dust in AGN has also been demonstrated
     and discussed by \cite{honig13} and more recently \cite{asmus19}. The model of
     \cite{mattila2018} for Arp~299-B (see Fig.~\ref{fig:Arp299B-AT1} in Sect.
     \ref{sec:timedomain}), the model of \cite{kool2020} for IRAS~23436+5257 and the
     model of \cite{pitchford2019} for a $z\sim 1$ FeLoBAL quasar also include a
     component of polar dust.
     
     In the starburst and AGN torus models discussed above, it is reasonable to assume
     that the energy source is located at the center of the GMC or torus. In order to
     model the UV to millimetre emission of a disc galaxy or a host galaxy with a
     spheroidal distribution, we need to solve the radiative transfer problem in the
     case where the energy source is mixed with the dust. As discussed above, this
     problem was solved for a spheroidal distribution by \cite{ks94} and later
     \cite{sk07}. GRASIL solves this problem for either a spheroidal or disc geometry.
     In the disc case, an approximation is used for scattering. \cite{efstathiou20} also
     presented a model for a spheroidal distribution which is an evolution of the model
     of \cite{err03} that  incorporates a population synthesis model, as in the case of
     GRASIL. The model of \cite{efstathiou20} has already been incorporated in the SED
     fits of LIRGs with the MCMC code SATMC in \cite{herrero-illana17}.  Other methods
     of computing the emission from a disc galaxy have been developed by
     \citet{popescu00} and \citet{baes03} using the code SKIRT. An interesting
     application of SKIRT was to compute simulated spectra for half a million galaxies
     resulting from the EAGLE cosmological simulation \citep{camps18}.
 
 Solving the equation of radiative transfer numerically is computationally very
 expensive, and currently it is in general not feasible to fit the SED by computing in
 `real' time the spectrum of a galaxy by varying the parameters of the model(s) for each
 component. For example, \cite{johnson13} experimented with integrating GRASIL in their
 MCMC code SATMC and found that fitting a single galaxy required several days on a
 single CPU.  The use of pre-computed grids or libraries of models for each component
 considerably speeds up the fitting procedure. This approach has been used successfully
 for fitting the SED of a galaxy with a combination of starburst and AGN torus models
 (e.g., \citealt{verma02,farrah02,farrah03,efstathiou13,efstathiou14}).
 \cite{herrero-illana17} and \cite{efstathiou20} showed that a fit with SATMC can be
 obtained in 10--15 minutes if libraries of starburst, AGN torus and spheroidal models
 are fed into the MCMC code SATMC. In this case, we need a different library for the
 spheroidal galaxy component for galaxies at different redshifts as the model spectra
 depend on the age of the galaxy. For example, \cite{herrero-illana17} in their study of
 local LIRGs used a spheroidal library computed at $z=0.1$ whereas \cite{efstathiou20},
 who modelled a galaxy at $z\approx 4.3$, used a library computed at $z=4.4$. In both
 cases the starburst and AGN libraries were assumed to be independent of redshift.

The typical visual extinction assumed by models of the three main components of emission
discussed in this section are $A_V \lesssim 20$ for the diffuse dust component of the
host galaxy, $A_V \sim 10$--$100$ for the star-forming regions or GMCs and $A_V$
$\gtrsim 100$ for the AGN torus. The extinctions assumed for diffuse dust are consistent
with the estimates of extinction from near-IR studies discussed in Sects.~2.1 and 2.2.2.
Note that observers usually calculate the extinction assuming a `screen' geometry
whereas the models usually assume mixing of the stars and dust and therefore higher
total extinctions. Some of the Compact Obscured Nuclei (CONs) discussed in Sect.~4.7 may
correspond to the most optically thick star-forming regions assumed in the models. Such
star-forming regions can produce the deep silicate absorption features discussed in
section 2.3. Note, however, that AGN tori observed edge-on can also produce deep
silicate absorption features (e.g., \citealt{efstathiou14}). In a typical LIRG we
observe the combined emission from all three components and so we may be dealing with
emitting regions with a distribution of visual extinctions which is not easy to infer by
studying the total emission from the galaxy. The SED decomposition methods discussed in
the following subsections can give estimates of the visual extinction of each component.

\subsection{Energy balance methods}
  
Energy balance methods are computationally fast and have been used to fit large samples
of galaxies resulting from multi-wavelength surveys of the sky \citep{smith12, driver18,
malek18}. Energy balance methods deal in an approximate way  with the reprocessing of
starlight from the optical/UV to the IR and sub-mm, which relies on the principle of
conservation of energy.  Any energy balance method basically follows the following three
steps. First, the method `builds' the intrinsic spectrum of the stellar population in
the model galaxy, by assuming a star-formation and metallicity history for the galaxy,
an IMF and possibly other parameters. The method then assumes an attenuation law and
other parameters that convert the intrinsic spectrum of starlight to the observed one.
The attenuation law is not simply the extinction law of the dust model assumed, and its
form may also be determined by other factors such as the geometry of the mixing of stars
and dust. In the final step, the total energy difference between the intrinsic and
observed spectra is converted to IR and sub-mm emission assuming a model for optically
thin dust emission, e.g., \cite{draine07}. 
  
  The first energy balance method was developed by \citet{err03} in order to fit the UV
  to sub-mm SEDs of the first galaxies detected in the 8 mJy SCUBA survey
  \citep{scott02, fox02, ivison02}. Modern widely used implementations of the method are
  the  CIGALE \citep{Noll09,Boquien19} and  MAGPHYS \citep{dacunha08} codes. CIGALE is
  actually a hybrid code, as it incorporates the AGN torus model of \cite{fritz06}.
  \cite{berta13} also presented an adapted version of MAGPHYS that includes the AGN
  torus emission.
 
  There are two main limitations of energy balance methods. First, it is very unlikely
  that all of the complexity of the geometry of a galaxy can be included in the
  attenuation law. Second, the method cannot model parts of the galaxy such as the
  deeply obscured young stellar populations which may be completely obscured in the
  optical/UV due to their high optical depth, e.g. CONs (see section 4.7) . For such
  regions it is necessary to carry out a radiative transfer calculation. Recently,
  \cite{buat19} also  discussed some limitations of energy balance methods. 
 
\subsection{Examples of SED decomposition methods}
\label{sec:sed-decomposition}

\begin{table}
	\centering
\caption{Examples of SED de-composition methods with at least two components}
	\label{tab:example_table}
	\begin{tabular}{ll} 
		\hline
		  {\bf Methods}   &    {\bf Link}    \\
		                  &                  \\
		  {\bf Energy balance}  &  \\
		                  &    \\
           \cite{err03}  &     \\
           CIGALE \citep{Noll09}    &     cigale.lam.fr   \\
           MAGPHYS \citep{dacunha08}  &     www.iap.fr/magphys/  \\
		         &               \\
		 {\bf Radiative transfer} &  \\
		         &      \\
           CYGNUS \citep{farrah03,efstathiou20}   &   ahpc.space/cygnus/    \\
           GRASIL \citep{silva98, vega08,lofaro15}      &  www.ascl.net/1204.006  \\
           SATMC  \citep{johnson13}    &   github.com/sethspjohnso/satmc     \\
           \cite{siebenmorgen15}   &   www.eso.org/~rsiebenm/agn\_models/  \\
		         &               \\
		  {\bf Hybrid}   & \\
		       &     \\
           \cite{AlonsoHerrero2012}   &      \\
            AGNfitter \citep{Calistro_Rivera16}         &   ascl.net/1607.001    \\
             SATMC  \citep{johnson13}    &   github.com/sethspjohnso/satmc     \\
            SED3D \citep{berta13}        &                                   \\
		\hline
	\end{tabular}
\end{table}

In this section we discuss various methods of SED decomposition. We also present a
summary of the various methods in Table~\ref{tab:example_table}. Because of space
limitations, we do not present an exhaustive discussion, but aim to give an idea of the
range of approaches that have been adopted to date.  
    
\citet{verma02}, \citet{farrah02}, and \citet{farrah03} were the first papers that
carried out starburst/AGN decomposition of the SEDs of ULIRGs and HLIRGs using a
statistical method, in particular $\chi^2$ minimization. The authors used the grids of
AGN torus and starburst models of \citet{err95} and \citet{errs00} respectively
demonstrating a significant contribution by both components to the IR luminosity but
with the AGN fraction increasing with luminosity.  \citet{vega08} fitted the near-IR to
radio SEDs of a sample of around 30 LIRGs and ULIRGs with either a starburst, or  a
combination of starburst and AGN torus. They computed starburst models with GRASIL, and
AGN torus models with the method of \citet{gd94}.  MAGPHYS is an energy balance code
developed by \citet{dacunha08}, which was adapted to include warm dust from an AGN
component, using AGN torus models from \citet{fritz06}.  It has been widely used to fit
large samples of galaxies resulting from extragalactic surveys, e.g., \citet{driver18}.
The method of \citet{AlonsoHerrero2012} is a hybrid method that combines the CLUMPY
models for the AGN torus \citep{nenkova08} and templates for the starburst. We discuss
in more detail the method of \citet{AlonsoHerrero2012} and their main results in the
next subsection. 

\citet{johnson13} presented a versatile MCMC SED fitting code which can either fit an
SED with libraries of models or with a `synthesis' routine that can describe any model,
e.g., GRASIL. SATMC can also determine simultaneously a photometric redshift for the
galaxy using all the data used in the SED fit.  \citet{efstathiou14} fitted the SED of
the ULIRG IRAS~08572+3915 by $\chi^2$ minimization using libraries of the tapered discs
of Efstathiou \& Rowan-Robinson and the starburst models of \cite{es09}. The IRS
spectroscopy from Spitzer was included in the fitting. This study demonstrated the
importance of correcting the luminosity of the AGN torus due to its anisotropic
emission. With this correction IRAS~08572+3915 is predicted to be the most luminous
galaxy in the $z < 0.1$ Universe.  \cite{siebenmorgen15} used libraries of their
two-phase torus models and  the starburst models of \cite{sk07} to fit a number of
galaxies with SATMC. The original energy balance code CIGALE \citep{Noll09} has also
been adapted to include the AGN torus model of \citet{fritz06}. We therefore list CIGALE
as both an energy balance and a hybrid method.  AGNfitter \citep{Calistro_Rivera16} is
an MCMC code that uses theoretical, empirical, and semi-empirical models to characterize
both the nuclear and host galaxy emission simultaneously. We classify AGNfitter as a
hybrid code. The method described in \citet{efstathiou20}, which was also used in
\cite{herrero-illana17}, is the only method to date that fits the panchromatic emission
of galaxies including the emission of the host galaxy using exclusively radiative
transfer models.
    
  \subsection{Estimating the AGN contribution to $L_{\rm IR}$ of local LIRGs using SED decomposition}\label{subsubsec:AGN}

The sensitivity of the Spitzer and AKARI spectroscopic observations opened for the first
time the possibility of detecting elusive AGN as well as quantifying the AGN
contribution to the observed mid-IR and IR luminosity in large samples of LIRGs. Apart
from the detection of high excitation lines (Section~\ref{subsubsec:fine}), the PAH
emission and the shape of the mid-IR emission have been used as diagnostics to identify
the presence of AGN. This is because AGN tend to show PAH emission with lower EW and
steeper mid-IR emission continuum than star-forming galaxies. 

For a volume-limited
sample of LIRGs, \cite{AlonsoHerrero2012} decomposed the Spitzer/IRS
$5$--$38\,\mu{\rm m}$ 
spectra of LIRGs using the CLUMPY torus models \citep{nenkova08}, to represent the AGN mid-IR
emission and starburst templates. They detected an AGN
component in approximately half of their sample and derived a combined optical
and mid-IR AGN detection rate of $\sim 70\%$. 
The fraction of buried AGN (that is, those not classified as Seyferts in the optical) using 
AKARI $2.5$--$5\,\mu{\rm m}$ spectroscopy is about 15\% 
\citep{Imanishi2010} and similar to that 
derived from Spitzer \citep{AlonsoHerrero2012}.
However, this fraction in
LIRGs is much lower than in local ULIRGs, and is related to the higher nuclear extinctions in ULIRGs.

The CLUMPY torus models \citep{nenkova08} used for the spectral decomposition by
\cite{AlonsoHerrero2012} are normalized to the AGN bolometric luminosity. Thus these
authors inferred a total AGN bolometric contribution to the IR luminosity in local LIRGs
of $5^{+8}_{-3}\%$, whereas in  local ULIRGs the total AGN bolometric contribution was
estimated to be in the range 25--40\% \citep{Veilleux2009, Nardini2010}. Indeed, in the
majority of local LIRGs the AGN contributes less than 5\% and only in about 10\% of
local LIRGs the AGN contributes more than one-quarter of the total IR luminosity.  These
contributions are smaller than those of ULIRGs, which also increase with $L_{\rm IR}$
\citep{Nardini2010}. In summary, despite the fact that an AGN might be present in more
than 70\% of local LIRGs, its bolometric contribution is small and thus the bulk of
their IR luminosity is produced in SF related processes. 

\subsection{Models of high redshift LIRGs and ULIRGs}  

SED modelling of nearby LIRGs serves as a comparison sample against which to test any
development that is later on applied to fit the SED of high-$z$ LIRGs and ULIRGs.  In
the local Universe, the IR emission of normal star-forming galaxies with no AGN can be
understood in terms of two components \citep{rrc89}: starburst emission associated with
optically thick GMCs illuminated by recently formed stars and cirrus emission associated
with diffuse and cold dust ($T < 30$ K) illuminated by the interstellar radiation field.

\cite{rr97} made the first attempts to fit the SEDs of high-$z$ LIRGs in the ISO survey
of the Hubble Deep Field.  \cite{hughes98} presented the first SED fit of a SMG
(HDF850.1) detected in the SCUBA survey of the Hubble Deep Field that followed.  These
were the first works to apply starburst radiative transfer models and `cirrus' models to
a sample of high-$z$ LIRGs and SMGs using the models of \cite{errs00} and \cite{err03}.
In the latter work, \cite{err03} carried out a study of the SEDs of a sample of around
20 SMGs detected in the SCUBA 8 mJy survey \citep{scott02,fox02}, for which it was
possible to establish optical and near-IR associations from the positions obtained with
mm- or radio interferometry \citep{ivison02}. 

\cite{err03} fitted the SEDs using two different libraries of models.  The first library
used an early version of an energy balance method, and assumed a stellar population
whose SFR declined exponentially with time. These models were referred to as `cirrus'
models, although they are essentially models for what are now known as `Main Sequence'
galaxies (\cite{Noeske07}, \cite{elbaz2007}, see Fig.~\ref{fig:pereira} in
Sect.~\ref{sec:intro}).  The second library of models  included a starburst component in
the last 50 Myr of the history of the galaxy. 
\cite{err03} showed that the SED of high-$z$ SMGs could be explained reasonably well
with pure `cirrus' models, or a combination of `cirrus' and starburst models.  Despite
the scarce data in the rest-frame far-IR to discriminate between the two possible
models, \cite{err03} found for the first time evidence for a significant contribution
from cold diffuse dust in the sub-mm band, in most of these extremely luminous objects.
Similary, \cite{taylor05} explored the properties of galaxies discovered in the FIRBACK
$175\,\mu{\rm m}$ ISO survey using the libraries of starburst and cirrus models of
\cite{errs00} and \cite{err03} to fit their SEDs, and concluded that most of the FIRBACK
galaxies were dominated by cirrus emission. Later on, \cite{es09} carried out a study
similar to that of \cite{err03} on a dozen SMGs in the redshift range $1.2 < z < 3.4$
for which Spitzer IRS spectroscopy was available. This (pre-Herschel) study also
demonstrated  that these galaxies had a significant contribution to their bolometric
luminosity from `cirrus' emission.  Furthermore, the availability of the Spitzer spectra
made also possible the identification of 3 galaxies with a significant contribution from
an AGN torus. Studies of submillimeter galaxies  detected by Herschel (e.g.,
\citealt{rodighiero11}, \citealt{elbaz11}) allowed the classification of galaxies in
terms of their mode of star formation, the `Main sequence' mode or the starburst mode.
The `Main sequence' mode is found to be more dominant in redshifts up to $z \sim 3.5$
whereas the starburst mode dominates at higher redshifts.

\begin{figure}[htb!]
\center
\includegraphics[width=8.cm]{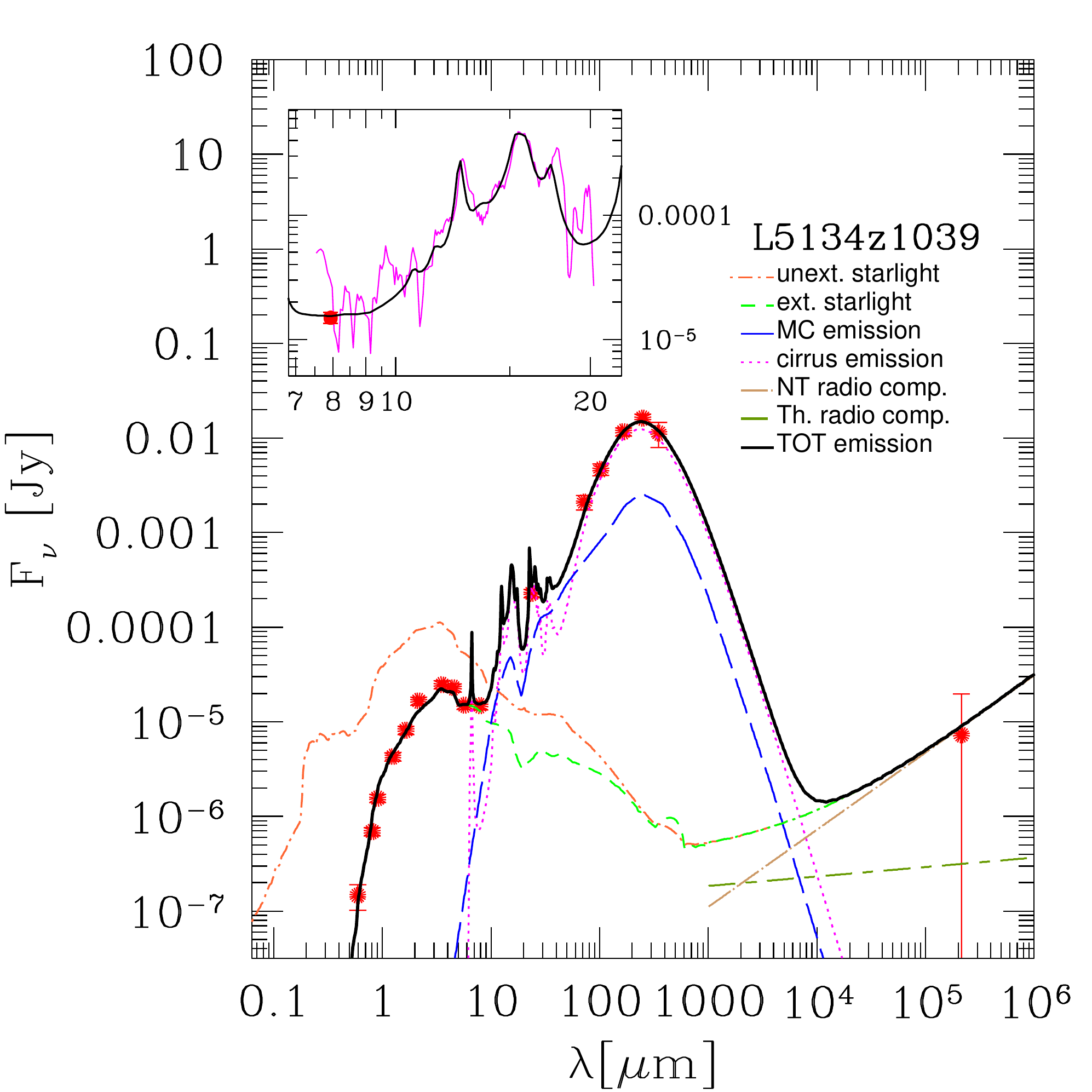}
\caption{Fit to the spectral energy distribution of L5134 at $z=1.039$ \citep{lofaro15} with the GRASIL model \citep{silva98} which solves the radiative transfer problem for a disc or a spheroidal galaxy. GRASIL has been thoroughly tested in samples of nearby
galaxies, and has also been applied to samples of galaxies at higher redshifts. In this specific model, `cirrus' emission dominates the luminosity of L5134.}
\label{fig:L5134}
\end{figure}

For example,  \cite{lofaro13,lofaro15} applied the GRASIL model to a sample 
 of 31 LIRGs at around $z=1$ and ULIRGs at $z=2$ in GOODS-South and determined the star formation history of the galaxies as well as current star formation rates, stellar and gas masses. 
GRASIL could fit well photometry and spectroscopy data (see Fig.~\ref{fig:L5134}), and 
\citet{lofaro15} concluded that this sample consists mainly of `Main Sequence' galaxies with very few objects requiring a recent starburst. No contribution from an AGN component was included in the SED fitting.
Similarly,  \cite{rr04,rr05,rr10,rr14,rr16,rr18} carried out a series of studies of high-$z$ galaxies discovered in ISO, Spitzer and Herschel surveys. The authors used the radiative transfer models for starbursts and `cirrus' of \cite{errs00} and \cite{err03} to  explore the physics of high-$z$ IR and SMGs, and determined their contribution to the star formation density of the universe up to $z=6$. One of the major conclusions of this study was that although there is reasonable agreement of IR and UV star formation rate density (SFRD) estimates  for $z < 3$, the IR SFRD is significantly higher than UV estimates at $z = 3$--$6$. 
    
Alhough determining SFRs and its value across cosmic time is probably the single most
valued variable, it is not the only useful application of SED model fitting.  For
example, \cite{driver18} used MAGPHYS to determine stellar and dust masses, and dust
corrected star formation rates for hundreds of thousands of galaxies over the redshift
range $0 < z < 5$, corresponding to the last 12 Gyr of the history of the Universe.
Another application is the use of SED fitting to identify gravitational lens candidates.
\cite{malek18} fitted the SEDs of around 40,000 HELP galaxies with photometry from the
optical/UV to the sub-mm using CIGALE, identified around 350 possible gravitational
lenses, and carried out a statistical study of the application of different attenuation
laws. As first discussed by \cite{rr14}, a gravitational lens can be identified by an
anomalous SED which shows an unusually high $500\,\mu{\rm m}$ to $24\,\mu{\rm m}$ colour
for the photometric redshift determined from its optical SED (e.g., see Fig.~16 of
\citealt{rr14}).  Finally, SED model fitting is being used nowadays to precisely
determine the redshift of obscured quasars and galaxies. For example,
\cite{efstathiou20} used  HELP data to discover a hyperluminous obscured quasar at
$z\approx 4.3$ in the  COSMOS field \citep{scoville07}, using the MCMC code SATMC. 
This galaxy has similar properties to the Hot Dust Obscured Galaxies (Hot DOGs)
discovered by WISE at lower redshifts but is only the second such object discovered at
$z > 4$. 

\section{Time-domain observations of LIRGs} 
\label{sec:timedomain}

Time-domain studies of LIRGs have revealed powerful supernova (SN) factories within
their innermost nuclear and circumnuclear regions where massive stars with relatively
short lifetimes are being formed and consequently explode as SNe at high rates (e.g.,
\citealt{pereztorres09a,mattila2012,bondi12,kool2018}). On the other hand, most LIRGs
are known to harbour at least one supermassive BH in their nuclei (e.g.,
\citealt{mazzarella2012AJ}) and therefore also variability linked to the accretion onto
this BH can be expected. Such variability can be either intrinsically related to an AGN,
or in some cases to tidal disruption of a star by the BH. Consequently, the emission
from LIRGs cannot be expected to stay constant over time and substantial variability can
be anticipated on time scales of months to years.

\subsection{Supernovae} It is now well established that in the circumnuclear regions of
LIRGs the bursts of star formation result in high-rates of core-collapse SNe, the
end-points in the evolution of massive stars initially with masses of at least $\sim8\,
M_{\odot}$. Thermonuclear (type Ia) SNe on the other hand occur with a much longer time
delay from the formation of the progenitor stars than core-collapse SNe, and hence
enhanced rates are not expected in LIRGs with typical starburst ages $\lesssim$100 Myr. 

Assuming a constant SFR and an immediate conversion of the short-lived ($\lesssim$50
Myr) massive stars to core-collapse SNe, we can relate the SN rate ($\Re$; yr$^{-1}$)
and the SFR ($M_{\odot} {\rm yr}^{-1}$) assuming the IMF $\Phi$ between $m_{\rm low}$
and $m_{\rm up}$, and the initial mass range (m$_{min}$, m$_{max}$) of the core-collapse
SN progenitors are known (e.g., \citealt{condon92,dahlen1999}).

\begin{equation} \label{eq:ccsnrate}
\frac{\rm \Re}{\rm SFR} = \frac{\int^{m_{\max}}_{m_{\min}}\Phi (m)dm}
{\int^{m_{\rm up}}_{m_{\rm low}}m\,\Phi (m)dm}
\end{equation}

For example, assuming a Salpeter IMF ($\alpha =-2.35; m^\alpha$) with cut-offs of 0.1
and 120 $M_{\odot}$ and core-collapse SN progenitors between 8 and 30 $M_{\odot}$,
yields $\Re = 6.3\times10^{-3}\,{\rm SFR}\, M_{\odot}^{-1}$. Therefore, LIRGs with
typical SFRs in the range $\sim 20-200$ $M_{\odot}$ yr$^{-1}$ (adopting Eq.
\ref{eq:sfr_ir}; but see also Sect. \ref{sec:sed-modelling}) can be expected to be
prolific SN factories, with corresponding core-collapse SN rates of the order of
0.13-1.3 yr$^{-1}$.  If we use a Kroupa IMF instead ($\alpha_1 = -0.3$ for $0.01 \leq
M/M_\odot < 0.08$; $\alpha_2 = -1.3$ for $0.08 \leq M/M_\odot < 0.5$; $\alpha_3 = -2.3$
for M$\geq 0.5\,M_\odot$), $\Re = 9.8\times10^{-3}\,{\rm SFR}\, M_{\odot}^{-1}$, about
54\% higher than for a Salpeter IMF.  However, in the case of LIRGs the assumption of a
constant SFR is unlikely to be applicable, and the age of the starburst can also be
comparable to the life times of the massive stars exploding as core-collapse SNe.
Therefore, also the core-collapse SN rate will depend strongly on the SFH and the age of
the starburst, and therefore requires a detailed modelling of the properties of the
individual galaxies (see Sect. \ref{sec:sed-modelling}).

Core-collapse SN rates for LIRGs can also be estimated via the galaxy's radio and/or IR
properties and relations exist between the SN rate and the galaxy radio and IR
luminosities. For example, \citet{huang1994} presented a relation between the SN rate
and the galaxy's integrated non-thermal radio continuum luminosity making use of 8.4 GHz
VLA observations of the population of compact radio sources in the nearby starburst
galaxy M~82. In addition, near-IR spectroscopic and narrow-band imaging observations can
provide useful diagnostics for individual galaxies. \citet{rosenberg2012} studied a
sample of nearby starburst galaxies and LIRGs and found a strong correlation between the
SN rate and the [Fe~II] 1.26 $\mu$m line luminosity, as expected given the association
of the [Fe~II] emission with efficient destruction of the interstellar dust grains by
the shocks in SN remnants. Furthermore, \citet{mattila01} provided an empirical relation
based on comparison between SN rate estimates and IR luminosities for nearby dusty
starburst galaxies: \begin{equation} \label{eq:ccsnrate_empirical} {\rm \Re} = 2.7 \,
\left(\frac{L_{\rm IR}}{10^{12}\,L_{\odot}}\right)\,{\rm yr}^{-1} \,, \end{equation}
where $L_{\rm IR}$ is the 8--1000$\mu$m luminosity. Therefore, adopting this relation
one can expect core-collapse SN rates in the range 0.27--2.7 yr$^{-1}$ for galaxies with
$L_{\rm IR}$ in the $10^{11}$ to $10^{12}$ L$_{\odot}$ range. However, Eq.
\ref{eq:ccsnrate_empirical} does not either take into account the effects of the
starburst age, SFH, nor the contribution of older stellar populations or AGN to the
heating of the interstellar dust that is responsible for the galaxy's IR luminosity.
Instead, more accurate estimates can be obtained for individual galaxies making use of
detailed radiative transfer modelling of the galaxy's entire IR SED (see Sect.
\ref{sec:sed-modelling}). For example, \citet{mattila2012} estimated SN rate of
$\sim$0.76 yr$^{-1}$ for Arp~299-A and $\sim$0.33 yr$^{-1}$ for Arp~299-B based on
radiative transfer modelling of their IR SEDs, in agreement with estimates based on high
resolution radio observations (see below). These compact regions are expected to host
almost 60\% of all the core-collapse SNe of Arp~299 with a total SN rate for Arp~299 of
$\sim$1.9 yr$^{-1}$ \citep{mattila2012}.

Summarising, direct detection of core-collapse SNe can be used as an independent tracer
of the recent star formation activity and comparison between the SN rate and SFR may in
principle even provide useful information on the IMF in LIRGs and ULIRGs. Evidence for a
top-heavy IMF in galaxies with the highest SFR densities has been previously obtained
both locally and at high-$z$ for example making use of isotopic ratios from radio
observations (see Sect.~\ref{sec:molecular}) with important implications to parameters
governing galaxy formation and evolution. On the other hand, the conversion between SN
rate and SFR can also be altered substantially by including the effects of stellar
evolution in close binary systems and uncertainties in the actual range of the
core-collapse SN progenitor masses (\citealt{smartt2015PASA}).

\begin{figure}
\center
\includegraphics[width=\textwidth]{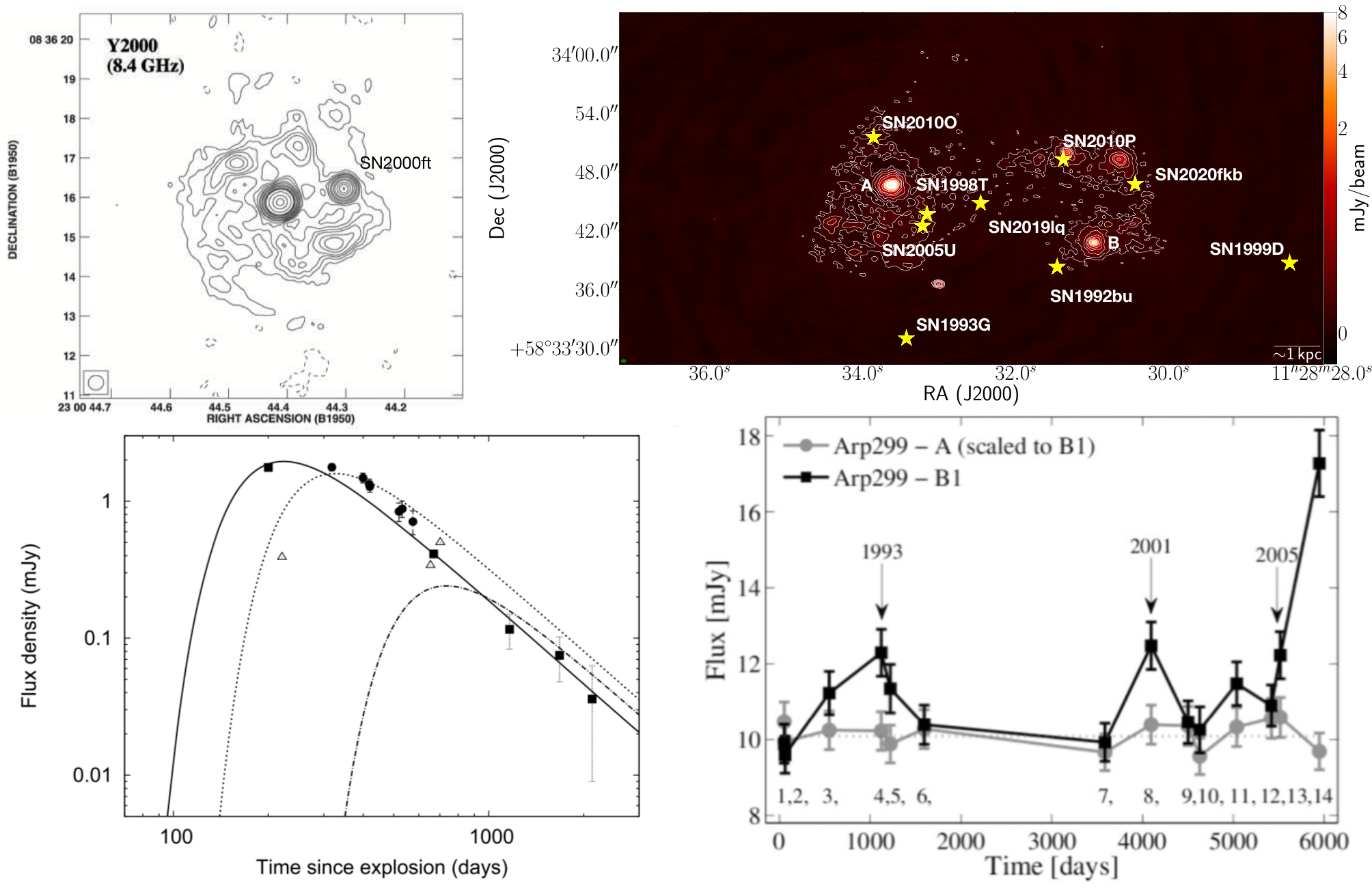}
\caption{\textbf{Top left.} 8.4 GHz VLA continuum contour map ($0.39"\times0.37"$ beam size) of the nuclear regions of the LIRG NGC 7469, showing the discovery of the radio SN 2000ft about 500 pc west from the galaxy nucleus (from \citealt{colina01}). \textbf{Bottom left.} Observed radio light curves of SN~2000ft over almost six years along with the light curve fits  at 8.4\,GHz (filled squares; solid line), 5.0\,GHz (filled circles; dotted line) and 1.6\,GHz (upper limits; open triangles; dashed–dotted line) from \citet{pereztorres09b}. The radio emission follows a rather steep decline typical of core-collapse SNe exploding in `normal’ spiral galaxies with the radio emission being powered by interaction with the pre-SN stellar wind rather than with a dense ISM. \textbf{Top right.} 8.4 GHz continuum JVLA image of Arp 299, showing the identifications of the different nuclei and the locations of a number of recent SNe within the circumnuclear regions. (Image courtesy of Na\'im Ram\'irez-Olivencia). \textbf{Bottom right.} 8.4 GHz VLA light curves of the Arp 299 nuclei A and B1 between 1990 and 2005. The significant brightenings shown by Arp 299-B1 in 1993 and 2001 were identified as moderately luminous radio SNe \citep{romero-canizales11} most likely resulting from normal core-collapse SN explosions. The brightening in 2005 corresponds to Arp 299-B AT1 (see Sect. 6.2.2). Image from \citet{romero-canizales11}.}
\label{fig:2000ft}
\end{figure}

\subsubsection{Radio studies} Radio observations are not affected by dust extinction.
Moreover, significant radio emission is expected from core-collapse SNe in contrast to
thermonuclear SNe as the interaction of the SN ejecta with a CSM gives rise to a
high-energy density shell, which is Rayleigh-Taylor unstable and drives turbulent
motions. These amplify the existing magnetic field in the pre-SN wind, and efficiently
accelerate relativistic electrons, thus enhancing the emission of synchrotron radiation
at radio wavelengths \citep{chevalier82}. The starburst activity in the circumnuclear
regions of LIRGs ensures the presence of a high number of massive stars, so
core-collapse SNe are expected frequently in these regions. Some of these SNe have a
sufficiently dense CSM resulting in bright radio SNe and many of them can be expected to
evolve into compact SN remnants as a result of interaction with a dense ISM
\citep{chevalier01}.

Such radio SNe have been detected in local LIRGs by interferometric radio observations
with sub-arcsecond angular resolution. For example, \citet{colina01} reported the VLA
discovery of the radio SN 2000ft within the circumnuclear regions ($\sim$500 pc
projected distance from the nucleus) of NGC 7469 (see Fig.~\ref{fig:2000ft}). Its peak
radio luminosity corresponds to $\sim$15\% of the 8.4\,GHz flux density of the entire
nucleus. \citet{alberdi06} and \cite{pereztorres09a} analysed the radio light curves of
SN 2000ft covering six years of evolution and found that a CSM typical of a red
supergiant progenitor star can adequately explain its observed properties (see
Fig.~\ref{fig:2000ft}). Furthermore, \citet{romero-canizales11} used archival 8.4\,GHz
VLA observations ($\sim$0.8" spatial resolution) of Arp 299 in multiple epochs over a
period of 11 years to study radio variability of the entire nucleus B1. Based on their
analysis they detected two transient events (in 1993 and 2001) in which the nucleus
brightened significantly at $\sim$20\% level, which could be explained by moderately
luminous radio SNe resulting from normal core-collapse SN explosions (see
Fig.~\ref{fig:2000ft}).

\begin{figure}
\center
\includegraphics[width=8cm]{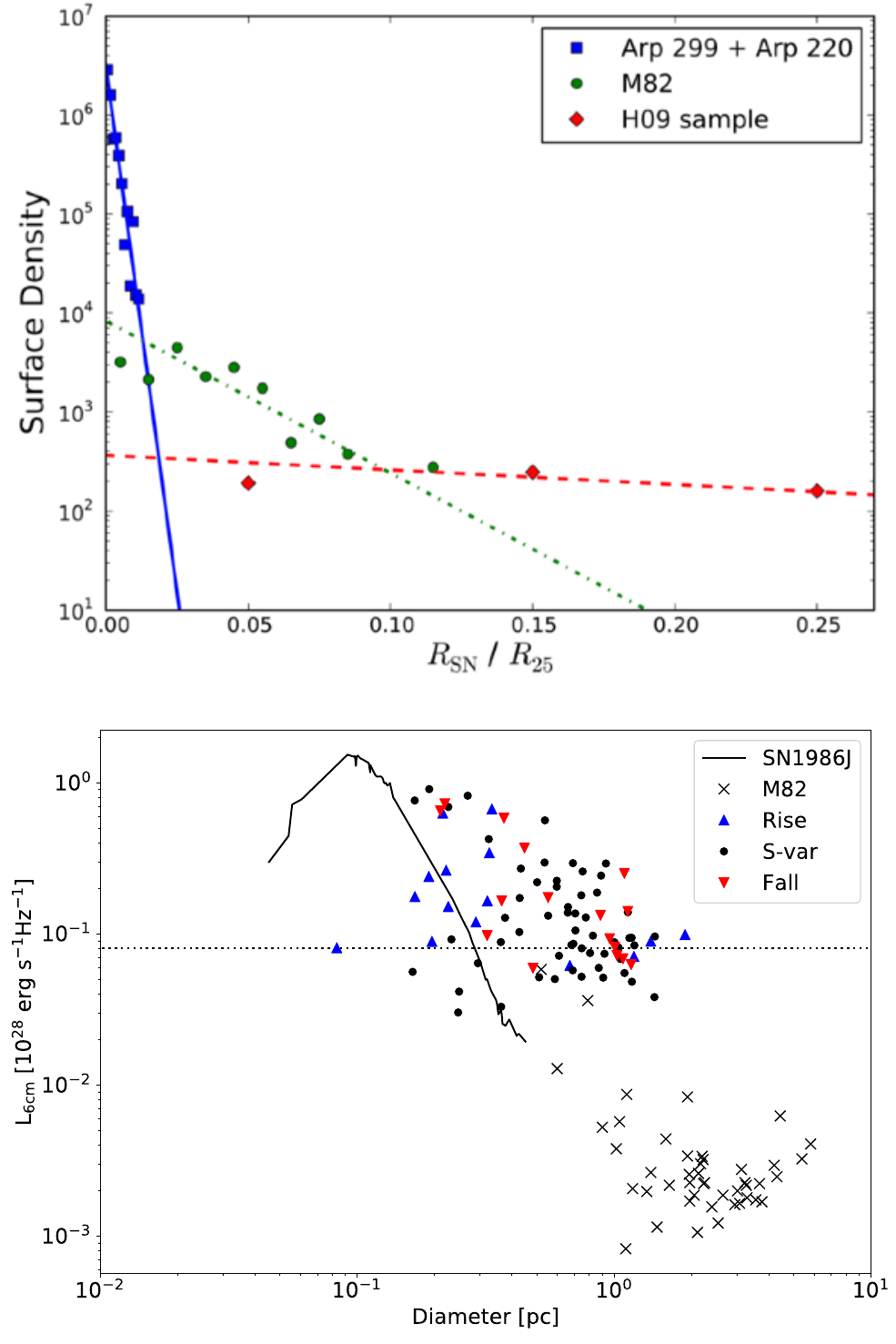}
\caption{\textbf{Top.} The radial distributions of radio SNe and SN remnants of Arp 299-A and Arp 220 (blue squares), and M 82 (green circles), obtained by \citet{herrero-illana12}, compared with the sample of SNe observed in spiral galaxies from \citet{hakobyan09} (red diamonds). Here, the radial distances are in units of the isophotal radius at which the surface brightness is 25 mag/arcsec$^{2}$ in the Johnson B filter. \textbf{Bottom.} Radio spectral luminosity vs. diameter for the compact radio sources within the nuclear regions of Arp~220 compared with the SN remnants in the nuclear regions of the nearby starburst galaxy M~82 and the luminous radio SN 1986J. The figure is a modified version from \citet{varenius2019}.}
\label{fig:RSN}
\end{figure}

Radio SNe can be detected even in the innermost ($\leq$100 pc) regions of nearby LIRGs
and ULIRGs, by means of high angular resolution ($\leq 0.01$ arcsec), high-sensitivity
($\leq$0.05 mJy) VLBI observations (e.g., \citealt{gallimore04,pereztorres09a,
varenius2019}; see also Sect. \ref{sec:radio-high-resolution}).  Core-collapse SN rates
have been estimated for the nuclei of Arp 299 based on high spatial resolution VLBI
radio observations.  \citet{neff04} identified five compact radio sources within the Arp
299 nucleus A, using VLBA observations. Based on their measured sizes and observed flat
or inverted radio spectra, they interpreted these as likely radio SNe or young SN
remnants.  \citet{pereztorres09b} and \citet{ulvestad09} used EVN and VLBA observations,
respectively, to reveal the existence of more than 20 compact radio sources within a
$\sim$100 pc diameter region around Arp 299-A. In addition, \citet{ulvestad09} reported
four within a $\sim$30 pc region around Arp 299-B1. Both groups interpreted these
sources as radio SNe and/or young SN remnants; \citet{ulvestad09} also noted that the
ratio of the source counts in the two nuclei was approximately equal to the ratio of
their predicted SN rates. \citet{pereztorres09b} and \citet{bondi12} monitored the
evolution of the compact sources in the innermost nuclear regions of Arp~299-A with the
EVN (see Fig.~\ref{fig:arp299a-vlbi}), and found a total of 26 compact radio sources.
The high flux densities and small sizes of most of the compact sources correspond to
brightness temperatures that rule out thermal radio emission. Therefore, the observed
radio emission must be non-thermal and is consistent with the sources being a mixed
population of radio SNe and young SN remnants. There was clear evidence for at least two
new radio SNe in the observations separated by two years and significant variability was
detected for several objects (see Fig.~\ref{fig:arp299a-vlbi}), therefore, implying a
lower limit for the core-collapse SN rate in the A-nucleus of $\gtrsim$ 0.8 yr$^{-1}$.
The authors conclude that the luminosities are consistent with the sources resulting
from the explosions of normal core-collapse SNe of types II-P, II-L and IIb. 

Although not a LIRG, it is interesting to discuss Arp 220, the nearest ($D_L = 79$ Mpc)
ULIRG (L$_{\rm IR} = 1.6 \times 10^{12}\, L_{\odot}$). VLBI observations over 20 years
have been used to study the population of very luminous compact radio sources within its
innermost $\sim$80 pc ($\sim$0.2") regions around the east and west nuclei (e.g.,
\citealt{smith98, rovilos05, lonsdale06, parra07}). The radio luminosities of the
sources are an order of magnitude higher than seen for the population of compact radio
sources within the nuclear regions of Arp 299, and must therefore also have a
non-thermal origin in radio SNe and young SN remnants (interestingly, the precise
location of the putative AGN in Arp 220 is not yet known). A high core-collapse SN rate
within the nuclear regions is a direct consequence of a high SFR of this galaxy (see Eq.
\ref{eq:ccsnrate}). Recently, \citet{varenius2019} reported an analysis of the
population of almost 100 compact radio sources in Arp 220 (see Fig.~\ref{fig:RSN})
yielding an observed rate of the most luminous radio SNe of $\sim$0.2 yr$^{-1}$.
Assuming that about 5\% of the core-collapse SNe occurring in Arp 220 show strong enough
interaction with a dense surrounding medium to result in such luminous radio SNe (which
is similar to the fraction of core-collapse SNe in normal galaxies observed to produce
so called type IIn SNe as a result of interaction with a dense CSM) their estimated rate
corresponds to a total core-collapse SN rate of around 4 yr$^{-1}$. Adopting
Eq.~(\ref{eq:ccsnrate_empirical}), this rate is consistent with the IR luminosity of Arp
220. Thus, it seems that the core-collapse SN population of Arp 220 does not
significantly differ from the one observed in normal galaxies in producing a similar
fraction of events with a strong interaction with the surrounding CSM (classified as
type IIn in normal galaxies). \citet{varenius2019} therefore suggested that a standard
IMF may explain the number of bright radio SNe they have observed in Arp 220. This,
however requires that all type IIn SNe in Arp 220 reach high radio luminosities, and
this is different from observations of SNe in normal field galaxies \citep{VanDyk96}.
These results therefore highlight the fact that VLBI studies of (U)LIRGs at different IR
luminosity ranges are needed to better understand the radio SN populations within their
innermost nuclear regions. 

\begin{figure}
\center
\includegraphics[width=\textwidth]{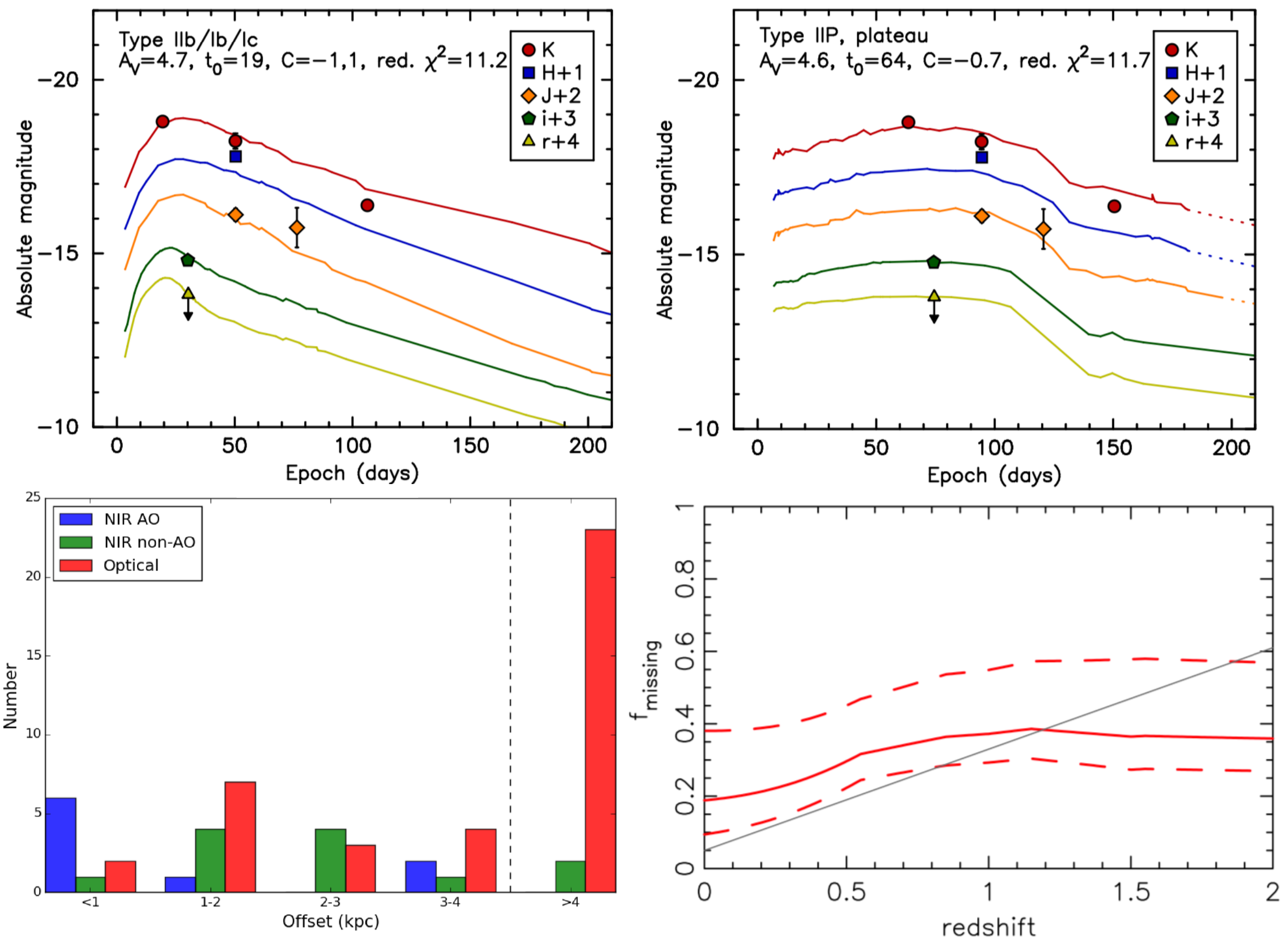}
\caption{\textbf{Top.} Light curve fits (solid lines) for observations (points) of SN 2015cb, which was discovered by AO assisted observations in the near-IR K-band at a projected distance of $\sim$600 pc from the nucleus of the LIRG IRAS 17138-1017. The light curves are best fitted with either a stripped-envelope core-collapse SN (left) or a type II-P SN in its plateau phase (right) with a line-of-sight extinction of $A_V$ $\sim$ 4.5 towards the SN \citep{kool2018}. \textbf{Bottom left.} Nuclear offset distribution for core-collapse SNe discovered in LIRGs using near-IR or optical observations \citep{kool2018}. The discoveries in the near-IR with AO instruments are shown in blue, in the near-IR in natural seeing condition in green, and in the optical in red. \textbf{Bottom right.} The fraction of core-collapse SNe missed by rest-frame optical searches as a function of redshift from \citet{mattila2012}. Red lines show the best estimate together with low and high missing fraction models as dashed lines. The solid black line is the missing fraction from \citet{mannucci2007}.}
\label{fig:IR_SNe}
\end{figure}

\citet{herrero-illana12} examined the radial distribution of the surface density of
radio SNe and young SN remnants in the innermost nuclear regions of Arp 220, Arp 299 and
M 82 (see Fig.~\ref{fig:RSN}), and found that the radial distributions followed an
exponential distribution ($\Sigma^{SN}=\Sigma_0^{SN}\exp{(-r/h_{\mathrm{SN}})}$) with
scale lengths  $h_{\rm SN}\sim$20--30 pc in Arp 220 and Arp 299, much smaller than the
$\sim$140 pc obtained for the nearby starburst galaxy M~82. \citet{herrero-illana12}
also found that the radial distribution of SNe for the nuclear disks in Arp 299-A and
Arp 220 is consistent with a power-law surface density profile
($\Sigma^{SN}=\Sigma_0^{SN}(r/r_\mathrm{out})^{-\gamma}$) of exponent $\gamma = 1$, as
expected from detailed hydrodynamical simulations of nuclear disks \citep{kawakatu08}.
The results by \citet{herrero-illana12} support scenarios where a nuclear disk of size
$\lesssim$100 pc is formed in (U)LIRGs. In particular, these authors suggested that the
nuclear disk in Arp~299-A is sustained by gas pressure, in which case the accretion onto
the supermassive black hole could be lowered by SN feedback.  \citet{kangas2013}
compared the spatial distribution of the reported core-collapse SNe (all outside the
innermost nuclear regions) in IR-bright galaxies and normal spiral galaxies, and found
evidence for a significantly smaller normalized scale length for the surface density of
SNe in the former of ($0.23^{+0.03}_{-0.02}$) vs.\ ($0.29 \pm 0.01$) R$_{25}$ for the
latter, reflecting a more centrally-concentrated population of massive stars in
IR-bright galaxies.

\subsubsection{Infrared studies} Not all core-collapse SNe make sufficiently luminous
radio sources to be detectable at the distances of the most nearby LIRGs and ULIRGs
motivating further their detection and study also at near-IR and optical wavelengths.
Optical searches for SNe (e.g., \citealt{richmond1998}) in starburst galaxies and LIRGs
have detected SNe at rates only similar to those found in ``normal'' galaxies. This is
not surprising as much of the star formation in starburst galaxies and LIRGs is obscured
by large amounts of dust and therefore also the SNe exploding in these regions are not
accessible by optical observations. However, the combination of strongly reduced
extinction ($A_K = 0.112 A_V$; \citealt{Rieke1985}) and the sensitivity of ground-based
observations make near-IR $K$-band (2.2\,$\mu$m) imaging well suited for the detection
and study of SNe in starburst galaxies and LIRGs (e.g., \citealt{vanburen1989,
grossan1999, mattila01}). In contrast to continuum radio emission, which originates as a
result of the interaction between the SN ejecta and the surrounding CSM, the near-IR
emission can originate from the gas in the SN ejecta, and hot dust in the CSM or in the
SN ejecta (e.g., \citealt{mattila2008}). Over the past 25 years there have been a number
of attempts to detect SNe in LIRGs making use of seeing-limited (FWHM$\sim$1") $K$-band
observations \citep{vanburen94, grossan1999, maiolino2002, mannucci2003, mattila2004},
resulting mostly in discoveries of SNe outside the LIRG (circum)nuclear regions
suffering from modest amounts of extinction. Most recently, \citet{miluzio2013} carried
out a search covering repeat observations of $\sim$30 LIRGs. Their results were
consistent with up to 75\% of the SNe remaining 'hidden' because of a combination of
reduced search efficiency and high extinction within the nuclear regions.

Because of the concentration of star formation (see Sect.~\ref{sec:infrared} -
\ref{sec:radio}) and thus of the SNe (see also Fig.~\ref{fig:RSN}) within the central
regions in LIRGs and ULIRGs, high spatial resolution is also crucially important for the
SN detection in these regions. At near-IR wavelengths this could be achieved either with
observations with HST \citep{cresci2007} or with ground-based AO assisted imaging
observations. AO assisted near-IR observations making use of an 8-meter class telescope
typically yield a point spread function with FWHM$\simeq$0.1". In addition, the use of
image subtraction techniques is necessary for the detection of the likely faint SNe
against the often bright and complex nuclear background (see Fig.~\ref{fig:Birdimage}).
$K$-band searches using natural guide star AO observations with the ESO VLT
\citep{mattila2007}, single conjugate laser guide star AO observations with the Gemini-N
telescope \citep{kankare08, kankare2012, ryder2014} as well as multi conjugate AO
observations with the Gemini-S telescope \citep{kool2018} have revealed a number of SNe
also within the innermost nuclear regions of nearby LIRGs that were not detectable in
any optical SN searches (see Fig.~\ref{fig:IR_SNe}). The strongly reduced extinction in
$K$-band has allowed the detection of circumnuclear SNe with host galaxy extinctions as
high as $A_V \sim 16$ as estimated based on the SN's near-IR colours and light curve
information \citep{kankare08}.

A systematic mid-IR monitoring of a large number of nearby galaxies was carried out by
the SPIRITS survey (see \citealt{Jencson2019}) that also produced the IR discoveries of
two core-collapse SNe in a nearby LIRG, albeit outside the circumnuclear regions.
However, we note that the combination of the sensitivity and spatial resolution of
Spitzer was not optimal for an efficient search of obscured SNe within the nuclear
regions of LIRGs and ULIRGs which are mostly too distant for this. Therefore, the
contribution of Spitzer for this topic remained modest.

 Detailed follow-up studies including also spectroscopic observations have been
 performed only in the case of a small number of SNe (e.g., \citealt{kankare2012,
 mattila2013, miluzio2013, kangas2016}, Kankare et al. submitted) that occurred in IR
 bright starburst galaxies and LIRGs. This is due to the often high line-of-sight
 extinctions and also contrast issues against the bright nuclear or circumnuclear
 background emission making detailed observations of faint SNe challenging. For the
 events suffering from more substantial extinctions and/or located at small
 galactocentric distances the follow-up observations have typically concentrated on
 near-IR photometry (e.g., \citealt{mattila2007, kankare08, kool2018}) and in some cases
 also high resolution radio observations \citep{pereztorres07, romero-canizales12a,
 romero-canizales14}. In the absence of spectroscopy, the near-IR light curve
 information, colours and absolute magnitudes have been used to determine the likely SN
 types and line-of-sight extinctions. A wide range of host galaxy extinctions have been
 found between $A_V \sim 0$ and 16 corresponding to $A_K < 2$ making the SN detection
 still feasible at near-IR wavelengths. In all cases the near-IR properties have been
 found consistent with a core-collapse SN origin and in a few cases a radio detection
 has also definitely confirmed the core-collapse SN nature (e.g.,
 \citealt{pereztorres07, kankare08, pereztorres08a}). For example, SN 2015cb was
 discovered by AO assisted observations in the near-IR K-band at a projected distance of
 $\sim$600 pc from the nucleus of the LIRG IRAS 17138-1017 by \cite{kool2018}. They
 found the SN light curves, colours and absolute magnitudes are consistent with a
 core-collapse SN (of type II-P or type IIb/Ib/Ic) with a line-of-sight extinction of
 $A_V$ $\sim$ 4.5 towards the SN (see Fig.~\ref{fig:IR_SNe}). In the case of SN 2015cb
 near-IR spectroscopy was also attempted but could not be used to constrain the SN type.

\citet{anderson2011} analyzed the circumnuclear SN population in Arp 299 finding a
relatively high fraction of stripped envelope SNe (Types Ib, Ic and IIb) relative to the
usually more common Type II SNe. They suggested that this excess could be explained by
the young age of circumnuclear star formation in Arp 299 such that we would be
witnessing here the explosions of the most massive stars (with the shortest lifetimes)
formed. Alternatively, they suggested that this result might be explained by a top-heavy
IMF favoring the formation of the most massive stars in Arp 299. More recently, Kankare
et al. (submitted) analysed the circumnuclear (projected galactocentric distances less
than 2.5 kpc) population of core-collapse SNe in Arp 299 and other nearby LIRGs and
compared the SN subtypes with starburst ages obtained using radiative transfer modelling
of their IR SEDs (Sect. \ref{sec:sed-modelling}). They found that the excess of stripped
envelope SNe in Arp 299 is naturally explained by the young starburst age without the
need to invoke a top-heavy IMF,  as also suggested by \citet{varenius2019} for Arp~220.

The IR luminosities of all the LIRGs and ULIRGs listed in the IRAS Revised Bright Galaxy
Sample \citep{sanders03} correspond to an intrinsic rate of core-collapse SNe of about
250 yr$^{-1}$ \citep{kool2018}. However, there has been only 60 core-collapse SNe in
these galaxies reported between 1968 and 2015 and of these 48 since year 2000
\citep{kool2018}. About 40\% of these SNe were discovered in the near-IR (see
Fig.~\ref{fig:IR_SNe}); however, within the innermost nuclear regions even AO assisted
near-IR observations are not sensitive to the most highly obscured transients, and radio
VLBI observations are required for their detection and study. Therefore, SN searches and
studies carried out at IR and radio wavelengths effectively complement each other.

\begin{figure}
\center
\includegraphics[width=\textwidth]{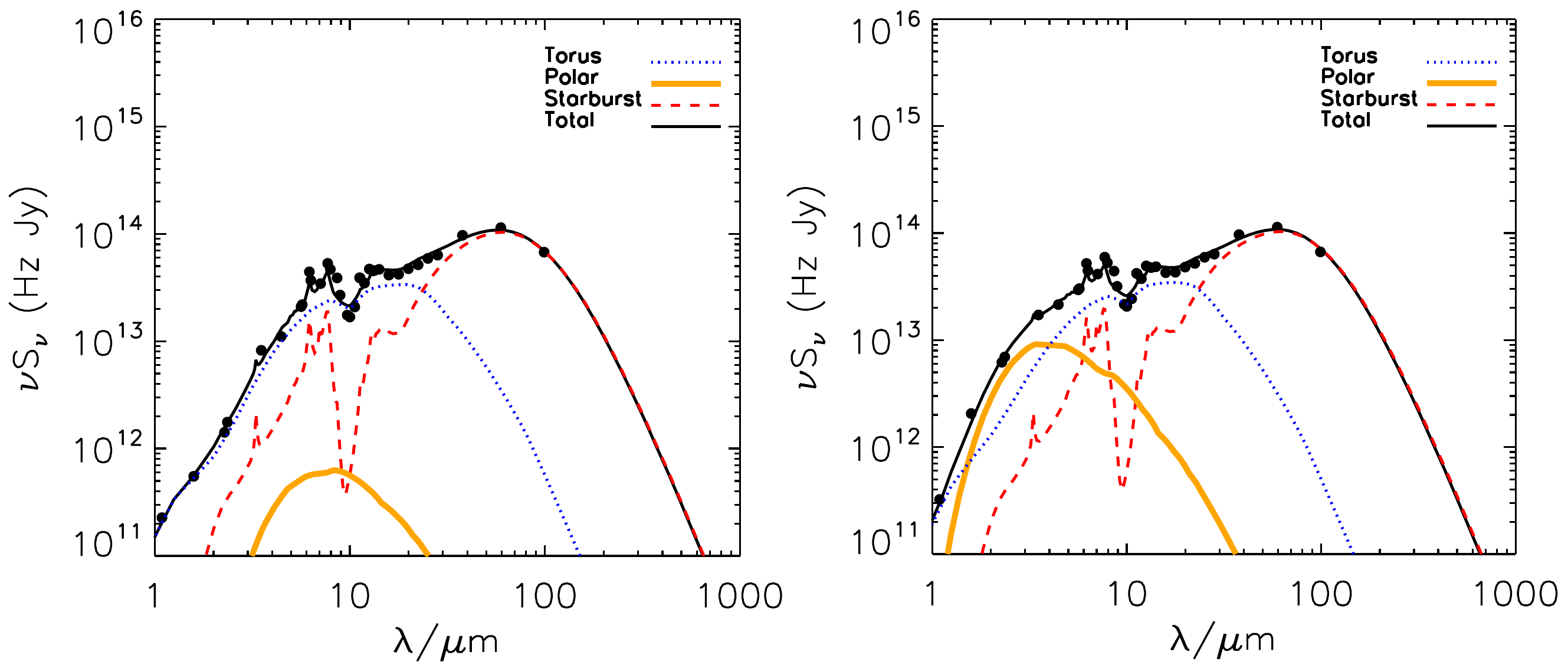}
\includegraphics[width=\textwidth]{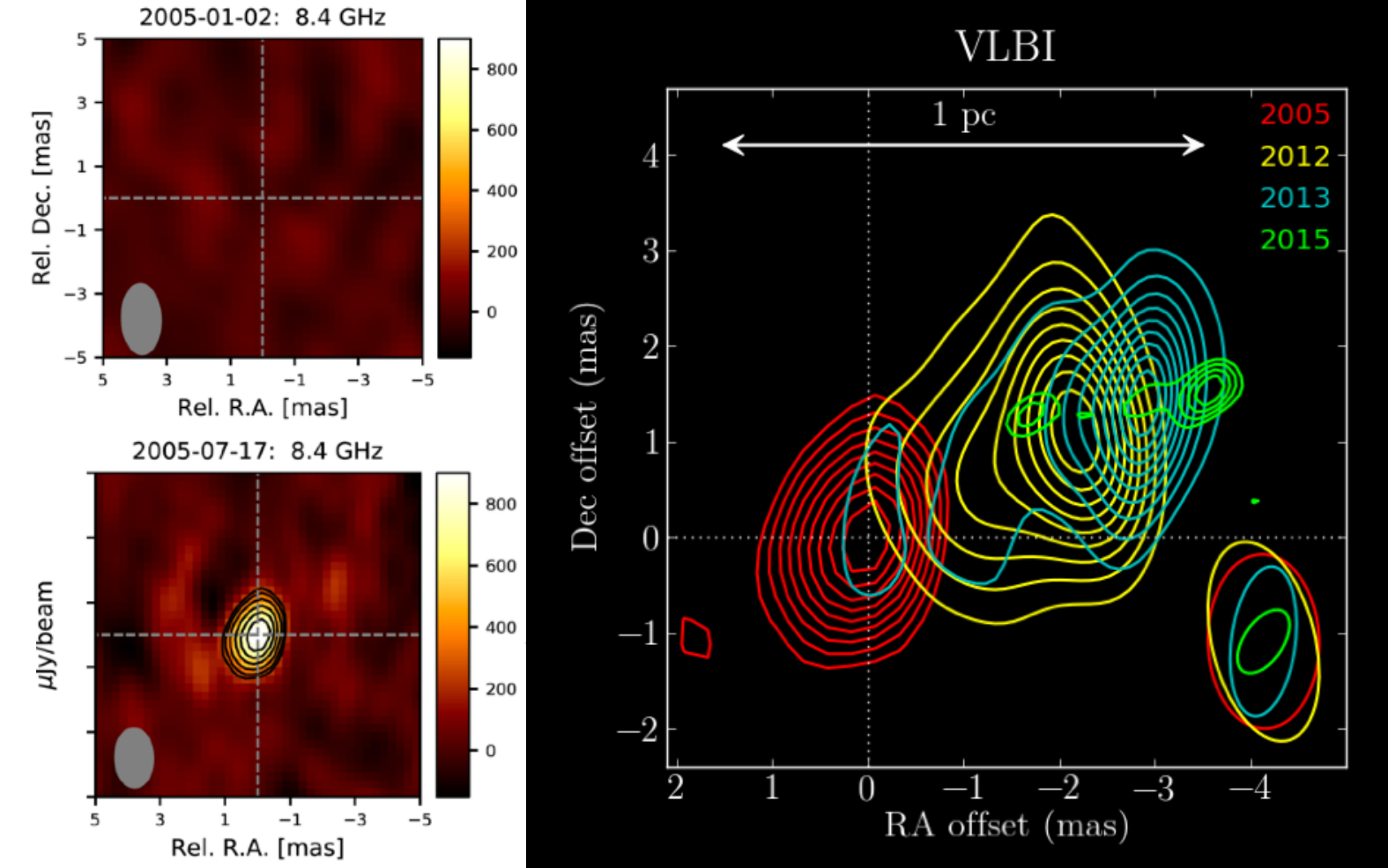}
\caption{\textbf{Top.} The evolution of the IR SED of Arp 299-B AT1. The SED of the Arp 299 nucleus B before the outburst (left) and at 734 days after the first IR detection (right) shown together with radiative transfer models for a starburst component (dashed line), an AGN dusty torus (dotted line), a polar dust component (thick solid line), and the sum of these components shown as a thin solid line. \textbf{Bottom.} The first observations of a resolved radio jet in a TDE by VLBI. The initially compact 8.4\,GHz radio source develops into an expanding jet structure as a fraction of the accretion power is channelled into a relativistic outflow. Figures from \citet{mattila2018}.}
\label{fig:Arp299B-AT1}
\end{figure}

Measuring the evolution of core-collapse SN rate as a function of redshift can provide an important consistency check for the cosmic star formation history. However, the measurements (e.g., \citealt{dahlen2012}) have been based on rest-frame optical observations and the number of SNe missed at these wavelengths in LIRGs and ULIRGs at different redshifts needs to be corrected for. For example, \citet{horiuchi2011} found the cosmic core-collapse SN rate to be a factor of two lower than that predicted from the cosmic SFR. Even in the local Universe, the SN population of LIRGs and ULIRGs is not yet well characterised due to the combination of large dust extinctions and difficulties in detecting the often faint sources against the bright and complex nuclear background. Furthermore, the fraction of the star formation (and hence also SNe) hidden from optical observations in LIRGs and ULIRGs increases rapidly toward redshift $z \sim 1$ (see Fig.~\ref{fig:magnelli11}). The fraction of SNe missed by rest-frame optical observations as a function of redshift has been studied previously by \citet{mannucci2007} and more recently by \citet{mattila2012} (see Fig.~\ref{fig:IR_SNe}). In the latter study the authors estimated that up to 83\% of the SNe in the local LIRGs (such as Arp 299) and close to 100\% in the local ULIRGs (such as Arp 220) have been missed by the previous searches at optical wavelengths. Adopting the contributions of LIRGs and ULIRGs to the co-moving SFR density from \citet{magnelli11} and these fractions of missing SNe in LIRGs and ULIRGs they estimated that the fraction of SNe missed by optical volumetric surveys increases from a local value of $\sim$19\% to $\sim$38\% at $z = 1.2$ and then remains roughly constant up to $z = 2$. \citet{dahlen2012} reported core-collapse SN rates from deep HST/ACS observations in $I$-band. Making use of the correction factors for the missing SNe they found that the measured CCSN rates between $z = 0$ and 1.3 are consistent with those expected from the cosmic SFR.

\subsection{Supermassive black hole induced variability}

\subsubsection{Active galactic nuclei} AGN are powered by accretion onto the
supermassive BH and show variability on a wide range of time scales. Well known examples
of nearby LIRGs hosting an AGN and showing significant variability from hard X-rays
\citep{Nardini2017} to UV \citep{Sukanya2018} to near-IR \citep{koshida2014} include NGC
6240, NGC 1068 and NGC 7469. In particular, time lag measurements between the flux
variations in the $V$- (optical continuum emission from the accretion disk) and $K$-band
(thermal re-radiation by hot dust within the torus) has allowed mapping the innermost
radii of the dusty tori in a number of galaxies including also LIRGs (e.g.,
\citealt{koshida2014}). 

\subsubsection{Tidal disruption events}
In contrast to other accretion-induced variability, in a stellar tidal disruption event (TDE) a star is torn apart by the tidal forces close to the supermassive BH, generating a bright flare of X-ray, UV and optical radiation (e.g., \citealt{rees1988,komossa2015}). 
Recently, N body simulations of major mergers of galaxies have predicted an increased TDE rate by two orders of magnitude compared to isolated galaxies \citep{Li2019} for a period of $\sim$10 Myr. Therefore, a strongly increased rate of detection of TDEs in LIRGs can be expected since $\sim$50\% of LIRGs are found to be mergers (Sect.~\ref{sec:intro}) with two or more nuclei. However, the large dust extinctions mean that most of these TDEs are likely to be missed by the searches working in the UV-optical domain, similar to the case of core-collapse SNe discussed above.

TDE candidates occurring in relatively dust free nuclear environment are already quite
routinely discovered by optical wide field surveys (e.g., \citealt{vanvelzen2020}).
However, TDEs occurring within the nuclear regions of dusty galaxy mergers and LIRGs
have remained elusive despite their high predicted rates in such environments. Recent
studies have now identified TDEs also in the extremely dusty nuclei of LIRGs. An optical
TDE candidate was identified in the nucleus of the LIRG IRAS F01004-2237 by
\citet{tadhunter2017}. It was detected as a result of repeated optical spectroscopic
observations of a sample of 15 LIRGs over a period of 10 years. The authors find the
interpretation of the transient as a TDE the most plausible explanation to explain the
strong and variable broad He I 5876 and He II 4686 lines in their spectrum. Its optical
absolute peak magnitude of $M_V = -20.1$ is similar to those observed for previous
optical TDE candidates but its light curve decline is substantially slower. The optical
event in IRAS F01004-2237 was followed by a bright and slowly evolving IR echo, i.e.,
transient's UV-optical light absorbed and re-radiated in the IR by dust \citep{dou2017}.
Interestingly, the total energetics of the transient inferred from mid-IR observations
is above $10^{52}$ erg, similar to the case of the TDE in Arp 299. Based on the
interpretation of the transient in IRAS F01004-2237 as a TDE, \citet{tadhunter2017}
suggested that LIRGs may host TDEs at orders of magnitude higher rates ($10^{-2}$ TDE
LIRG$^{-1}$ year$^{-1}$) than observed in the general galaxy population
($10^{-5}$--$10^{-4}$ TDE LIRG$^{-1}$ year$^{-1}$). However, \citet{trakhtenbrot2019}
have identified also other transients similar to the one in IRAS F01004-2237 and find
their origin to be instead in longer-term events of intensified accretion into the
supermassive BH rather than in TDEs of stars.

\citet{mattila2018} reported the discovery and multi-wavelength follow-up of a transient
event coincident with the nucleus B1 of Arp 299 that was extremely luminous in the IR
($M_K {\rm (peak)} = -22.9$) and radio but remained elusive at optical and X-ray
wavelengths (see Fig.~\ref{fig:Arp299B-AT1}). The transient, dubbed Arp~299-B~AT1, was
discovered as a result of systematic monitoring of a small sample of nearby LIRGs in the
near-IR $K$-band for SNe. This event turned out to be extremely energetic (above
10$^{52}$ erg radiated in the IR) and evolved slowly over 10 years of observations.
Arp~299-B~AT1 was found consistent with an extremely energetic TDE based on the
properties of a resolved radio jet detected thanks to deep, high spatial resolution VLBI
observations. The IR SED, and its evolution, for Arp~299-B~AT1 were found to be
consistent with absorption and re-radiation of the TDE's UV/optical light by dust in the
polar regions of the AGN torus (see Fig.~\ref{fig:Arp299B-AT1}). 

More recently, \citet{kool2020} reported the discovery and multi-wavelength follow-up of
an extremely luminous ($M_K{\rm (peak)} = -22.7$) nuclear transient in the LIRG IRAS
23436+5257. The transient, AT 2017gbl, was luminous in the IR and radio but much fainter
in the optical due to substantial dust extinction. 
The authors found a TDE by the supermassive BH at the centre of the northern nucleus of
IRAS 23436+5257 the most plausible scenario given the combined IR and radio evolution of
AT 2017gbl. \citet{mattila2018} and \citet{kool2020} suggested that Arp 299-B~AT1 and
AT~2017gbl might be just the tip of the iceberg of a missed population of TDEs: many
similar events may have occurred but remained hidden within the dusty nuclei of LIRGs.
Based on the detection of AT~2017gbl as a result of their AO-assisted near-IR monitoring
of $\sim$40 nearby LIRGs, \citet{kool2020} estimated the rate of AT 2017gbl-like events
in LIRGs to be $10^{n}$ TDE LIRG$^{-1}$ year$^{-1}$ where $n=-1.9^{+0.5}_{-0.8}$.
Furthermore, this population could have been far more numerous at high redshifts where
LIRGs and ULIRGs were more common (\citealt{magnelli11}; see Fig.\ref{fig:magnelli11}). 

\subsection{Lessons learned from time domain observations} High spatial resolution radio
and near-IR observations have been the critical breakthrough in revealing SNe in LIRGs.
These SNe have remained largely undetected by observations in the optical and even in
near-IR $K$-band under natural seeing conditions. It has become clear that both AO
assisted near-IR and high spatial resolution interferometric radio observations
(unaffected by dust extinctions) are necessary in order to provide a complete picture of
the SN activity in these dust obscured regions.

Recent studies have revealed both prolific SN factories and a previously unobserved
population of extremely IR luminous and energetic transients likely associated with
tidal disruption of stars within the nuclei of LIRGs. Such transients can outshine the
entire galaxy nucleus at $\sim$2--5 $\mu$m range where hot dust dominates over the
quiescent LIRG emission over a significant period of time. Therefore, LIRGs and ULIRGs
clearly cannot be considered to remain constant in flux, with a risk of
misinterpretation of the data if such outbursts are frequently present there.

The implied very high rates of core-collapse SNe and TDEs in LIRGs are of obvious great
interest for studies of galaxy formation and evolution. While TDEs can provide an
important means of feeding the supermassive BHs, both TDEs and SNe can have an important
role in driving outflows and regulating star formation if they are common within the
LIRG nuclei and the resulting injections of large energies are efficiently coupled to
the surrounding gas.

\section{A case study of the nearby LIRG Arp 299} 
\label{sec:casestudies}

\begin{figure}
  \includegraphics[scale=0.65]{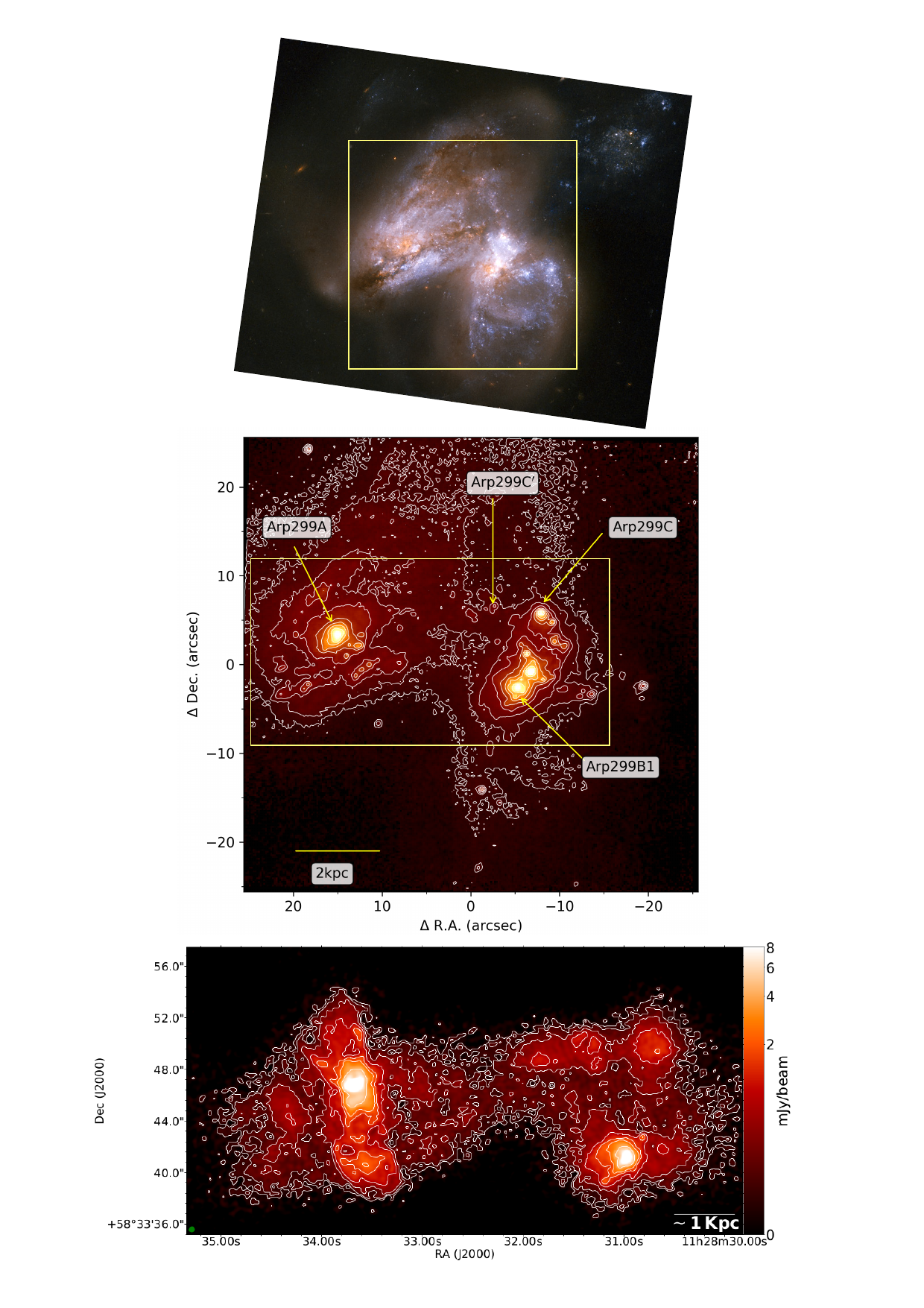}
  \caption{Optical, IR and radio images of the merger system Arp~299.
    \textbf{Top.} Colour-composite HST/ACS optical ($\lambda \approx 435+814 nm$) image showing a complex
    system of dust lanes and compact high surface brightness `knots'. \textbf{Middle.}
    HST/NICMOS near-IR image at $\lambda \approx 1.6\,\mu{\rm m}$, which traces mostly the stellar emission
    (see Sect.~\ref{subsec:stellar}). The main IR-emitting sources in the system are
    marked. Figure adapted from \cite{almudena00}.  \textbf{Bottom.} 150 MHz LOFAR image
    at an angular resolution of $\sim 0.45$ arcsec, which traces mostly the non-thermal
    synchrotron radio emission of the system (see Sect.~\ref{sec:radio-continuum}).
The peak of brightness corresponds to the nucleus of
Arp~299-A (the brightest knot to the left), whose powerful starburst activity is driving
a nuclear starburst-driven outflow \citep{ramirez18}. Image courtesy of Ram\'irez-Olivencia. }
  \label{arp299a-nicmos}
  \label{fig:casestudy}
\end{figure}

Arp~299 is one of the nearest examples of a LIRG with an IR luminosity almost qualifying
it as a ULIRG. However, it is particularly well suited as a case study example for
hosting a wide range of phenomena discussed throughout this review.  It consists of two
interacting galaxies that are in an intermediate merger stage \citep{Haan2011} and whose
nuclei are separated by approximately 5\,kpc (Fig.~\ref{fig:casestudy}). It has both
highly obscured nuclear starbursts with their associated SN factories, and less obscured
circumnuclear SF on much larger scales. It also hosts a Compton-thick AGN in the Western
component, observed directly in hard X-rays \citep{dellaceca02, ballo04}. Its proximity
allows for detailed studies of the link between the merger process, nuclear obscuration,
molecular gas distribution and gas properties and supermassive BH accretion and
feedback. Finally, Arp 299 is clearly above the main sequence of star forming galaxies
in the local Universe (see Fig.~\ref{fig:pereira}), similar to starbursting ULIRGs at $z
\sim 1$--$2$. Furthermore, \citet{almudena09} have shown the integrated mid-IR spectrum
of the whole Arp~299 system to resemble those of $z \sim 1.5$ ULIRGs making it also a
useful laboratory of the conditions where the bulk of SF took place at high-$z$. In this
section we aim to summarise some of the most important findings on Arp~299, and discuss
what we can learn on LIRGs in general from these findings, rather than attempting to
give a complete overview of the literature.

Arp~299 has been the target of a large number of extensive studies for almost four
decades already. \citet{Gehrz1983} first noticed that the Arp~299 system consisted of
two distorted spiral galaxies that were interacting so closely as to
overlap\footnote{There has been some confusion with the naming of the galaxies that make
  up the Arp~299 system.  \citet{HerreroPhDT} gives a detailed account of this ``tale of
  confusion''. Arp~299-A has often been incorrectly called IC~694, although
  \citet{yamaoka98} had already pointed out that IC 694 was actually a small, unrelated
E/S0 galaxy located to the northwest of the merging system of galaxies, while NGC 3690
refers to entire system. In this review, we have adopted the name Arp~299 (rather than
NGC 3690) to refer to the entire system.}.  They presented the first mid-IR and radio
continuum maps of Arp~299, and identified the distinct regions A, B, C and C$^\prime$,
which dominated the emission at these wavelengths (Fig.~\ref{arp299a-nicmos} and top
panel of Fig.~\ref{fig:arp299a-vlbi}). By inspecting the IR and radio flux densities and
SEDs, \citet{Gehrz1983} interpreted that the emission from these regions arose from the
effects of large numbers of massive stars in a starburst requiring $\gtrsim
10^{9}\,M_\odot$ of massive stars to have been formed during the starburst episode,
which resulted in several SNe per year. They also found that the compact, flat-spectrum
radio source in region A could not be explained by a starburst similar to those in the
other regions, and suggested that A would be the actual galaxy nucleus of Arp~299.
\citet{telesco85} presented the first near-IR maps of the system and noted the near-IR
nuclei to be substantially offset from the centers of the optical emission. They found
the near-IR colours of these regions to be among the most extreme observed for
extragalactic sources, which required a combination of effects that included
interstellar extinction and emission of hot dust.  Later near-IR studies with higher
spatial resolution were able to see the region B breaking up into two components, B1 and
B2, with the former one being coincident with the mid-IR and radio continuum emission
peaks and the latter one coinciding with the apparent optical nucleus
\citep{WynnWilliams1991, Satyapal1999, almudena00}.

\subsection{Molecular gas in Arp~299}

Arp~299 is very rich in molecular gas \citep[e.g.,][]{sargent91,aalto97,sliwa12} with a
total molecular gas mass estimated to $8.5 \times 10^9$ \msun\ \citep{aalto97}. The
implied gas surface density in the central regions is huge, about 7 g\,cm$^{-2}$
for the Arp~299-A nucleus. The molecular gas is to a large degree concentrated on the
two nuclei, but there is still a respectable reservoir of molecular gas distributed on
larger scales which may be funnelled to feed activity at the final gravitational centre
once the two galaxies merge. In particular, molecular gas is concentrated also in the
region where the two disks are overlapping. The highest CO surface brightness emission
is however found in the two nuclei, and emission from dense gas tracers such as HCN
peaks here, in particular in the eastern nucleus \citep{aalto97, imanishi06}.

Lower resolution (single dish) observations reveal an elevated large $^{12}$CO/$^{13}$CO
1--0 line intensity ratio of $\gtrsim 20$ in Arp~299 \citep[e.g.,][]{aalto95,casoli92}.
Aperture synthesis imaging shows that the $^{12}$CO/$^{13}$CO 1--0 ratio varies within
the system, with ratios increasing towards the two nuclei, in particular in Arp~299-A
\citep{aalto97}. This may be caused by reduced line opacities in the $^{12}$CO and
$^{13}$CO lines due to high gas temperatures and/or large turbulent line widths. 
Herschel observations of the CO ladder \citep{rosenberg14} suggest higher gas
temperatures in Arp~299-A than in other regions of the merger. Line ratios will also be
impacted by the $^{12}$C/$^{13}$C abundance ratios (see Sect.~\ref{subsec:isotopes}).
\citet{falstad17} note that the starburst in Arp~299~A has not led to an enhancement of
$^{18}$O in the nucleus. This may be due to the IMF or evolutionary stage of the
starburst and/or that the inflowing gas affects the abundances.

Recent Herschel OH and H$_{\rm 2}$O studies suggest the presence of a compact
obscured nucleus in Arp~299-A \citep{falstad17}. A reversed P-Cygni profile implies that
molecular gas is infalling onto the nucleus, suggesting a state of rapid evolution. Herschel does not detect a molecular outflow, despite the presence of a starburst wind
seen in radio continuum emission (\citet{ramirez18} and Fig.\ref{fig:casestudy}). Very
recent NOEMA observations however reveal a dense and compact molecular outflow from
Arp~299-A. Emission from more complex molecules such as C$_2$H and HC$_3$N is also
found, together with vibrationally excited HCN, in the nucleus (Falstad et al in prep.).
Arp~299 hosts both OH megamaser and 22\,GHz H$_2$O maser emission, which is an unusual
combination \citep[e.g.,][]{baan90,polatidis01,tarchi11}.

\subsection{Extreme star formation activity across Arp~299}

The brightest component at mid-IR and radio wavelengths is the nuclear region of the
eastern galaxy Arp~299-A (Fig.~\ref{fig:casestudy}), which accounts for $\sim$50\% of
the total IR luminosity of the system \citep{charmandaris02, almudena00}, and
$\sim$70\% of its 5\,GHz radio emission \citep{neff04}. Numerous H~II regions and
relatively young star clusters populate the system near the star-forming regions, which
implies that star formation has been occurring at a high rate for the past $\sim$10 Myr
\citep{almudena00, Randriamanakoto2019}. Given that Arp~299-A accounts for a large
fraction of the IR emission in Arp~299, it is the region that is most likely to contain
new SNe. 

The nuclear region of the western component Arp~299-B1+B2 (Figure~\ref{arp299a-nicmos})
accounts for approximately 20\% of the IR luminosity of the system
\citep{charmandaris02}. There is compelling evidence for the presence of an AGN at
Arp~299-B1 (see next section). Nevertheless, on scales of up to a few kpc this nucleus
also appears to be surrounded by H~II regions as well as young and relatively evolved
star clusters detected with HST and AO-assisted ground-based observations
\citep[see][]{almudena00, AlonsoHerrero2002,Randriamanakoto2019}. 

At the interface between the two galaxies, Arp~299-C and Arp~299-C$^\prime$ (located
approximately 2\,kpc to the north of Arp~299-B1) are extended regions undergoing young
and vigorous star forming activity, probably as a result of the interaction between the
two galaxies. There is also evidence of significant amounts of dust heated by a young
star cluster or clusters, with some indications of the presence of ultracompact H~II
regions \citep{almudena09}. Observations with high angular resolution at different
wavelengths reveal that the Arp~299-C+C$^\prime$ complex is resolved into a number of
H~II regions and star clusters \citep{aalto97, almudena00, AlonsoHerrero2002,
soifer2001, Randriamanakoto2019}. Overall, the Arp~299-C+C$^\prime$ region contributes
nearly 30\% of the IR luminosity of this system \citep{charmandaris02} and contains a
large fraction of the youngest star clusters detected in this interacting system
\citep{Randriamanakoto2019}.

\subsection{Evidence for AGNs in Arp~299-A and Arp~299-B1}
\label{agn-sb}

The dusty nuclear regions of LIRGs can be heated by an intense starburst or an AGN, or a
combination of both. Disentangling the contribution of the starburst and AGN can be
achieved using radiative transfer modelling of the IR SED (see
Sect.~\ref{sec:sed-modelling}), direct imaging of the innermost few pc at radio
wavelengths, using very high-angular resolution VLBI observations (see
Sect.~\ref{sec:radio}), or by hard X-ray observations.

Contemporaneous VLBI imaging of the innermost central 8-pc region of Arp~299-A with the
EVN at 1.7 and 5.0\,GHz has demonstrated the existence of several compact components,
forming a complex. \citet{pereztorres10} showed that the morphology, spectral index,
radio luminosity, and radio-to-X ray luminosity ratio of this complex was consistent
with that of a low-luminosity active galactic nucleus (LLAGN), which ruled out the
possibility that it consists of a chain of young radio SNe and SN remnants in a young
super star cluster. \citet{pereztorres10} therefore concluded that A1 is the long-sought
AGN in Arp~299-A.

A Compton-thick AGN in the B1 nucleus has been detected directly with hard X-ray
observations collected by the BeppoSAX observatory in 2001 \citep{dellaceca02, ballo04}.
The estimated X-ray luminosity for the AGN in the 0.5--100\,keV range was $1.9\times
10^{43}\,{\rm erg\,s}^{-1}$ \citep{ballo04} and its bolometric luminosity $\sim 5.5
\times 10^{43}\,{\rm erg\,s}^{-1}$ \citep{PereiraSantaella2011}. A column density of
$\sim 3 \times 10^{24}\,{\rm cm}^{-2}$ was derived from the hard X-ray observations
towards the AGN \citep{dellaceca02}. The presence of an AGN in the B1 nucleus of Arp~299
was also suspected from a high excitation ionization cone seen in optical IFU
observations \citep{garcia06}.

Recently, the AGN in Arp~299-B1 made itself known also through a tidal disruption of a
star by the supermassive BH launching a radio jet and illuminating polar dust clouds,
making them shine in the IR \citep{mattila2018}.  Furthermore, \citet{mattila2018}
reported compact 2.3\,GHz emission detected with VLBI in January 2005, which was
spatially very close to the VLBI position of Arp~299-B AT1 but before its radio
detection. They favoured this pre-outburst 2.3\,GHz emission to correspond to the core
of the AGN in Arp~299-B1. In this scenario, the initial 8.4\,GHz emission observed for
Arp~299-B AT1 in July 2015 and shifted at 8.4\,GHz by a linear offset of $\sim$ 0.35 pc
with respect to the 2.3\,GHz pre-outburst peak can be explained by the well-known
core-shift effect which is not unexpected in an AGN.

Radiative transfer model fits to the Spitzer/IRS spectra and IR SEDs of Arp~299 have
been reported in \citet{mattila2012}, \citet{herrero-illana17} and \citet{mattila2018},
and are shown for Arp~299-B1+B2 in Fig.~\ref{fig:Arp299B-AT1}. For nucleus A, no AGN
component was required to fit the SED. However, for nucleus B1+B2 \citet{mattila2018}
found a 24\% contribution by an AGN to the total IR (8--1000\,$\mu$m) luminosity, in
good agreement with the findings of \citet{AlonsoHerrero2012, AlonsoHerrero2013}. High
angular resolution IR studies of Arp~299-B1 indicate that the AGN torus is seen almost
edge-on \citep{AlonsoHerrero2013} with an extremely high extinction of A$_{V}$ $\sim$
460 magnitudes towards the AGN which is also consistent with the large column density
derived from the hard X-ray observations.

Using high angular resolution mid-infrared spectroscopy with the CanariCam instrument on
the Gran Telescopio Canarias, \cite{AlonsoHerrero2013} showed that on scales of 120\,pc
Arp~299-A is deeply embedded in dust. They also detected strong nuclear star formation.
However, the decreased EW of the 8.6 and $11.3\,\mu$m PAH features compared to the
larger scale Spitzer/IRS spectra suggested the presence of a strong mid-infrared
continuum. Assuming that this dust is heated by an AGN, \cite{AlonsoHerrero2013}
estimated that the AGN in Arp~299-A would be about five times less luminous than that at
Arp~299-B1, in agreement with the findings by \citet{mattila2012}. 

\subsection{Extremely prolific supernova factory in Arp~299}

The IR luminosity of Arp~299 indicates (Eq. \ref{eq:ccsnrate_empirical}) a very high
core-collapse SN rate of $\sim$ 2 yr$^{-1}$, which is one of the highest expected in
local galaxies. This is supported by the frequent SN discoveries at optical and near-IR
wavelengths within the circumnuclear regions of the Arp~299 system. Over the last two
decades a total of five core-collapse SNe\footnote{SNe 2005U (type IIb), 2010O (Type
Ib), 2010P (Type IIb), 2019lqo (Type IIb), and 2020fkb (Type Ib); note that SN 2018lrd
reported close to the position of SN 2020fkb has been identified as a false detection
due to image subtraction residuals (Ting-Wan Chen, private communication).} have been
reported in Arp~299. \citet{kankare2014} presented a detailed optical and near-IR study
of SNe 2010O and 2010P that occurred in the circumnuclear regions of Arp~299-A and
within the region C', respectively. These two core-collapse SNe of types Ib and IIb were
found to have exploded by chance within only a few days of one another and suffered from
host galaxy extinctions of A$_{V}$ $\sim$ 2 and 7 mag, respectively. Furthermore,
Kankare et al. (submitted) studied the type IIb SN 2019lqo with A$_{V}$ $\sim$ 2 mag and
the type Ib SN 2020fkb with A$_{V}$ $\sim$ 1 mag, which were discovered in the central
regions of components Arp~299-A and Arp~299-C, respectively. Although such detailed
studies of the SNe occurring within the circumnuclear regions are possible, the optical
and near-IR observations are likely to miss a significant fraction of the SNe occurring
within the innermost $\sim$200 pc nuclear regions of Arp~299-A and Arp~299-B due to the
very high values of extinction and the lack of the necessary angular resolution. The
detection of SNe within these strongly obscured nuclear regions is very challenging even
with AO assisted observations at near-IR wavelengths, and requires radio VLBI
observations.

Radio VLBA observations of Arp~299-A carried out during 2002 and 2003 resulted in the
detection of five compact sources \citep{neff04}, one of which (A0) was identified as a
young SN. Further VLBI observations of the nuclear regions of Arp~299-A with the EVN and
the VLBA resulted in the discovery of a large population of radio SNe and SN remnants
\citep{pereztorres09a, ulvestad09, bondi12} (see also Fig.~\ref{fig:arp299a-vlbi}).
\citet{bondi12} obtained a core-collapse SN rate $\gtrsim $0.8\, yr$^{-1}$ for Arp~299-A
based on a three-year long VLBI monitoring. Thus, radio observations at high angular
resolution and high sensitivity are currently the only viable way of directly detecting
core-collapse SNe in these strongly obscured regions with the potential of providing,
independent of any model, the core-collapse SN and star formation rates (see Sect.
\ref{sec:timedomain}). Furthermore, it is interesting to note that high spatial
resolution radio observations do not support the SN population within the innermost
nuclear regions of Arp~299-A differing from the core-collapse SNe observed in normal
field galaxies.

The component A0 (see Fig.~\ref{fig:arp299a-vlbi}) was also detected by
\citet{pereztorres10} at 5\,GHz, but not at 1.7\,GHz, from contemporaneous observations
in 2009. This implies the existence of a foreground absorbing H~II region, which
precludes its detection at frequencies $\lsim$1.7 GHz due to the large particle density
of this foreground screen. It is remarkable that the radio SN exploded at a mere
(projected) distance of two parsecs from the putative AGN in Arp~299-A, which makes this
SN one of the closest to a central supermassive black hole ever detected. This result
may also be relevant to accreting models in the central regions of galaxies, since it is
not easy to explain the existence of very massive, SN progenitor stars so close to an
AGN. While seemingly contradictory, this could explain the low-luminosity of the AGN in
Arp~299-A.  In fact, since massive stars shed large amounts of mechanical energy into
their surrounding medium, thereby significantly increasing its temperature, those
massive stars might hinder the accretion of material to the central black hole, which
could in turn result in a less powerful AGN than usual. 

Summarizing, Arp~299-A has long been believed to be a pure starburst likely hosting a
LLAGN. For Arp~299-B1 there is now clear evidence of a relatively luminous AGN from
radio, IR and hard X-ray wavelengths. These results suggest that both starburst and AGN
are frequently associated phenomena in galaxy mergers. However, disentangling between
these two can be very challenging and requires extremely detailed observations at
different wavelengths and over multiple epochs.

\section{Summary and outlook} 
\label{sec:summary}

The facilities available for IR observations over the previous years have offered an
extremely useful combination of high angular resolution thanks to ground-based AO
assisted instruments (in the near- and mid-IR) and HST (in the near-IR) with the broad
spectral coverage of the other space-based facilities (e.g., AKARI, Spitzer, and
Herschel).  In particular, high angular resolution imaging studies with the HST and AO
assisted facilities in the near-IR have revealed a large population of (super) star
clusters in LIRGs, some of which are completely obscured in the UV and optical. 
A number of mid-IR diagnostics including the presence of high excitation emission lines
(mostly the mid-IR [Ne~V] lines) together with other optical and X-ray indicators have
revealed that the majority of local LIRGs host an AGN. However, in general the AGN of
local LIRGs do not account for a large fraction of the IR emission.
Future space-based IR facilities are expected to provide a leap forward in studies of
dust obscured star formation in the local Universe and more importantly at high
redshifts,
in particular JWST to be launched in 2021.
Furthermore, the Multi-AO Imaging Camera for Deep Observations
(MICADO\footnote{\url{http://www.mpe.mpg.de/ir/micado}}) on the ESO Extremely Large
Telescope (ELT; becoming available around 2025) is expected to provide a point-source
sensitivity that is comparable to JWST but a spatial resolution about a factor of six
better at near-IR wavelengths. This very substantial leap in angular resolution will
allow detailed studies of the circumnuclear star formation and nuclear activity that is
currently possible only for the most nearby LIRGs well beyond the local Universe. The
Roman Space Telescope (RST\footnote{\url{https://www.stsci.edu/roman}}) working at
0.5-2\,$\mu$m will provide HST-like spatial resolution but over a field of view about
100 times larger. It is planned to be launched in the mid-2020s and will provide HST
quality images for the millions of LIRGs and ULIRGs discovered by the IR and submm
surveys allowing the study of their morphology and physics, also helping with
cross-matching them with optical sources and providing photometric redshift estimates.
It is important to note that a significant fraction of submillimetre galaxies are
currently undetected in the optical. The SPace Infrared-telescope for Cosmology and
Astrophysics (SPICA\footnote{\url{https://spica-mission.org/}}), is another next
generation IR (12-350\,$\mu$m) observatory that, if built, will likely be the only
facility capable of tracing the evolution of the obscured star-formation rate and
black-hole accretion rate densities from the peak of their activity back to the
reionisation epoch (i.e., $3 \lesssim z \lesssim 6$--$7$).

Radiative transfer models for the IR emission as well as the methods for comparing the
models with the observed SEDs have improved significantly in recent years.  Further
refinements will focus on improving the self-consistency of these models and in treating
better the geometry of the emitting and absorbing regions in galaxies. For example, a
number of complex models for the AGN torus are currently available which deal with a
smooth, clumpy or 2-phase geometry. Similar methods should be developed for a compact
starburst consisting of an ensemble of GMCs centrally illuminated by young stars or for
a model which treats the combined emission and absorption of a starburst and an AGN
torus.
Methods that fit the SED of a part of a galaxy will also be extremely useful in the JWST
era in order to take advantage of its constraining power to disentangle the
star-formation and AGN contributions.

Radio continuum emission covering $\lesssim$1.0--100\,GHz ($\sim$30\,cm--3\,mm) is
powered by a mixture of thermal free-free (bremsstrahlung), non-thermal synchrotron, and
thermal dust emission and disentangling those components in LIRGs is a challenging task,
and requires multi-frequency data at high-spatial resolution (Sect.
~\ref{sec:radio-continuum}).  The use of VLBI at cm-wavelengths provides pc-scale
resolution for galaxies up to a distance of $D \sim$ 250 Mpc, allowing the
identification of in particular the AGN and SN factories in the local universe.
Currently, continuum radio surveys of nearby LIRGs at high spatial resolution have have
been done with the VLA (e.g., GOALS),
e-MERLIN\footnote{\url{http://www.e-merlin.ac.uk/}} and the EVN (e.g., the Luminous
Infrared Galaxy Inventory; LIRGI\footnote{\url{http://lirgi.iaa.es/}}).
Future radio continuum surveys to be undertaken with the SKA (becoming available in
mid-2020s) and its precursors (MeerKAT\footnote{\url{https://www.sarao.ac.za/}} and
ASKAP\footnote{\url{https://www.atnf.csiro.au/projects/askap/index.html}}), covering the
frequency range of $\sim$(350\,MHz--3\,GHz; i.e., 90\,cm--10\,cm),
will provide $\mu$Jy sensitivities with exquisite image fidelity over a wide range of
spatial scales for all nearby galaxies. 
If SKA will be extended in frequency up to 33 GHz [and/or the next generation Very Large
Array (ngVLA\footnote{\url{https://ngvla.nrao.edu/}}) is realized], rest-frame
observations at this frequency band may act as a reliable method to measure the SFRs of
galaxies at increasingly high redshift, without the need of ancillary radio data to
account for the non-thermal emission. 

LIRGs are often very rich in molecular gas, which feeds star formation and AGN activity.
Tracer lines are required to probe the mass, dynamics and distribution of the molecular
gas but other methods are emerging, for example the 850 $\mu$m dust continuum, and fine
structure lines of atomic carbon. High-resolution CO studies of LIRGs with
mm-interferometers such as SMA, NOEMA and ALMA, have shown a variety of morphologies and
dynamics in LIRGs.  The number of available mm/sub-mm diagnostic lines is steadily
increasing, which permits to study the chemistry of molecular gas.  For example, we can
track how starbursts evolve in different environments, as well as separate a buried AGN
from embedded star formation.  Another important signature of nuclear activity in LIRGs
is outflows. Winds and feedback often occur in the form of molecular outflows that carry
large amounts of cold molecular gas out from the centre of the LIRG.  Since molecular
outflows may evict significant masses of gas out of the galaxy, or the gas may return to
the system to fuel another growth spur, they are important to our understanding and
modelling of galaxy evolution.  New mm/submm instruments will allow us to study the
physical conditions and even chemistry in the outflows, important in determining the
mass outflow rates, the origin and ultimate fate of the gas in the outflow as well as
which mechanism is driving the outflow.

Time domain studies of LIRGs have revealed powerful SN factories within their
circumnuclear and innermost nuclear regions.  Furthermore, most LIRGs are known to
harbour at least one supermassive BH in their nuclei and variability linked to the
accretion onto this BH is expected, in some cases also caused by tidal disruption of a
star by the BH.  While TDEs can provide an important means of feeding the supermassive
BHs, both TDEs and SNe can have an important role in driving outflows and regulating
star formation within the LIRG nuclei.  Such luminous transients can also easily
outshine the entire LIRG nucleus at IR and radio wavelengths which clearly cannot be
considered to remain constant in flux, with a risk of misinterpretation of the data
obtained at a single epoch of observation.
It appears that the SN properties in LIRGs
are consistent with those observed in normal galaxies indicating that
the IMF of the progenitor stars
appears to be consistent with a Salpeter-like IMF observed for stars in normal galaxies.
However, these results are still based on a rather limited sample of SNe discovered in a
small number of nearby LIRGs and also directly measured SN rates in LIRGs remain
uncertain. Furthermore, recent observations have suggested that LIRGs may host also TDEs
at orders of magnitude higher rates than observed in the general galaxy population. Many
of these events are likely to suffer from a substantial amount of dust extinction,
remaining hidden within the dusty nuclei of LIRGs. This population could be far more
numerous at higher redshifts, where LIRGs and ULIRGs are more common. The JWST and later
on RST and also ELT/MICADO will be very well suited to constrain the rates and
properties of SNe and TDEs in LIRGs also beyond the local Universe. Furthermore, the
future availability of wide-field VLBI together with the increase in sensitivity will
allow much more efficient studies of radio SNe and TDEs in local LIRGs, especially with
the advent of the SKA in the next years.

The future observational facilities will allow a leap forward in the IR and radio
studies of star formation and nuclear activity in LIRGs, near and far, shedding light on
how galaxies formed and evolved through cosmic time.  They will produce a complete
census of star-formation and AGN activity as a function of galaxy mass, morphology and
spectral type, black-hole mass and luminosity, becoming the cornerstone of
multi-wavelength studies of the local Universe.

\begin{acknowledgements}
We thank an anonymous referee for many useful comments and suggestions that have significantly improved our manuscript. We also thank Stuart Ryder for a careful reading of the whole paper, and Duncan Farrah, Erkki Kankare, Eskil Varenius and Na\'im Ram\'irez-Olivencia for comments on different sections of the paper. We also thank Na\'im Ram\'irez-Olivencia for the LOFAR image of Arp~299 in Fig.~\ref{arp299a-nicmos}, in advance of publication, and  Miguel Pereira-Santaella, Eric Murphy, and Eskil Varenius for producing new versions of their original figures (Figs.~\ref{fig:pereira}, \ref{fig:radio-sed-FIRC}-left, and \ref{fig:RSN}-bottom, respectively).

MPT acknowledges financial support from the State Agency for Research of the Spanish
MCIU through the Center of Excellence Severo Ochoa award to the Instituto de
Astrof\'isica de Andaluc\'ia (SEV-2017-0709) and through grant PGC2018-098915-B-C21
(MCI/AEI/FEDER, UE).  SM acknoweledges support from the visitor and mobility program of
the Finnish Centre for Astronomy with ESO (FINCA), funded by the Academy of Finland
grant nr 306531. AA-H acknowledges support through grant PGC~2018-094671-B-I00
(MCIU/AEI/FEDER, UE). AA-H work was done under project No.~MDM-2017-0737 Unidad de
Excelencia ``Mar\'{\i}a de Maeztu'' - Centro de Astrobiolog\'{\i}a (INTA-CSIC).  SA
gratefully acknowledges support from an ERC Advanced Grant 789410 and from the Swedish
Research Council.  AE acknowledges support from the Cyprus Research \& Innovation
Foundation (GRATOS; EXCELLENCE/1216/0207) and the European Space Agency (CYGNUS; ESA
contract 400126896/19/NL/MH). 

\end{acknowledgements}

\bibliographystyle{spbasic}      

\bibliography{lirgs}   

\printindex

\end{document}